\documentclass[12pt]{article}

\usepackage{verbatim, color, amssymb, amsmath}					
\usepackage{amsthm}		
\usepackage{bm}
\usepackage{dsfont}			
\usepackage{multirow}
\usepackage[mathscr]{euscript}
\usepackage{fancyhdr}
\usepackage{enumitem}
\usepackage{graphicx}
\usepackage{geometry}
\usepackage{xcolor}
\usepackage[hidelinks]{hyperref}
\usepackage{bbm}
\usepackage[section]{placeins}

\usepackage{tabularx, booktabs}
\newcolumntype{Y}{>{\centering\arraybackslash}X}

\usepackage{subfig}
\usepackage{float}
\usepackage{lineno}

\usepackage{setspace}
\usepackage{tikz}
\usetikzlibrary{arrows}
\usepackage{wrapfig}

\usepackage{natbib}
\usepackage{lscape}

\newtheorem*{Proof*}{Proof}

\def\bbR{\mathbb{R}}
\def\bbE{\mathbb{E}}

\def\N{{\cal N}}

\def\var{\hbox{var}}

\def\diag{\hbox{diag}}

\def\wh{\widehat}
\def\wt{\widetilde}

\def\diag{\hbox{diag}}
\def\log{\hbox{log}}

\def\trans{^{\rm T}}

\def\th{^{th}}

\def\Ga{\hbox{Ga}}

\def\Normal{\hbox{Normal}}

\def\wh{\widehat}

\def\b1f{{\mathbf f}}

\def\bH{{\mathbf H}}

\def\bW{{\mathbf W}}

\newcommand{\etam}{\mbox{\boldmath $\eta$}}

\newcommand{\btheta}{\mbox{\boldmath $\theta$}}

\renewcommand\footnoterule{\kern-3pt \hrule \textwidth 2in \kern 2.6pt}

\def\boxit#1{\vbox{\hrule\hbox{\vrule\kern6pt \vbox{\kern6pt \textcolor{blue}{#1}\kern6pt}\kern6pt\vrule}\hrule}}

\def\authorfootnote#1{{\let\thefootnote\relax\footnotetext{#1}}}

\allowdisplaybreaks

\begin{document}
	\thispagestyle{empty}
	\baselineskip=28pt
	
	\begin{center}
		{\LARGE{\bf
				Bayesian Mixed Multidimensional Scaling\\ for Auditory Processing
		}}
	\end{center}
\baselineskip=12pt
\vskip 2mm

\begin{center}
		Giovanni Rebaudo$^{1}$ (giovanni.rebaudo@unito.it)\\
        Fernando Llanos$^{2}$ (fllanos@utexas.edu)\\
        Bharath Chandrasekaran$^{3}$ (bchandra@northwestern.edu)\\
        Abhra Sarkar$^{4}$ (abhra.sarkar@utexas.edu)
		\vskip 3mm
		$^{1}$ESOMAS Department, University of Torino and Collegio Carlo Alberto
		\vskip 4pt
		$^{2}$Department of Linguistics, University of Texas at Austin
		\vskip 4pt 
		$^{3}$Department of Communication Science and Disorders, Northwestern University
		\vskip 4pt $^{4}$Department of Statistics and Data Sciences, University of Texas at Austin
\end{center}
\vskip 8mm
\begin{center}
	{\Large{\bf Abstract}} 
\end{center}

The human brain distinguishes speech sounds by mapping acoustic signals into a latent perceptual space.
This space can be estimated via multidimensional scaling (MDS), preserving the similarity structure in lower dimensions.
However, individual and group-level heterogeneity, especially between native and non-native listeners, remains poorly understood.
Prior approaches often ignore such variability or cannot capture shared structure, limiting principled comparisons.
Moreover, the literature often focuses on latent distances rather than the underlying features themselves.
To address these issues, we develop a Bayesian mixed MDS method that accounts for both subject- and group-level heterogeneity, allows for the recovery of unique, identifiable latent features, facilitating their biological interpretability, while also determining the effective dimensionality of the latent space in an automated, data-adaptive manner.
Simulations and an auditory neuroscience application demonstrate how these features reconstruct observed distances and vary with individual and language background, revealing novel insights.
\baselineskip=12pt
\vskip 8mm


\par\medskip\noindent
\underline{\bf Short/Running Title}: 
Bayesian Mixed Multidimensional Scaling


\clearpage\pagebreak\newpage
\pagenumbering{arabic}
\newlength{\gnat}
\setlength{\gnat}{16pt}  
\baselineskip=\gnat

\section{Introduction}
Understanding how the brain represents behaviorally relevant information such as speech is a fundamental question in auditory neuroscience.
Speech signals are inherently multidimensional, with subtle variations in frequency, intensity, and timing information distinguishing different categories \citep{caclin2005acoustic, caclin2006separate}.
These acoustic subtleties need to be robustly encoded by the auditory system and mapped onto existing speech representations (e.g., in native speakers) or emergent representations (e.g., in non-native learners).
A major goal for auditory neuroscientists is to decipher how these attributes are differentially represented in the brain and how these neural representations are shaped by different auditory experiences across the lifespan.
One critical step to elucidate these questions is to quantify the degree of similarity (or dissimilarity) between neural representations of sounds within the same sensory-perceptual representational space.
This step is critical because the degree of dissimilarity between neural representations of speech sounds must reflect linguistically relevant differences across languages.
For instance, native speakers of Mandarin Chinese rely on subtle syllable-level pitch changes to decode word meanings, exhibiting a more nuanced neural encoding of pitch contours (or lexical tones) than native speakers of languages that are not tonal \citep[e.g.,][]{feng2021distributed, raizada2010linking, krishnan2010effects, bidelman2011enhanced}.

Multidimensional scaling (MDS) \citep{torgerson1952multidimensional, torgerson1958theory, carroll1980multidimensional, borg2005modern} and its extension to individual difference scaling (INDSCAL) \citep{carroll1970analysis} are two dimensionality reduction techniques that have been extensively used to quantify the degree of similarity between neural representations of sensory stimuli captured with modern neuroimaging technologies.
Given data on dissimilarity matrices between pairs of stimuli (e.g., in our work, speech sounds), MDS and INDSCAL can extract lower-dimensional latent features from which a denoised version of the observed dissimilarities can be reconstructed.
In the classical literature, these quantities are usually described in geometric terms: the point coordinates of the stimuli in the low-dimensional space form the configuration, and the coordinate axes are called dimensions.
In our Bayesian formulation, we treat these coordinates as unobserved random variables endowed with a prior and a posterior distribution, and we refer to them as latent features to emphasize this latent-variable interpretation.
Throughout this paper, latent features therefore denote the coordinate vectors representing the stimuli in the low-dimensional space, endowed with a probability model.
More specifically, the latent features represent the original stimuli as points in a low-dimensional space such that the pairwise (usually Euclidean) distances computed in the reduced space (i.e., from the latent features) match the observed pairwise dissimilarities as well as possible.

Standard MDS methods, when applied to a single dissimilarity matrix, extract common latent features that allow the reconstruction of that matrix.
Since they are restricted to handling two-way pairwise dissimilarity data, when multiple subject-specific dissimilarity matrices are available, they are often reduced to a single group-level matrix via complete pooling (e.g., by averaging the distances element-wise) and analyzed with standard MDS, or analyzed by running separate MDS for each subject, followed by post-hoc alignment of the resulting configurations across subjects.
In contrast, INDSCAL can accommodate three-way data comprising pairwise dissimilarities for different individuals, allowing for the extraction of individual latent features and denoised distances by re-weighting the common latent features for each subject \citep{carroll1970analysis}.
INDSCAL can therefore be used to simultaneously achieve dimensionality reduction, data visualization, and, when appropriate, meaningful interpretation of the inferred latent features.
MDS-type models have also been extended to alternating least squares scaling \citep[ALSCAL,][]{takane1977nonmetric, young1978alscal} and proximity scaling \citep[PROXSCAL,][]{de1980multidimensional}, which minimize the standardized residual sum of squares \citep[STRESS,][]{shepard1962analysis, kruskal1964multidimensional, kruskal1964nonmetric}, a measure of scaled difference between observed distances and their model reconstructed values. Originally introduced for continuous data, such extensions made MDS more broadly applicable to include non-metric data as well as missing values.
MDS has gradually transformed from a descriptive geometric optimization method into a rigorous statistical inference framework, shifting the objective function from minimizing an empirical fit measure, like Kruskal's STRESS, to maximizing a formal likelihood function, which combines the underlying true distance in the latent MDS space and a random error term \citep{ramsay1977maximum, takane1981nonmetric}. Asymmetric Mahalanobis distance-based formulations have also been considered \citep{bradlow2000little}.
Furthermore, MDS-type models based on latent utility functions have been extensively developed for binary choice and ordered preference data for psychometric analysis in marketing applications. Here, the utilities are constructed using choice and individual-specific location vectors in the MDS latent space. Two main strategies exist for this construction: the `vector' approach uses the inner product, and the `unfolding' approach uses the squared Euclidean distances \citep[][etc.]{zinnes1977single, desarbo1989stochastic, desarbo1998bayesian, park2008hierarchical, desarbo2008model, fong2010bayesian}.

For the rest of the article, we focus mainly on MDS methods for metric-valued continuous data, aligning with the requirements of our motivating application.
More recently, such methods have seen widespread use in the neuroscience literature, becoming a staple technique over the last few decades.
The notable appeal of MDS in this field may lie in its ability to process distance or dissimilarity matrices directly, thereby obviating the often-involved step of complex modeling of the raw, high-dimensional data sets. 
\cite{mesgarani2014phonetic, di2015low, zinszer2016semantic, khalighinejad2017dynamic, feng2019role, feng2021distributed, llanos2021neural} have used MDS to align acoustic and neural representations of sounds in a common space for comparison.
In this body of literature, a higher degree of alignment between acoustic and neural representations is interpreted as a signature of a more robust neural encoding of stimulus signals.
Here, by ``alignment'' we mean the correspondence between the space in which acoustic stimuli can be represented and the latent space in which the brain represents them. 
These works suggest that the degree of alignment between neural and/or acoustic representations of sounds can be investigated using a reduced set of latent features derived from the primary acoustic cues used to perceive sounds.
For example, the perception of Mandarin tones seems to be based on two major acoustic dimensions: pitch height (high vs low pitch) and pitch direction (rising vs falling pitch) \citep{chandrasekaran2007neuroplasticity, gandour1978perception}.
Some work \citep{gandour1978crosslanguage, gandour1983tone} has also suggested a possibly relevant third dimension, namely the magnitude of the slope in the pitch contour.
In the last decades, MDS and INDSCAL have also become popular tools to assess the degree of structural alignment between acoustic and neural representations of speech sounds \citep{raizada2010linking, chandrasekaran2007neuroplasticity, chandrasekaran2007mismatch}.
Much of the attractiveness of these tools lies in their ability to represent global dissimilarity patterns between complex neural signals in a way that is easy to visualize and interpret.
While MDS is not exclusively conceived as a technique for dimensionality reduction, using a reduced number of dimensions, prior neuroscience work \citep{feng2021emerging, mesgarani2014phonetic, llanos2021neural} has been able to capture fine-grained linguistically-relevant differences in sound processing, 
which suggests that theoretical principles of dimensionality reduction inform the representation of speech sounds in the brain.

Current metric MDS approaches in neuroscience are, however, often used as mere visualization tools.
Indeed, joint modeling and uncertainty quantification across groups (e.g., native vs.\ non-native listeners) are not addressed by the existing methods.
Furthermore, most methods do not natively support inference on the number of latent features.

The Bayesian paradigm possesses the machinery to resolve these issues, though they have yet to all be comprehensively addressed in the current literature.
Following the foundational work by \cite{oh2001bayesian}, who established a framework for performing inference and uncertainty quantification on latent coordinates via Markov chain Monte Carlo (MCMC), the Bayesian MDS literature has expanded to include scalable extensions for hyperbolic geometries \citep{praturu2022bayesian, liu2024bayesian} and massive-scale phylogenetic applications \citep{holbrook2021massive}.
However, these models generally assume population homogeneity, rendering them insufficient for characterizing the complex neuroplasticity and varied speech-sound encoding found across diverse language groups.
While domain-specific adaptations have emerged, such as the bioinformatics model by \cite{nguyen2017bayesian}, they are often restricted to a single latent dimension and fail to account for individual or group-specific variation in the underlying features.
Even hierarchical approaches designed to capture heterogeneity, such as the model for orchestral interpretations of Beethoven symphonies by \cite{yanchenko2020hierarchical}, situate variability at the level of denoised distances rather than within the unobserved latent features that induce those distances.
Consequently, existing methods lack the multi-dimensional flexibility and feature-level hierarchical structure necessary to quantify the nuanced, group-specific latent representations required for our motivating neuroscience applications.

Indeed, inference about the latent features is challenging due to identifiability issues; while the reconstructed distances are identifiable, the latent features are unique up to transformations, i.e., rotations, translations, and reflections.
While these indeterminacies may be overlooked for simple visualization, anchoring the model becomes essential for meaningful interpretation and comparison.
Specifically, in Bayesian frameworks, the posterior remains invariant under these transformations, necessitating explicit strategies for valid posterior summarization and uncertainty quantification.
Common approaches to resolve this symmetry include imposition of strong priors to favor a specific orientation \citep[e.g.,][]{desarbo1998bayesian},
or utilizing post-hoc alignment to map posterior draws to a common reference configuration \citep[e.g.,][]{oh2001bayesian}; however, such strategies may not fully resolve more complex symmetries, such as signed permutations of the latent dimensions.
Alternatively, employing hard coordinate constraints to lock the configuration during estimation can provide a more concrete mathematical resolution \citep{bradlow2000little, park2008hierarchical}; however, this rigidity may lead to sampler inefficiency, leading to poor MCMC mixing and slow convergence.
\citet{lin2019bayesian} addressed identifiability by minimizing a loss function post-MCMC, allowing the sampler to move freely, avoiding the bottlenecks of hard constraints, while ensuring that the final latent features are unique and interpretable.

Building on these prior works, while addressing some of their aforementioned limitations, we propose a flexible Bayesian mixed MDS model for studying dissimilarity across subjects and groups.
Our proposal allows performing biologically interpretable inference and straightforward uncertainty quantification for both individual and group-specific distances across stimuli, taking into account the individual idiosyncrasies as well as the heterogeneity between native and non-native Mandarin listeners at the level of the unobserved features.
Our proposal thus takes into account two different sources of heterogeneity and does this at the latent feature level and not at the latent distance level.
Moreover, by weighting the latent features, our proposal allows the latent axes to be identified uniquely and not be affected by rotation invariance, in contrast to standard MDS and in line with the classical INDSCAL.
Importantly, this allows inferring biologically interpretable latent dimensions when such interpretation is meaningful.

Our proposed approach can also infer the effective number of relevant features in a semi-automated data-adaptive way. 
This is achieved by using cumulative shrinkage priors (\citealp{bhattacharya2011sparse}) that impose increasing shrinkage on higher-dimensional latent features.
By starting with a generous number of features, and then eliminating redundant higher-dimensional features which are shrunk toward near-zero values on-the-fly using an adaptive MCMC algorithm \citep[following][]{roberts2009examples}, this approach obviates the need for computationally intensive methods, such as refitting models for different latent dimensions or executing trans-dimensional MCMC steps (e.g., \citealp{richardson1997bayesian}). 
Finally, we employ a post-processing procedure, adapting to recent developments to address similar issues in Bayesian latent factor models \citep{papastamoulis2022identifiability}, which maintains excellent sampling efficiency while fully resolving the identifiability issues, including translations and signed permutations of the values of the latent dimensions across MCMC samples, enabling meaningful inference on the latent features coherently across groups and subjects.

We apply the proposed model to assess group-level differences in the neural representation of Mandarin Chinese tones between native Mandarin speakers and monolingual native English speakers.
Neural representations of Mandarin tones were extracted from a dataset of frequency-following responses (FFR) to Mandarin tones \citep{llanos2017hidden}.
The FFR is a scalp-recorded electrophysiological brain component that reflects phase-locked activity from ensembles of neurons along the central auditory nervous system.
When the brain is stimulated with a periodic sound, neurons in the auditory system synchronize their oscillatory activity by firing at the same phase of each cycle in the stimulus waveform.
This synchronized phase-locked activity is aggregated by the scalp-recorded FFR, which thus mimics the temporal structure of the sound with a high degree of fidelity.
Prior cross-linguistic work \citep{krishnan2005encoding, llanos2017hidden, reetzke2018tracing} has shown that the FFR reflects language experience-dependent plasticity.
Specifically, Mandarin lexical tones are more faithfully represented in the FFR of native speakers of Mandarin Chinese, relative to native speakers of English \citep[e.g.,][]{krishnan2005encoding}.
Therefore, Mandarin listeners are expected to convey bigger differences between FFRs to Mandarin tones than non-native speakers of Mandarin Chinese \citep{llanos2017hidden, reetzke2018tracing}.
Motivated by these neuroscientific experiments and related prior literature, our proposed mixed MDS methodology allows scientists to map the observed pairwise FFR distances to a lower-dimensional common feature space, 
simultaneously enabling the evaluation of heterogeneity among language groups and subjects, presenting a principled statistical approach for comparing groups and individuals, 
while also allowing the visualization of the geometry.
Notably, in our formulation, the latent axes of the reduced space are all uniquely identifiable, which makes inference and interpretation of the features biologically very meaningful.

An alternative approach to obtaining a low-dimensional representation of the FFR time series might be through traditional latent factor models (LFMs).
See, e.g., \cite{aguilar2000bayesian}, albeit in a financial time series context.
LFMs assume that observed time series can be represented as linear combinations of latent sources plus noise.
However, these assumptions may not hold when relationships are nonlinear, nonstationary, or fundamentally distance-based, as is often considered to be the case with the geometrically complex patterns of brain activity.
Additionally, while LFMs can recover latent components, they usually lack meaningful spatial arrangement, making intuitive visualization challenging. 
In contrast, MDS does not rely on any such restrictive assumptions on a specific generative model for the observed time series.
By operating directly on pairwise dissimilarities, MDS can be applied to a wide range of distance and dissimilarity matrices and provides low-dimensional embeddings that summarize complex similarity structures. 
It provides interpretable, low-dimensional visual embeddings that preserve large-scale geometric patterns, making it especially well-suited for exploratory and visualization-driven neurobiological analyses, which likely contributed to its popularity in such contexts.
These features are particularly important for our application as well, where the focus is on characterizing and visualizing differences in underlying auditory processing mechanisms from pairwise dissimilarities between the corresponding FFRs. 
In particular, our proposal is tailored to dissimilarity data with multiple sources of heterogeneity (subjects and groups) and, motivated by the auditory neuroscientific application, targets low-dimensional Euclidean representations for dissimilarities.

The rest of this article is organized as follows.
Section~\ref{sec: data} provides additional background on our motivating scientific experiments.
Section~\ref{sec: MMDS} details our novel multi-group mixed multidimensional scaling models.
Section~\ref{sec:postmain} outlines statistical and computational challenges and solution strategies.
In particular, Section~\ref{sec: MCMC} reports the proposed MCMC algorithm.
Section~\ref{sec: no of features} describes a strategy to select the number of features adaptively, 
Section~\ref{sec: feature identifiability} outlines the post-processing algorithm to solve the identifiability issues and allow inference on the latent features.
Section~\ref{sec: sim} discusses the results of the proposed method applied to some synthetic numerical experiments.
Section~\ref{sec: applications} presents the results of the proposed method applied to the aforementioned two language groups' neural dissimilarities data.
Section~\ref{sec: discussion} contains concluding remarks.

\section{Neural Dissimilarities Data}\label{sec: data}
\begin{figure}
	\centering
	\includegraphics[width=\linewidth]{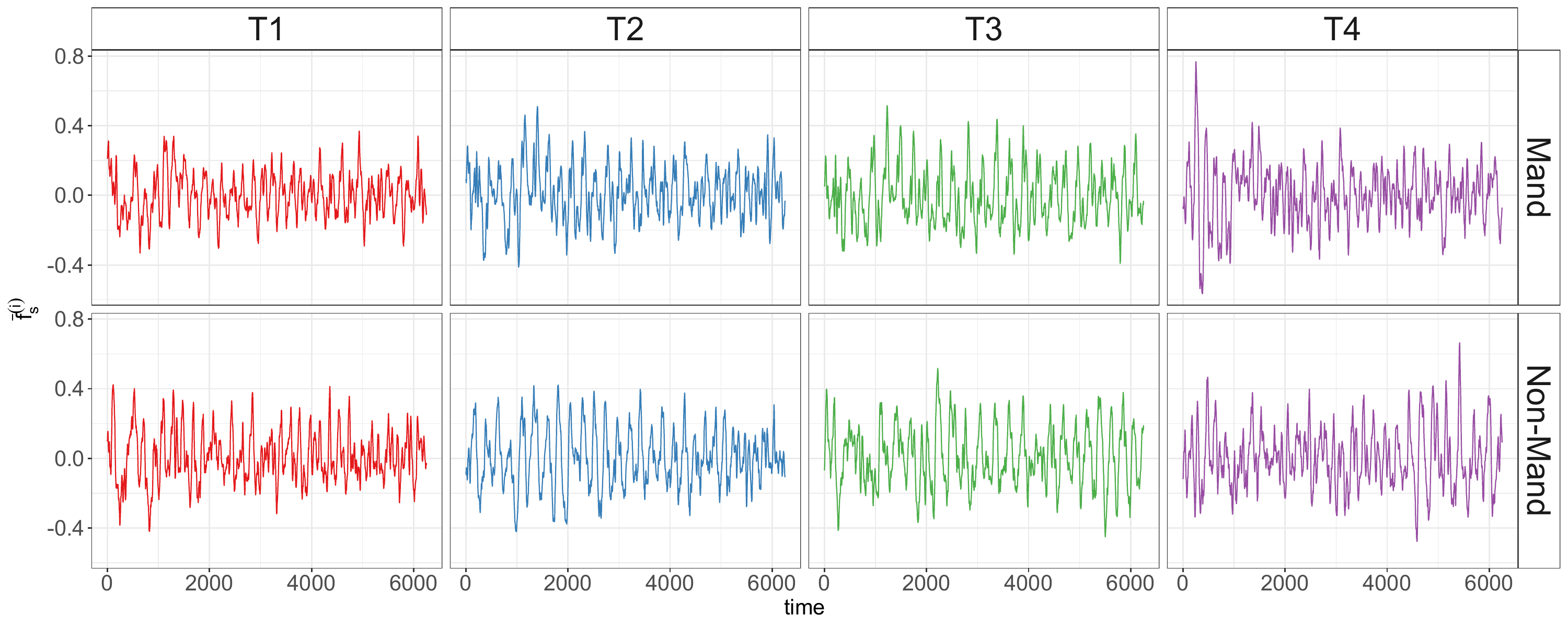}
	\caption{Tone neural data: Mean (intra-tone cross-measurements) FFR in the time domain from one Mandarin-speaking and one non-Mandarin-speaking listener.
    T1, T2, T3, and T4 denote the four Mandarin lexical tones: high level (T1), low-rising (T2), low-dipping (T3), and high-falling (T4).}
	\label{Fig:FFR_real}
\end{figure}

The neural dissimilarity matrices used in our analysis come from the previously published auditory neuroscience study \cite{llanos2017hidden}.
The brain activities of $n=28$ subjects, $n_{1}=14$ Mandarin speakers and $n_{2} = 14$ English (non-Mandarin) speakers, were recorded under exposure to different Mandarin tones.
Individual distance matrices between stimuli were computed as Euclidean distances between FFRs.
As a tonal language, Mandarin Chinese has four syllabic pitch contours or tones that are used to convey different lexical meanings.
For instance, the syllable ``ma'' can be interpreted as ``mother'', ``hemp'', ``horse'', or ``scold'' depending on whether it is pronounced with high level (T1), low-rising (T2), low-dipping (T3), or high-falling (T4) tones, respectively.
These four tones, pronounced by native Mandarin speakers, represented the stimuli.
Figure~\ref{Fig:FFR_real} provides context for the associated neural signals underlying our distance matrices.
As FFRs reflect fast, phase-locked oscillations (often at hundreds of Hz), they are traditionally presented in the time domain to demonstrate neural tracking of fine temporal structure.
In such representations, meaningful information is encoded in the rate of oscillation (neural pitch tracking) rather than peak amplitude.
While qualitative cues exist, such as T3's slower overall rate and T4's terminal deceleration, these differences are naturally subtle.

This limited visual separability of FFRs justifies our reliance on quantitative, distance-based comparisons of the entire response rather than waveform inspection.
Specifically, we focus on analyzing the Euclidean distances between neural representations of Mandarin tones computed from the subjects' brain activity measured via scalp-recorded FFRs, yielding for each subject an $S \times S$ (here, $S=4$) symmetric distance matrix with zeros on the diagonal.
FFRs were extracted from the dataset described in \cite{llanos2017hidden}.
They were acquired using 3 Ag-Cl electrodes placed on the vertex of the scalp (active channel), the left mastoid (ground), and the right mastoid (reference channel).
The neural activity captured by these electrodes was amplified and digitized with an actiCHamp system and a dedicated preamplifier (EP-preamp, gain 50x).
The FFR dataset included $1000$ artifact-free FFR trials per tone (i.e., $1000$ FFRs in response to $1000$ repetitions of each tone).
To account for different trial-by-trial noise levels in the neural representation of the brain between the two language groups across subjects and time, some rescaling of the observations is necessary \citep[see e.g.,][]{nili2014toolbox, feng2021emerging}, which helps eliminate internal variability.
Specifically, we compute the average FFR across $1000$ repeated measurements (i.e., trials) rescaled by the average (across stimuli) of the within (repeated measurements) standard deviations.

Formally, let $f^{(i)}_{s,m}(t)$ denote the FFR in the time domain, representing the brain activity of subject $i$ at time $t$ under the exposure to stimulus $s$ in the measurement $m$.
We denote by $\overline{f}^{(i)}_{s}(t) = \sum_{m=1}^{1000} f^{(i)}_{s,m}(t)/1000$ the average over repeated measurements (as shown in Figure~\ref{Fig:FFR_real}) and by $\text{SSE}^{(i)}_{s}(t) = \sum_{m=1}^{1000} \{f^{(i)}_{s,m}(t)-\overline{f}^{(i)}_{s}(t)\}^{2}$ the sum of squared errors at time $t$ for subject $i$ and stimulus $s$.
Thus, as a measure of noise for the measurements at time $t$ in subject $i$ under the exposure to stimulus $s$, we use the estimated standard deviation $\wh{\sigma}_{s}^{(i)}(t) = \{\text{SSE}^{(i)}_{s}(t)/(1000-1)\}^{1/2}$; and as a measure of individual dispersion at time $t$, the average of these standard deviations across stimuli $\wh{\sigma}^{(i)}(t)=\sum_{s=1}^{S} \wh{\sigma}_{s}^{(i)}(t)/S$.
Finally, we rescale the average empirical distances at time $t$ by these estimated noise variances that cannot be explained by the average FFR, i.e., $\wt{f}^{(i)}_{s}(t) = \overline{f}^{(i)}_{s}(t)/\wh{\sigma}^{(i)}(t)$.
By rescaling the FFRs between average brain activities by the corresponding language group's inherent noise levels in this manner, we allow for biologically meaningful comparisons of the brain's ability to distinguish different auditory stimuli across subjects.

\begin{figure}
	\centering
\includegraphics[width=0.8\linewidth]{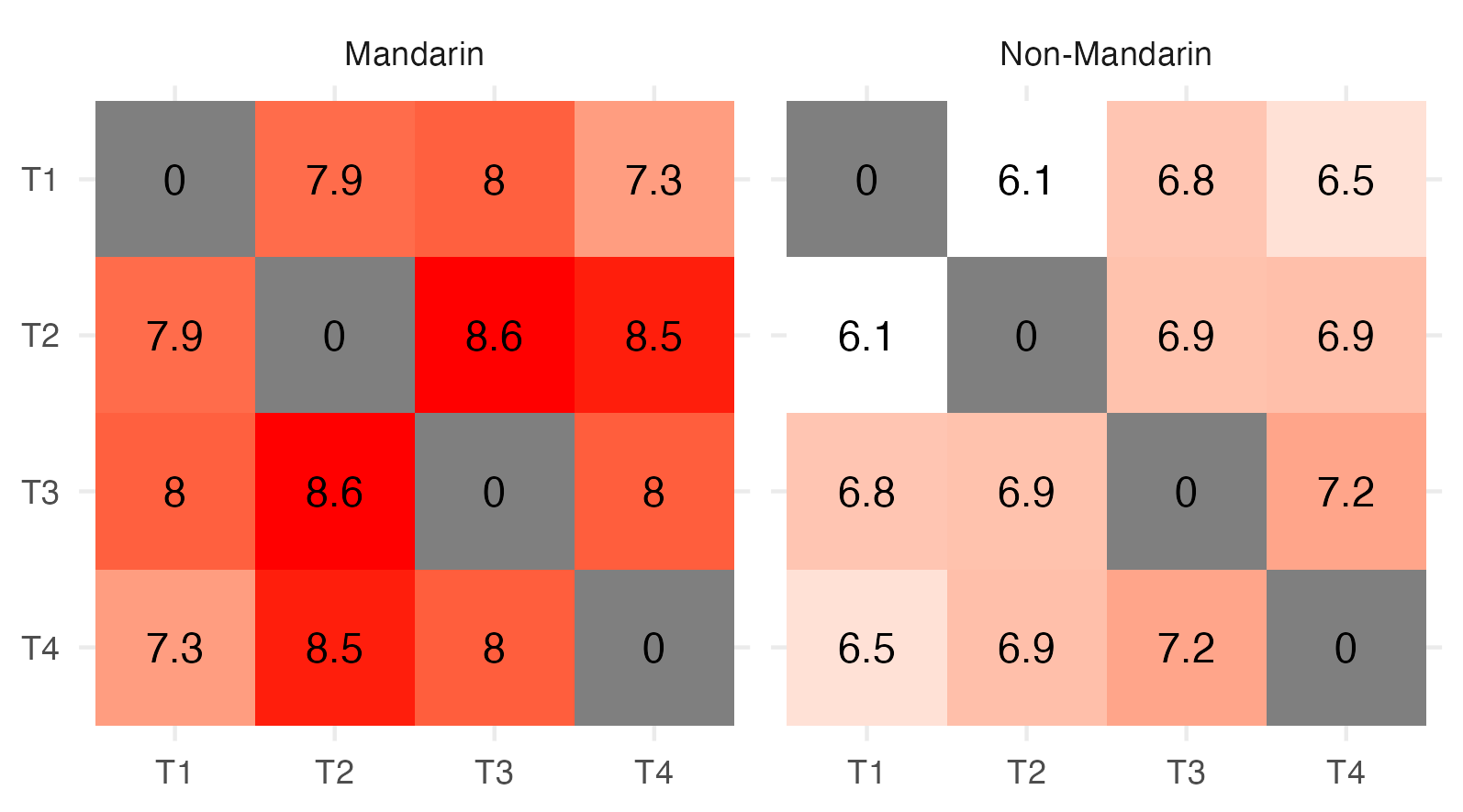}
	\caption{Tone neural distance data:
	Mean (intra-group cross-subject) FFR distances between different Mandarin tones, T1, T2, T3, and T4.
    }
	\label{fig: avg_Stimuli_dist}
\end{figure}

Figure~\ref{fig: avg_Stimuli_dist} shows the average distances between stimuli in the two groups.
We can see that, although there are group-specific characteristics due to language, e.g., the recorded Mandarin speaker distances are better differentiated across pairs of tones than the ones from the non-Mandarin group, there are also strongly shared patterns across these groups, e.g., the pairs $\{2,3\}$ and $\{2,4\}$ are well-separated pairs of tones while the pairs $\{1,4\}$ and $\{1,2\}$ are the closest in both groups.

Figure~\ref{fig: box_delta_s_i_real_check} reports the individual tone distances, showing that there is also substantial individual variability within groups.
Note also a higher within-group variability in the non-Mandarin group, as expected.

We are interested in learning the shared geometry of the lower-dimensional latent space representing these tones in the human brain across language groups and individuals, while also assessing how this space varies across these groups and individuals.

\section{Mixed Multidimensional Scaling} \label{sec: MMDS}
\paragraph{Modeling the Distances.}
In MDS, the observed distances between stimuli $s$ and $r$, namely $d_{s,r}, r<s, r=1,\dots, S-1, s=2,\dots,S$, are noisy measurements of latent distances $\delta_{s,r}$ computed in a lower $H$-dimensional feature space ($H <S$).
We therefore treat the observed dissimilarities $d_{j,s,r}^{(i)}$ as noisy,
strictly positive measurements of latent denoised distances $\delta_{j,s,r}^{(i)}$.

As in \cite{nguyen2017bayesian} and \cite{yanchenko2020hierarchical}, we assume a gamma likelihood for the observed distances to perform our model-based MDS.
However, motivated by the application, we allow for individual differences at the level of the latent features.
We also consider different variances in the likelihood terms to take into account the heterogeneity of the noise levels between the two groups in our data.

More precisely, we denote by $d_{j,s,r}^{(i)}$ the distance between the $s\th$ and the $r\th$ stimuli ($r < s$) for the $i\th$ individual from the $j\th$ group and we assign a gamma likelihood for the observed distances
\begin{equation}\label{eq:LikGA}
	d_{j,s,r}^{(i)} \overset{ind}{\sim} \Ga\left( \texttt{mean}= \delta_{j,s,r}^{(i)}, \, \texttt{var}= \sigma_{\epsilon,j}^{2}\right).
\end{equation}
In the context of the experiment described in Section~\ref{sec: data}, $j=1$ and $j=2$ refer to the Mandarin-speaking group and the non-Mandarin-speaking group, respectively.
The gamma likelihood is parametrized in terms of mean and variance to ease the interpretation of the latent parameters, and it is centered around the underlying individual denoised distances $\delta_{j,s,r}^{(i)}$, which can vary across subjects within and across the two groups.
The variance term $\sigma_{\epsilon,j}^{2}$ varies across groups, allowing for different noise levels in different groups.
The more standard shape and rate representation can be recovered as \texttt{shape $=$ mean$^{2}$/variance = $(\delta_{j,s,r}^{(i)})^{2}/\sigma_{\epsilon,j}^{2}$} and \texttt{rate $=$ mean/variance = $\delta_{j,s,r}^{(i)}/\sigma_{\epsilon,j}^{2}$}.

The underlying individual distance $\delta_{j,s,r}^{(i)}$ is a function of $H$ unobserved individual features $\eta_{j,s,h}^{(i)}$, such as
\begin{equation}\label{eq:shared_distGa_RE}
		\delta_{j,s,r}^{(i)} = \sqrt{\sum_{h=1}^{H} \left(\eta_{j,s,h}^{(i)}-\eta_{j,r,h}^{(i)} \right)^{2}},
\end{equation}
where the number of latent features $H$ is smaller than the number of original stimuli $S$.
Here $\etam_{j,s}^{(i)} = (\eta_{j,s,1}^{(i)},\dots,\eta_{j,s,H}^{(i)})\trans$ denotes the latent vector with the feature values for the $s^{th}$ stimulus for the $i^{th}$ individual belonging to the $j^{th}$ group.

\paragraph{On the Choice of the Gamma Likelihood.}
Note that in our hierarchical framework, the Gamma likelihood for the observed distances $d_{j,s,r}^{(i)}$ is conditional on latent distances $\delta_{j,s,r}^{(i)}$ derived from a lower-dimensional feature space.
Under this specification, the Euclidean structure enters through the conditional mean of $d_{j,s,r}^{(i)}$, while the sampling variances $\sigma_{\epsilon,j}^{2}$ accommodate varying noise levels across groups.
The likelihood thus implicitly accounts for an error term, $\epsilon$, characterizing the variability of the observed distances $d_{j,s,r}^{(i)}$ around the latent distances $\delta_{j,s,r}^{(i)}$, which is itself a function of the underlying feature parameters $\eta_{j,s,h}^{(i)}$, as defined in \eqref{eq:shared_distGa_RE}.
The model for these features $\eta_{j,s,h}^{(i)}$ and their associated priors are detailed in Sections \ref{sec: MMDS} and \ref{sec:priors}.
By operating strictly within this conditional hierarchical framework, we fulfill our inference goals while avoiding the derivation of the induced marginal distribution, which remains analytically intractable.

The choice of the Gamma distribution for the conditional likelihood provides a good framework for modeling dissimilarities on $(0,\infty)$, consistent with established practice in Bayesian MDS \citep[e.g.,][]{nguyen2017bayesian, yanchenko2020hierarchical}.
Although parametric, the Gamma family offers some advantages over more rigid alternatives: unlike the log-Normal distribution \citep[e.g.,][]{ramsay1977maximum,ramsay1982some}, which is constrained by heavy tails and sparse mass near zero, or the truncated-Normal specification \citep[e.g.,][]{oh2001bayesian}, which is often dominated by its behavior at the origin, the Gamma distribution can smoothly transition from Exponential-like to Gaussian-like shapes (as \texttt{shape $=$ mean$^{2}$/variance} increases) while maintaining a robust intermediate tail behavior.

While more flexible nonparametric approaches, such as mixture models, could be considered when larger sample sizes are available, in the present application, the sample size is relatively modest, and the empirical data exhibit means that substantially exceed their variances (Figure~\ref{fig: avg_Stimuli_dist}).
In this high-shape regime (\texttt{shape} $\gg 1$), the Gamma, log-Normal, and truncated-Normal distributions all converge toward a symmetric, Gaussian-like form, rendering the specific choice of parametric family largely inconsequential.
Overall, the parametric Gamma likelihood thus represents a parsimonious and robust choice for our current modeling requirements.

\paragraph{Modeling the Latent Features.}
Our modeling efforts concentrate henceforth on flexibly characterizing the latent features $\eta_{j,s,h}^{(i)}$.
Specifically, we model $\eta_{j,s,h}^{(i)}$ as a product of latent features shared across groups and subjects, $\eta_{s,h}$, and multiplicative coefficients specific to dimension $h$, group $j$, and subject $i$ within group $j$.
That is, we let 
\begin{equation}\label{eq: IndFeat}
	\eta_{j,s,h}^{(i)} = \wt{w}_{j,h}^{(i)} \, \eta_{s,h}.
\end{equation}
The random coefficients $\wt{w}_{j,h}^{(i)}$ vary across features $h$, allowing variation, and thus importance, of the dimensions across individuals between and within groups.

Finally, in order to learn the subjects' and the two groups' variations, we specify an inverse gamma distribution with hyperparameters $a_{w}$ and $b_{w}$ such that the multiplicative terms have a mean equal to $1$ and a large variance equal to $10$.
That is,
\begin{align}
	\begin{split}
	w_{j}^{-1} \overset{iid}{\sim}\text{Ga}\left(\texttt{shape}=a_{w}, \texttt{rate}=b_{w}\right),\\
	{w_{j,h}^{(i)}}^{-1} \overset{iid}{\sim}\text{Ga}\left(\texttt{shape}=a_{w}, \texttt{rate}=b_{w}\right).
	\end{split}
\end{align}
Here, with some abuse of notation, $w_{j,h}^{(i)}$'s represent the individual-level weights that do not vary by group; however, the index $j$ is still retained to distinguish individuals belonging to different groups (e.g., $w_{1,h}^{(1)}$ vs.\ $w_{2,h}^{(1)}$ denote the first individual in groups 1 and 2, respectively).
Also, we parametrize the gamma distribution in terms of shape and rate hyperparameters to facilitate updating these hyperparameters in the full conditionals.
If the multiplicative terms $w_{j}$ and $w_{j,h}^{(i)}$ are equal to $1$, the MDS model ignores group and individual differences, entailing the same latent feature values for the stimulus $s$ for all subjects within and across the two groups.

\paragraph{On the Choice of Shared Features.}
Studies on Mandarin lexical tones in both native and non-native listeners have shown their perception to rely on core pitch-related dimensions across language backgrounds \citep{gandour1978perception,gandour1983tone,gandour1978crosslanguage}.
Neurophysiological work further demonstrates that these dimensions are robustly represented in early auditory responses, such as FFRs, in both Mandarin and non-Mandarin listeners \citep{chandrasekaran2007mismatch}.

This motivates our shared-feature MDS framework, where group and individual differences are characterized as shifts in representational geometry rather than the use of distinct feature sets.
This approach aligns with the general consensus that these dimensions are universally relevant to auditory processing, while language experience modulates their relative weighting.
While relaxing the shared-feature assumption may suit other domains, our model's use of shared latent features with group and individual-specific weighting provides an interpretable framework, grounded in neurobiological theory, for comparing neural encoding across different language backgrounds.

\paragraph{On the Model for the Weights.}
In model \eqref{eq: IndFeat}, the weights $\wt{w}_{j,h}^{(i)}$ are not identifiable from the $\eta_{s,h}$ parameters without the imposition of scale constraints.
We discuss these constraints as part of our post-processing scheme in Section~\ref{sec: feature identifiability}.
Notwithstanding this specific form of non-identifiability, one may consider various strategies to impose further structure and interpretability on the weights.
Specifically, we can consider: 
\begin{enumerate}[label=(\alph*),itemsep=0pt]
\item ~unrestricted: $\wt{w}_{j,h}^{(i)}$, or
\item ~group-specific decomposition: $\wt{w}_{j,h}^{(i)} = w_{j}w_{j,h}^{(i)}$, or
\item ~group-and-dimension-specific decomposition: 
$\wt{w}_{j,h}^{(i)} = w_{j,h} w_{j,h}^{(i)}$, or 
\item ~group-and-(dimension-within-group)-specific decomposition: $\wt{w}_{j,h}^{(i)} = w_{j}w_{j,h}w_{j,h}^{(i)}$.
\end{enumerate}
As before, in models (b), (c), and (d), $w_{j,h}^{(i)}$'s represent individual-level weights that do not vary directly by group $j$, but the index $j$ is still retained to distinguish individuals from different groups.

Because the likelihood depends solely on latent distances, these specifications are stochastically equivalent, leaving internal structures unidentifiable without further constraints.
However, by imposing consistently defined identifying constraints across these parametrizations, the individual weight components can be uniquely recovered from one another.
For instance, the marginal effects remain well-defined in terms of the $\wt{w}_{j,h}^{(i)}$'s across all models:
Group-and-dimension-specific effects can be defined as $\overline{w}_{j,h} = \sum_{i=1}^{n_{j}}\wt{w}_{j,h}^{(i)}/n_{j}$.
Under model (c), this simplifies to $w_{j,h}$ given the constraint $\sum_{i=1}^{n_{j}}w_{j,h}^{(i)}/n_{j} = 1$.
Likewise, overall group-specific effects can be defined as $\overline{w}_{j} = \sum_{h=1}^{H}\sum_{i=1}^{n_{j}}\wt{w}_{j,h}^{(i)}/(H n_{j})$.
Under model (b), this equals $w_{j}$ given the constraint $\sum_{h=1}^{H}\sum_{i=1}^{n_{j}}w_{j,h}^{(i)}/(H n_{j}) = 1$.

In a Bayesian hierarchical framework, these distinct parametrizations do necessitate tailored prior specifications, which in turn dictate the mechanisms for information borrowing and shrinkage, and the efficiency of the MCMC sampling; generally, more granular decompositions will yield higher posterior uncertainty and require more robust implementation to manage the resulting computational complexity.
However, as we will see in Section~\ref{sec: feature identifiability}, using the constraints described above, the posterior distributions for weight components under alternative specifications of $\wt{w}_{j,h}^{(i)}$ can be readily derived (under implicit priors) from the MCMC samples of a primary model of choice.
Consequently, irrespective of the initial specification, the underlying representational geometry remains accessible via straightforward reparameterization of the posterior samples.

Nevertheless, the practical choice between these models can be guided by considerations of meaningful borrowing of information and computational ease and efficiency.
In this article, we adopted model (b) because it keeps the MCMC and the post-processing schemes simple, while inducing a higher correlation between individual distances $\delta$ within the same group.
As discussed above, the $w_{j,h}$'s can still be obtained using post-processing; see Section~4.3 and Figures~1 and 2 in the Supplementary Material.

\subsection{Model Identifiability} \label{sec: identifiability}
Our multiplicative individual effects make the latent feature space identifiable under rotations, enabling stronger interpretations of the latent features.
To see this, note that reweighting the latent features is equivalent to multiplying the feature matrix by a diagonal matrix of weights.
Therefore, if we rotate the axes (aside from the special case of permutations of the axes and the subclass of signed permutations that can be represented as rotation), the weight matrix would not be diagonal anymore, which is not permissible in our model.
Therefore, it allows the axes of the latent features to be uniquely identified and unaffected by rotation invariance.

More precisely, let $\bH^{(i)}_{j}=((\eta_{j,s,h}^{(i)})) \in \bbR^{S \times H}$, $i=1,\ldots,n_{j}, \, j=1,\ldots,J$ be the individual latent feature matrices, $\bW_{j}^{(i)}=\diag(\wt{w}_{j,1}^{(i)}, \ldots, \wt{w}_{j,H}^{(i)}) \in \bbR^{H\times H}$, $i=1,\ldots,n_{j}, \, j=1,\ldots,J$ be the corresponding diagonal weight matrices, and $\bH=((\eta_{s,h}))\in \bbR^{S \times H}$ be the shared latent feature matrix.
Then, we can rewrite $\bH^{(i)}_{j} = \bH \mathbf{W}_{j}^{(i)}$.
Now, let $\mathbf{R} \in \bbR^{H\times H}$ be a rotation matrix.
If we rotate the individual features as $\bH^{(i)}_{j} \mathbf{R}\trans = \bH \mathbf{W}_{j}^{(i)} \mathbf{R}\trans$, it will not change the implied latent distances $\delta_{j,s,r}^{(i)}$, but the new individual weight matrices $\mathbf{W}_{j}^{(i),\text{new}} = \mathbf{W}_{j}^{(i)} \mathbf{R}\trans$ are in general not diagonal anymore.
This is why the individual weights, which are shared across stimuli, make the latent feature space identifiable in INDSCAL, contrary to standard MDS analysis.
See \cite{carroll1970analysis} for further discussion.
Notably, the use of individual-specific dimension weights is not exclusive to INDSCAL; other prominent MDS frameworks, such as ALSCAL and PROXSCAL, can similarly accommodate individual differences by weighting different latent dimensions according to each subject's unique perceptual profile \citep[see, e.g.,][]{takane1977nonmetric,commandeur1993mathematical}.

That said, some non-identifiability persists since the features $\eta_{j,s,h}^{(i)}$ are still invariant to signed permutations and translations that preserve the latent distances $\delta$ (even if the axes are identifiable and cannot be arbitrarily rotated as discussed above).
We address these issues separately via a post-processing scheme discussed later in Section~\ref{sec: feature identifiability}.

Finally, we note that the latent distances $\delta_{j,s,r}^{(i)}$ and group-specific variance terms $\sigma_{\epsilon,j}^{2}$ are identifiable 
in the traditional sense of \cite{basu2004identifiability} as stated in \cite{swartz2004bayesian}.
Specifically, the identifiability of $\btheta =\{\eta_{j,s,h}^{(i)}, \, \sigma_{\epsilon,j}^{2}, \, i =1,\ldots, \, n_{j},j=1,\ldots, J, \, s=1, \ldots, S, \, h=1, \ldots, H\}$ up to the aforementioned transformations is equivalent to the identifiability of $\wt{\btheta}=\{\delta_{j,s,r}^{(i)}, \, \sigma_{\epsilon,j}^{2}, \, i =1,\ldots, \, n_{j},j=1,\ldots, J, \, s=1, \ldots, S, \, r=1, \ldots, S\}$.
And the latter follows from the moment equations 
\begin{align*}
    \bbE(d_{j,s,r}^{(i)} \mid \wt{\btheta}) = \delta_{j,s,r}^{(i)},~~~\var(d_{j,s,r}^{(i)} \mid \wt{\btheta}) = \sigma_{\epsilon,j}^{2}.
\end{align*}

\subsection{Prior Specification}\label{sec:priors}
To learn the shared latent features $\eta_{s,h}$, we adapt the multiplicative gamma process (MGP) prior \citep{bhattacharya2011sparse} for ordinary factor analysis that allows a stochastic ordering of the latent features in diminishing order of their importance.

Specifically, we assume a Gaussian distribution for the shared feature component $\eta_{s,h}$ whose variance varies across stimuli $s$ and dimension $h$
\begin{equation}
\eta_{s,h} \overset{ind}{\sim} \Normal\left(\texttt{mean}=0, \, \texttt{sd}=\phi_{s,h}^{-1/2} \, \tau_{h}^{-1/2}\right).
\end{equation}
The precision parameter of the shared component $\eta_{s,h}$ follows the MGP
\begin{eqnarray}\label{eq:ShrinkageGa}
	& \phi_{s,h} \overset{iid}{\sim} \Ga \left(\texttt{shape}=\nu/2, \texttt{rate}=\nu/2\right), \\
	& \tau_{h} = \prod_{l=1}^{h}\delta_{l}, \quad \delta_{l} \overset{ind}{\sim} \Ga\left(\texttt{shape}=a_{l}, \, \texttt{rate}=1\right),
\end{eqnarray}
where $a_{l}=a_{2}$ for $l \ge 2$.
The MGP prior shrinks the latent features increasingly towards zero as the dimension $h$ increases.
This implies that the prior favors a few relevant features, 
shrinking the rest to zero, 
while also inducing a probabilistic ranking of the relevant features such that, on average, the first feature explains more variability than the second one, and so on, coherent with the current understanding of the brain's behavior.

We follow \cite{durante2017note} recommendations for the choice of the hyperparameters of the MGP.
In particular, we set $a_{1}=2$ and $a_{2}=3$ since they induce the desired prior shrinkage.
More precisely, they induce an order in the probability of the dimensions in a neighborhood around zero \citep[Lemma 2 in][]{durante2017note}.
Section~4 of the Supplementary Materials provides further discussion on the choice of the hyperparameters.

Finally, we put an inverse-gamma prior to learn the group-specific variance terms 
\begin{equation}
	\sigma_{\epsilon,j}^{2} \overset{ind}{\sim} \mbox{Inv-Ga}\left(\texttt{mean}= \mu_{\sigma^{2}_{\epsilon,j}}, \, \texttt{var}=\sigma^{2}_{\sigma^{2}_{\epsilon,j}}\right).
\end{equation}

\subsection{Model Reparametrization} \label{sec:repara}
We can reparametrize the model in terms of the individual latent features $\eta_{j,s,h}^{(i)}$.
Integrating out the shared dimensions $\eta_{s,h}$, we obtain
\begin{equation}\label{eq: individual_features}
	\eta_{j,s,h}^{(i)} \sim \Normal\left(\texttt{mean}=0, \, \texttt{sd}= {w_{j,h}^{(i)}} \, w_{j} \, \phi_{s,h}^{-1/2} \, \tau_{h}^{-1/2}\right).
\end{equation}
This parameterization allows restating the probabilistic statements directly in terms of one of the main quantities of interest, namely the individual-specific latent features.
Note that, by the definition in \eqref{eq: IndFeat}, the $\eta_{j,s,h}^{(i)}$'s in \eqref{eq: individual_features} are dependent on each other since they share the same latent $\eta_{s,h}$'s.
That is, $\big(\eta_{j,s,h}^{(i)} \big)_{h},$ $i=1,\ldots,n_{j},j=1,\ldots,J,s=1,\ldots,S$ belong to the span of the vectors $\eta_{s,h}, s=1,\ldots,S$.
This is consistent with our goal to infer a common latent feature's space across groups as well as subjects.

\section{Posterior Inference} \label{sec:postmain}
Posterior inference is performed by sampling from an MCMC algorithm that exploits conditional conjugacy of the parameters when available.
When sampling from the full conditionals is not straightforward, we exploit adaptive Metropolis schemes \citep{roberts2009examples}.
The details are discussed in the next subsection.

\subsection{MCMC Algorithm}\label{sec: MCMC}
\noindent
\textbf{(1)} Sample $\delta_{1}$ from $p(\delta_{1} \mid \cdots)$
\begin{equation*}
	 = \Ga\bigg(\texttt{shape}=a_{1}+ \frac{SH}{2}, \, \texttt{rate}= 1+ \frac{1}{2} \sum_{l=1}^{H} \tau_{l}^{(1)} \sum_{s=1}^{S} \phi_{s,l} \eta_{s,l}^{2} \bigg).
\end{equation*}
\textbf{(2)} Sample $\delta_{h}$ from $p(\delta_{h} \mid \cdots)$
\begin{equation*}
	= \Ga\bigg(\texttt{shape}=a_{h}+ \frac{S(H-h+1)}{2}, \, \texttt{rate}= 1+ \frac{1}{2} \sum_{l=h}^{H} \tau_{l}^{(h)} \sum_{s=1}^{S} \phi_{s,l} \eta_{s,l}^{2} \bigg),
\end{equation*}
where $\tau_{l}^{(h)} = \prod_{t=1,\, t\ne h}^{l} \delta_{t}$, for $l=1,\ldots H$.\\
\noindent
\textbf{(3)} Sample $\phi_{s,h}$ from
\begin{equation*}
	p(\phi_{s,h} \mid \cdots) = \Ga\bigg(\texttt{shape}=\frac{\nu+1}{2}, \, \texttt{rate}= \frac{\nu+\tau_{h} \eta_{s,h}^{2}}{2}\bigg).
\end{equation*}
\textbf{(4)} Sample $\eta_{s,h}$ from
\begin{align*}
	\begin{split}
		p(\eta_{s,h} \mid \cdots) &\propto \N \left(\eta_{s,h} \mid \texttt{mean}=0, \, \texttt{sd} =\phi_{s,h}^{-1/2} \tau_{h}^{-1/2} \right)\\
		&\quad \times \prod_{j=1}^{J} \prod_{i=1}^{n_{j}} \prod_{r: r \ne s} \Ga\left( d_{j,s,r}^{(i)} \mid \texttt{mean}= \delta_{j,s,r}^{(i)}, \, \texttt{var}= \sigma_{\epsilon,j}^{2}\right).
	\end{split}
\end{align*}
We perform an adaptive MH step \citep{roberts2009examples} with a random-walk univariate Gaussian proposal.
\\
\textbf{(5)} Sample $\sigma^{2}_{\epsilon,j}$ from
\begin{align*}
	\begin{split}
		p\left(\sigma^{2}_{\epsilon,j} \mid \cdots\right) &\propto \mbox{Inv-Ga}\left(\sigma^{2}_{\epsilon,j} \mid \texttt{mean}= \mu_{\sigma^{2}_{\epsilon,j}}, \, \texttt{var}=\sigma^{2}_{\sigma^{2}_{\epsilon,j}}\right)\\
		&\quad \times \prod_{i=1}^{n_{j}} \prod_{s=2}^{S} \prod_{r=1}^{s-1} \Ga\left( d_{j,s,r}^{(i)} \mid \texttt{mean}= \delta_{j,s,r}^{(i)}, \, \texttt{var}= \sigma_{\epsilon,j}^{2}\right).
	\end{split}
\end{align*}
We perform an adaptive MH step with a random-walk Gaussian proposal for $\log(\sigma^{2}_{\epsilon,j})$.
\\
\textbf{(6)} Sample $w_{j,h}^{(i)}$ from
\begin{align*}
	\begin{split}
		p\left(w_{j,h}^{(i)} \mid \cdots\right) &\propto \mbox{Inv-Ga}\left(w_{j,h}^{(i)} \mid \texttt{shape}= a_{w}, \, \texttt{rate}=b_{w}\right)\\
		&\quad \times \prod_{s=2}^{S} \prod_{r=1}^{s-1} \text{Ga}\left( d_{j,s,r}^{(i)} \mid \texttt{mean}= \delta_{j,s,r}^{(i)}, \, \texttt{var}= \sigma_{\epsilon,j}^{2}\right).
	\end{split}
\end{align*}
We perform an adaptive MH step with a random-walk Gaussian proposal for $\log(w_{j,h}^{(i)})$.
\\
\textbf{(7)} Sample $w_{j}$ from
\begin{align*}
	\begin{split}
		p(w_{j} \mid \cdots) &\propto \mbox{Inv-Ga}\left(w_{j} \mid \texttt{shape}= a_{w}, \, \texttt{rate}=b_{w}\right)\\
		&\quad \times \prod_{i=1}^{n_{j}} \prod_{s=2}^{S} \prod_{r=1}^{s-1} \text{Ga}\left( d_{j,s,r}^{(i)} \mid \texttt{mean}= \delta_{j,s,r}^{(i)}, \, \texttt{var}= \sigma_{\epsilon,j}^{2}\right).
	\end{split}
\end{align*}
We perform an adaptive MH step with a random-walk Gaussian proposal for $\log(w_{j})$.
More precisely, in all adaptive MH steps, we use an adaptive random walk Metropolis kernel with Gaussian proposals, after transforming parameters to have support on $\bbR$ when needed.
Every $50$ iterations, we compute, for each parameter, the empirical acceptance rate over the last $50$ iterations.
If this acceptance rate is below the target value $0.44$, we decrease the corresponding proposal standard deviation by a factor $e^{-\delta_m}$; if it is above $0.44$, we increase it by a factor $e^{\delta_m}$.
Here $\delta_m = \min\{0.01, 1/\sqrt{m}\}$, with $m$ the current MCMC iteration, so the adaptations diminish over time.
This scheme targets an acceptance rate of $0.44$ for all scalar random walk proposals, which is a standard choice for efficient adaptive random walk Metropolis algorithms \citep{roberts2009examples}.

\subsection{Selecting the Number of Features}\label{sec: no of features}
Our Bayesian mixed multidimensional scaling model allows performing dimensionality reduction by setting $H < S$.
If we choose a conservative (i.e., larger than needed) upper bound $H^{+}$, the model allows us to give relevant importance to a few latent features $H<H^{+}$ and small importance to the remaining $H^{+}-H$ latent features.
Here, we think of $H$ as the effective dimensionality of the latent feature space (i.e., the number of latent dimensions) so that the contribution from adding additional dimensions in reconstructing a denoised version of the observed distance matrices is negligible.
However, running the MCMC for an upper bound $H^{+}$ larger than needed can be computationally inefficient.
Finding the number of relevant latent features $H$ can therefore be helpful in reducing costs.
Finding $H$ can also be of inferential interest itself.

To address this, we use a relative Frobenius error $D(t)$, similar in spirit to Kruskal's STRESS, defined as the average, across subjects $i$ and groups $j$, of the Frobenius distance between the observed $S \times S$ distance matrices $((d_{j,s,r}^{(i)}))$ and the corresponding $S \times S$ denoised latent distance matrices $((\delta_{j,s,r}^{(i)}))$ sampled at iteration $t$, divided by the Frobenius norm of the observed distance matrix.
We set, according to the noise level in the measurements, a threshold $D_{T}$.
In particular, we found that $D_{T}=0.1$ favors a parsimonious latent space that can reconstruct well all the individual observable distances in our motivating neural tone distance experiment, as shown in Figure~\ref{fig: box_delta_s_i_real_check}.
More generally, $D_{T}$ should be chosen to reflect the desired level of reconstruction accuracy for the observed distances and, when possible, calibrated using prior knowledge about the signal-to-noise ratio in the specific application.
We perform the following steps with probability $p(t)=\exp\{-(\alpha_{0}+\alpha_{1}t)\}$ at iteration $t$, where $\alpha_{0}\ge 0$ and $\alpha_{1}>0$ such that the adaptations occur often at the beginning of the chains, but decrease in frequency exponentially fast as the chain settles in.
At the $t\th$ iteration, if the current latent features are not sufficient to recover the distances well, i.e., $D(t)>D_{T}$, we set $H(t+1)=H(t)+1$ and add a latent feature $\eta_{s,H(t)+1}$.
Otherwise, if $D(t)<D_{T}$, we set $H(t+1)=H(t)-1$ and delete the feature $\eta_{s,H(t)}$.
When we add a feature, we sample the parameters from the prior distribution.

The adaptive method allows, in a single run, to perform posterior inference on the individual and group-level latent distances, $\delta_{j,s,r}^{(i)}$ and $\delta_{j,s,r}$, together with selecting the number of features, with the convergence of the chain guaranteed by the diminishing probability condition \citep{roberts2007coupling}.
If, as in our motivating auditory neuroscience application, one is interested in performing inference on the actual latent features, we can fix the number of active dimensions $H$ selected in a preliminary run of the MCMC chain or in an initial set of iterations of the chain after burn-in.
In this way, we have the same number of features, $H$, sampled in the final stages of the chain that we can use to perform posterior inference on the latent features after solving the identifiability issues as described in the next section.
Finally, as in \cite{bhattacharya2011sparse}, we can perform uncertainty quantification on the number of features around the point estimate of $H$, in our case, the median of the sampled values $H(t)$ after burn-in, via credible intervals.

Additional evidence on the empirical performance of this adaptive procedure is reported in Section~5 of the Supplementary Materials, where simulation studies show how it recovers the true latent dimension $H$ in application-motivated scenarios.

\subsection{Post-processing for Feature Identifiability} \label{sec: feature identifiability}
In MDS, inference on the latent features $\eta_{j,s,h}^{(i)}$ is challenging due to identifiability issues -- while the reconstructed distances are identifiable, the features themselves are not.
As discussed in Section~\ref{sec: identifiability},
our construction overcomes the rotation invariance issue of the latent dimensions that affects standard MDS methods.
However, while the set of the latent feature axes are now identifiable, the values of the latent features are still identifiable only up to translations and signed permutations of the axes (e.g., label switching of the dimensions and reflection).
This makes the posterior summaries of the sampled $\eta_{j,s,h}^{(i)}$'s, e.g., the posterior median, computed from the MCMC samples, not very meaningful.

To address this issue, we adapt recent post-processing ideas developed for Bayesian latent factor models by \cite{papastamoulis2022identifiability}.
First, we remove translation invariance by centering the shared coordinates so that $\sum_{s=1}^{S}\eta_{s,h}=0$ for every $h$, i.e., 
\[
\textstyle \eta_{s,h}\leftarrow \eta_{s,h}-\bar\eta_{\cdot,h},~~~\text{where}~ \bar\eta_{\cdot,h}=\frac{1}{S}\sum_{s=1}^{S}\eta_{s,h}.
\] 
Under the parametrization $\wt{w}_{j,h}^{(i)} = w_{j} w_{j,h}^{(i)}$, we then apply the corresponding translation to the individual product coordinates as 
\[
\eta_{j,s,h}^{(i)} \leftarrow \eta_{j,s,h}^{(i)} - w_{j} w_{j,h}^{(i)} \bar\eta_{\cdot,h},
\]
so that the fitted distances remain unchanged.
If the goal is inference on the latent distances and on the individual latent dimensions inducing them, together with comparison of the individual configurations across subjects and groups, then this centering step, combined with the signed permutation alignment described below, is sufficient.

Indeed, after centering, we resolve the remaining label switching and sign ambiguity by applying signed permutation transformations that align the latent dimensions across posterior draws.
These transformations are applied consistently to the shared coordinates $\eta_{s,h}$ and to the individual latent coordinates $\eta_{j,s,h}^{(i)}$, while the positive weights are affected only through the corresponding permutation of the dimension index $h$.
Based on these adjusted samples, we can compute meaningful posterior summaries for the latent geometry, such as posterior medians and credible intervals for the individual latent coordinates.

As discussed in Section~\ref{sec: MMDS}, for an additional summary of the contribution of each latent dimension within and across groups, it may also be useful to examine the weight structure directly.
To this end, we further normalize the fitted weights $\wt{w}_{j,h}^{(i)}$.
First, we remove the per-dimension scale indeterminacy induced by the transformation 
\[
\textstyle \eta_{\cdot,h} \mapsto c_{h}\eta_{\cdot,h}~\text{and}~\wt{w}_{j,h}^{(i)}\mapsto \frac{\wt{w}_{j,h}^{(i)}}{c_{h}}, ~~~\text{where} ~ c_{h} = \frac{1}{n}\sum_{j=1}^{J}\sum_{i=1}^{n_{j}}\wt{w}_{j,h}^{(i)}~\text{and}~ n=\sum_{j=1}^{J} n_{j},
\]
which enforces the normalization
$\frac{1}{n}\sum_{j=1}^{J}\sum_{i=1}^{n_{j}}\wt{w}_{j,h}^{(i)} = 1$ 
via the rescaling $\eta_{\cdot,h} \leftarrow c_{h}\eta_{\cdot,h}$ and $\wt{w}_{j,h}^{(i)} \leftarrow \wt{w}_{j,h}^{(i)}/c_{h}$.
Then, we reconstruct group-and-dimension-specific summaries as
\[
\textstyle w_{j,h}=\frac{1}{n_{j}}\sum_{i=1}^{n_{j}}\wt{w}_{j,h}^{(i)},
\qquad
\textstyle w_{j,h}^{(i)} = \frac{\wt{w}_{j,h}^{(i)}}{w_{j,h}},
\]
which implies the identifying constraint
$\frac{1}{n_{j}}\sum_{i=1}^{n_{j}} w_{j,h}^{(i)} = 1$.
Finally, we define the overall group-specific summary as the average of these dimension-specific effects as 
\[
\textstyle w_{j} = \frac{1}{H}\sum_{h=1}^{H} w_{j,h}.
\]
In this way, posterior draws obtained under the original model can be re-expressed post hoc in terms of group-and-dimension-specific weights $w_{j,h}$ and their group-level average $w_{j}$. 
These weights can approach zero if so warranted by the data, providing a practical way to assess the relative and overall utilization of different latent dimensions within and across groups.

Based on these adjusted samples, we can compute meaningful posterior summaries for inference for each subject that are comparable across subjects and groups, e.g., median posterior values and credible intervals.
Trace plots of the latent features before and after post-processing that solves the identifiability issues for synthetic data are shown in Figure~\ref{fig: Eta_i_pre_ps_fake_group2} below. 
Similar results for the real data application, along with 
the estimated posterior distributions of the weights $w_{j,h}$ and $w_{j}$, are included in Section 1 of the Supplementary Materials.

\subsection{MCMC Diagnostics}\label{sec: diagnostic}

The results reported in this article are all based on $10^5$ MCMC iterations with the initial $4\,000$ iterations discarded as burn-in.
The remaining samples were further thinned by an interval of $10$.
We programmed everything in \texttt{R} \citep{rcoreteam2024r}.
The analyses were performed on a MacBook Pro with Apple M2 Pro chip, 16 GB RAM, using \texttt{R} version 4.5.1.
For the real data set, the MCMC algorithm took approximately 17 minutes, and the post-processing algorithm took approximately 11 minutes.
Supplementary Materials provide further details and diagnostics.
The code is available as Supplementary Materials.

\section{Simulation Studies}\label{sec: sim}
In this section, we discuss the results of some synthetic numerical experiments.
In designing the simulation scenarios, we have tried to closely mimic our motivating `brain activities distances between tones' data set.
We thus consider distances between $S=4$ stimuli from $n=28$ subjects recorded in two groups with cardinalities $n_{1}=14$ and $n_{2}=14$.
We set the observed distances $d_{j,s,r}^{(i)}$ close to values that correspond to the estimated quantities (i.e., the posterior median of $\delta_{j,s,r}^{(i)}$) for the real data set, thus simulated distances can be reconstructed from a 3-dimensional latent space.

\begin{figure}[!ht]
	\centering
	\includegraphics[width=0.7\linewidth]{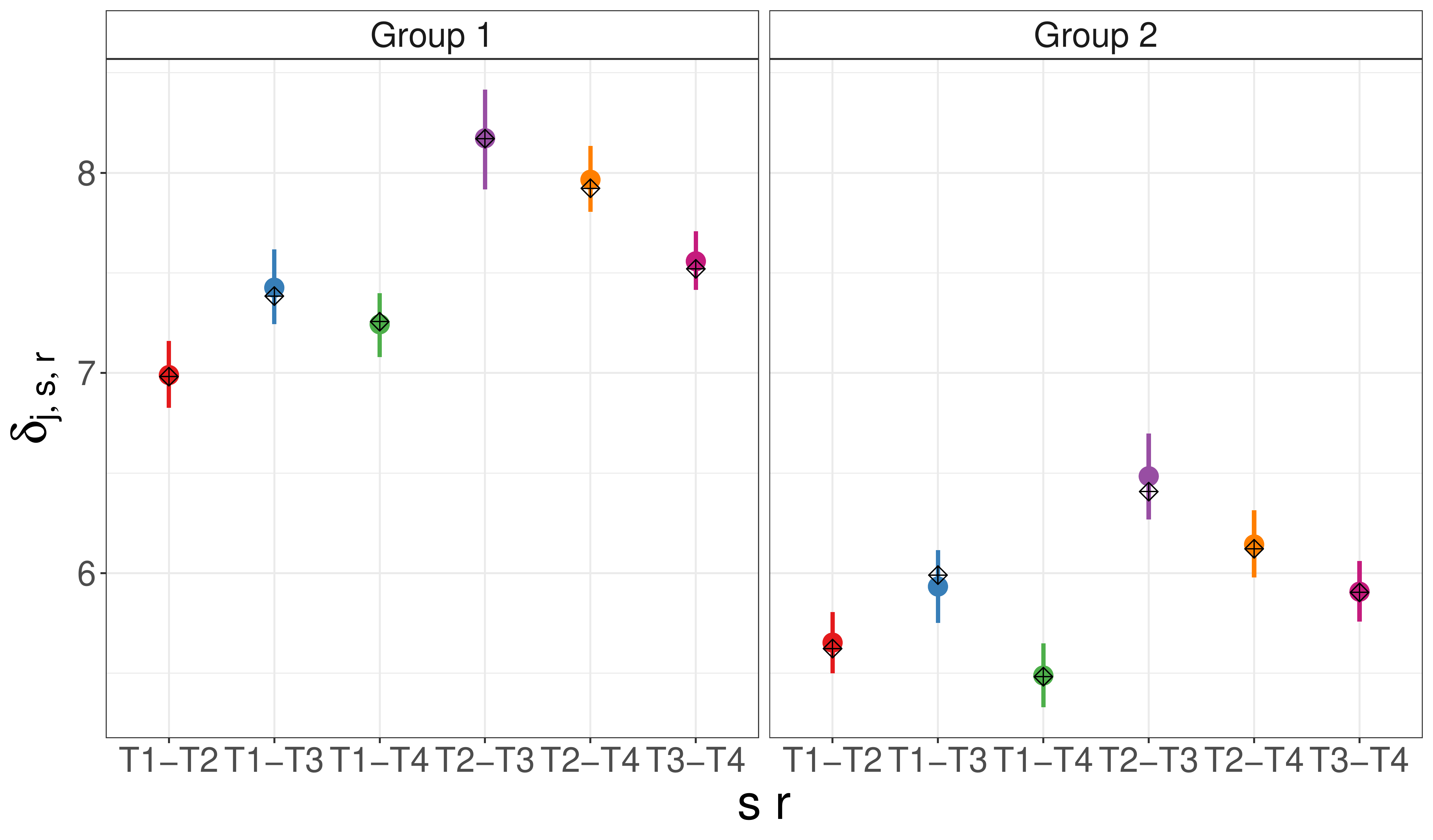}
	\vspace*{-15pt}
    \caption{Results for synthetic data. Posterior medians and 90\% credible intervals of the group-specific latent distances $\delta_{j,s,r}$.
    Black squares represent the median observed distances in each group.}
	\label{fig: delta_group_fake}
\end{figure}

\begin{figure}[!ht]
	\centering
	\includegraphics[width=0.5\linewidth]{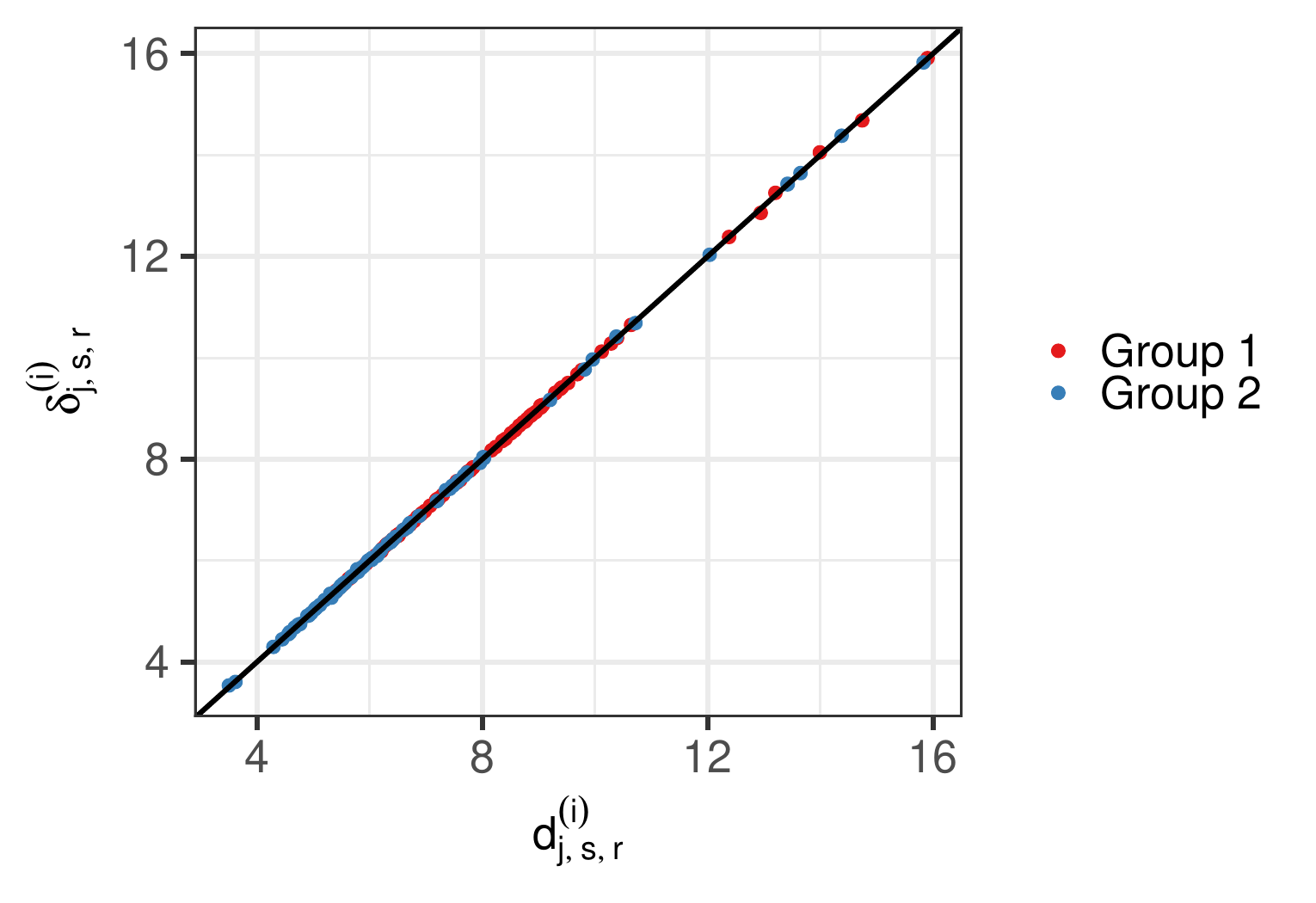}
	\vspace*{-15pt}
	\caption{Results for synthetic data.
    Observed distances $d_{j,s,r}^{(i)}$ vs.\ posterior medians of the denoised distances, $\delta_{j,s,r}^{(i)}$, reconstructed from three shared dimensions.}
	\label{fig: delta_s_i_fake_vs_obs_check}
\end{figure}

\begin{figure}[!ht]
	\centering
	\includegraphics[width=\linewidth]{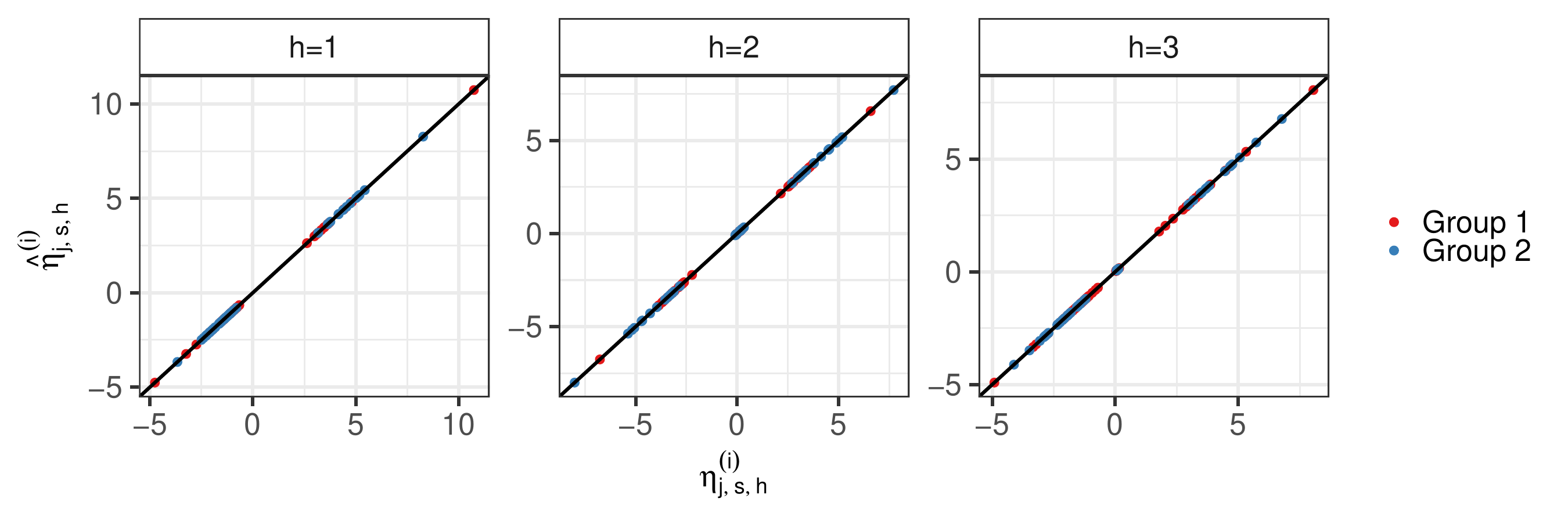}
	\vspace*{-35pt}
	\caption{Results for synthetic data.
    True latent features $\eta_{j,s,h}^{(i)}$ vs.\ their estimated posterior medians $\wh{\eta}_{j,s,h}^{(i)}$.}
	\label{fig: Eta_s_i_fake_vs_obs_check}
\end{figure}

Figure~\ref{fig: delta_group_fake} shows recovery of the distances at the group level.
One can, in principle, define corresponding group-level distances based on the shared features $\eta_{j,s,h}$.
However, due to the nonlinearity of the distance function, these would not correspond to the average of the individual-level distances within the group.
For this reason, we believe it is more meaningful to define group-level distances directly as the average of the corresponding individual-level distances, integrating out the random effects from equation \eqref{eq:shared_distGa_RE}.
Unfortunately, such integrals are not mathematically tractable in closed form, nor empirically calculable since the full populations are not accessible.
To circumvent this, in our illustrations, we approximate the group-level distance parameters $\delta_{j,s,r}$ using the corresponding sample averages of the individual parameters, i.e., $\delta_{j,s,r} = \sum_{i=1}^{n_{j}} \delta_{j,s,r}^{(i)}/n_{j}$, which serve as good proxies for the true $\delta_{j,s,r}$.
Figure~\ref{fig: delta_group_fake} suggests that the three shared latent features recover these $\delta_{j,s,r}$ quite well.
Moreover, Figure~\ref{fig: delta_s_i_fake_vs_obs_check} suggests that we recover the denoised version of the empirical distances quite well at the individual level.
Figure~\ref{fig: Eta_s_i_fake_vs_obs_check} shows that our method allows us to recover the underlying latent features very efficiently.
Here, we align the signs of the true and the estimated latent features for identifiability.
When it comes to permuting the dimension labels, however, we leave them undisturbed.
We can do this since our method assigns decreasing importance to the dimensions, consistent with the data-generating truth, and we are able to identify the order of the axes after post-processing.
Importantly, contrary to standard MDS methods, we do not need to arbitrarily rotate the latent space, since the directions are identifiable.

\begin{figure}[!ht]
	\centering
	\includegraphics[width=0.8\linewidth]{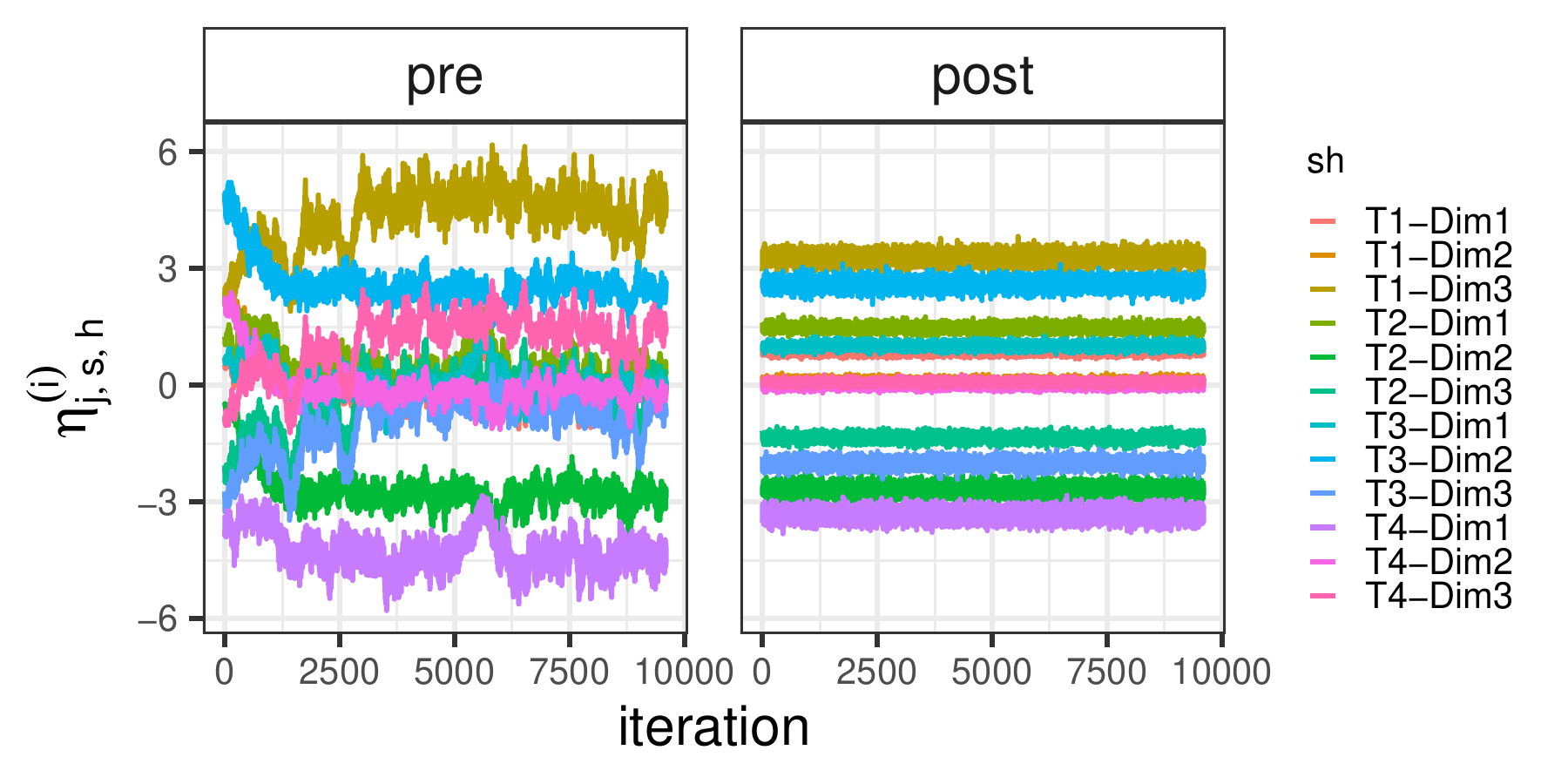}
	\vspace*{-15pt}
	\caption{Diagnostics for synthetic data.
 Trace plots of the individual features $\eta_{j,s,h}^{(i)}$ sampled in the first subject of group 2 pre- and post-processing.}
	\label{fig: Eta_i_pre_ps_fake_group2}
\end{figure}

MCMC diagnostics for the simulation experiments are similar to those shown for the real data analysis presented in the next section and hence are omitted.
Figure~\ref{fig: Eta_i_pre_ps_fake_group2} shows the trace plots of the individual latent features before and after applying the algorithm to solve the identifiability issue in the first subject.
We can see how the procedure described in Section~\ref{sec: feature identifiability} identifies the posterior samples of the different latent features, allowing meaningful inference and comparison at the level of the latent features $\eta_{j,s,h}^{(i)}$.


Section 5 in the Supplementary Materials provides additional simulations to systematically evaluate latent structure recovery in higher dimensions.
Specifically, we consider scenarios with $S \in \{4,8\}$ stimuli and $H_{\mathrm{true}} \in \{2,3\}$.
Our model demonstrates robust performance, accurately recovering both the true latent distances and the correct latent dimensions across replications.

\section{Application to Neural Tone Distances}\label{sec: applications}

In this section, we discuss the results produced by our method applied to the two groups' tone distance data described in Section~\ref{sec: data}.
Our inference goals, we recall, include understanding similarities in terms of stimulus distances between subjects within and across Mandarin and non-Mandarin listeners as well as comparing individual and group latent feature values.
The selected number of dimensions $H$ is $3$ since, after the burn-in, the adaptive criterion discussed in Section~\ref{sec: no of features} suggested sampling from $H=2$ in $5\%$ of the MCMC iterations and from $H=3$ in $95\%$ of the iterations.

\begin{figure}[!ht]
	\centering
	\includegraphics[width=0.65\linewidth]{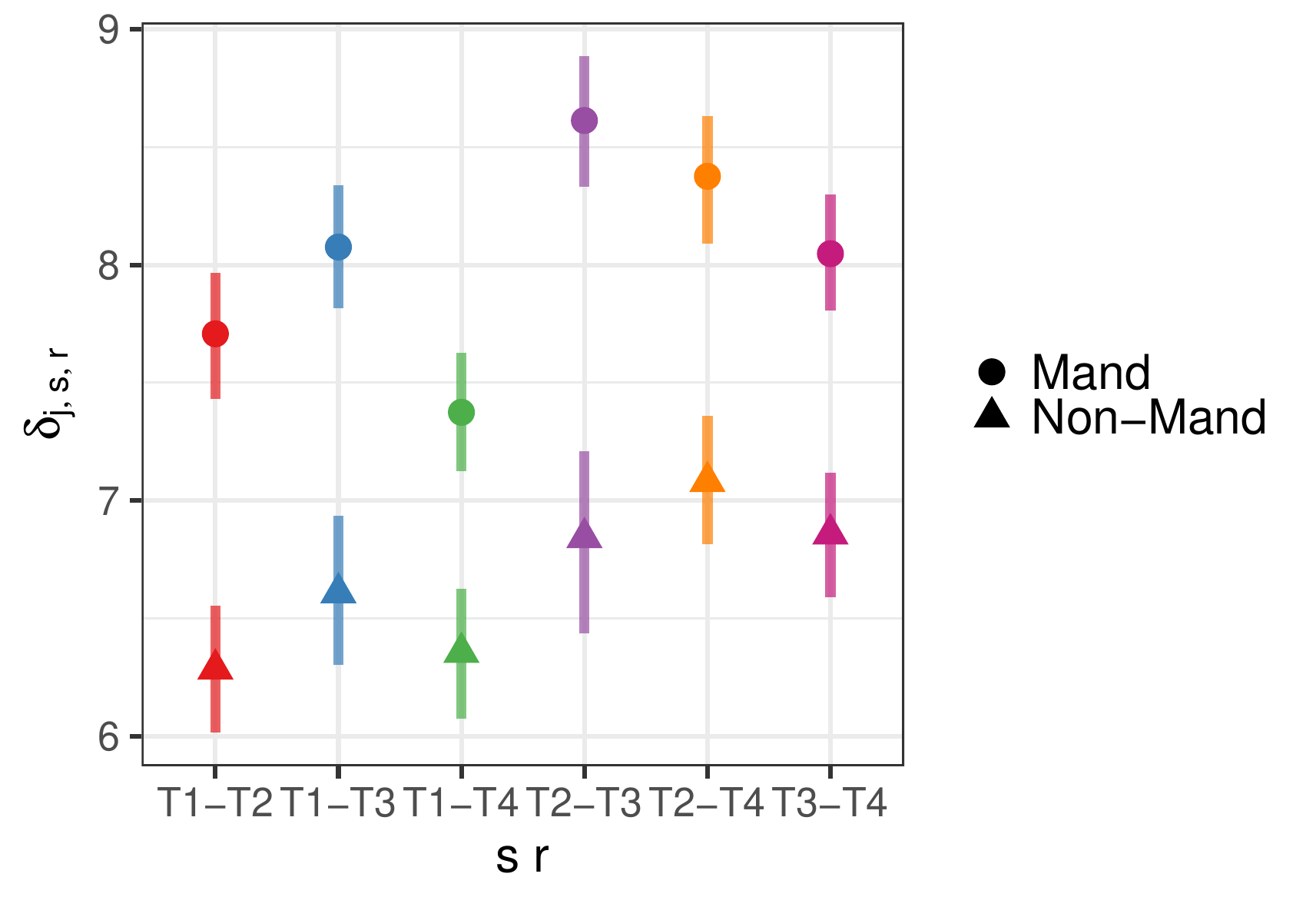}
	\vspace*{-15pt}
	\caption{Results for real data.
    Posterior medians and 90\% credible intervals of the group latent distances $\delta_{j,s,r}$ between stimuli.}
	\label{fig: box_avg_delta_s_real}
\end{figure}

Figure~\ref{fig: box_avg_delta_s_real} shows the posterior point estimates and $90\%$ credible intervals of the group distances $\delta_{j,s,r}$ between each of the $\binom{4}{2}=6$ pairs of tones.
We can see a similar ranking (across groups $j$) of the pairs of tones according to $\delta_{j,s,r}$.
Indeed, after discarding individual differences, in both groups, the pairs of Mandarin tones $\{2,3\}$ and $\{2,4\}$ exhibited the strongest degree of neural dissimilarity, whereas pairs $\{1,2\}$ and $\{1,4\}$ exhibited the strongest degree of neural similarity.
These findings are corroborated by empirical evidence from the preliminary analysis of the data in Section~\ref{sec: data}.
Note that although the rankings between the point estimates are similar, they are not exactly the same across the two groups, and that there is more uncertainty, quantified via the posterior credible intervals, in the non-Mandarin group due to greater variability across these individuals.
Moreover, consistent with recent neuroscience work \citep{llanos2017hidden, reetzke2018tracing}, we observe a better separation between neural representations of tones in the group of native listeners of Mandarin Chinese, relative to the group of non-native listeners.

\begin{figure}[!ht]
	\centering
	\includegraphics[width=0.8\linewidth]{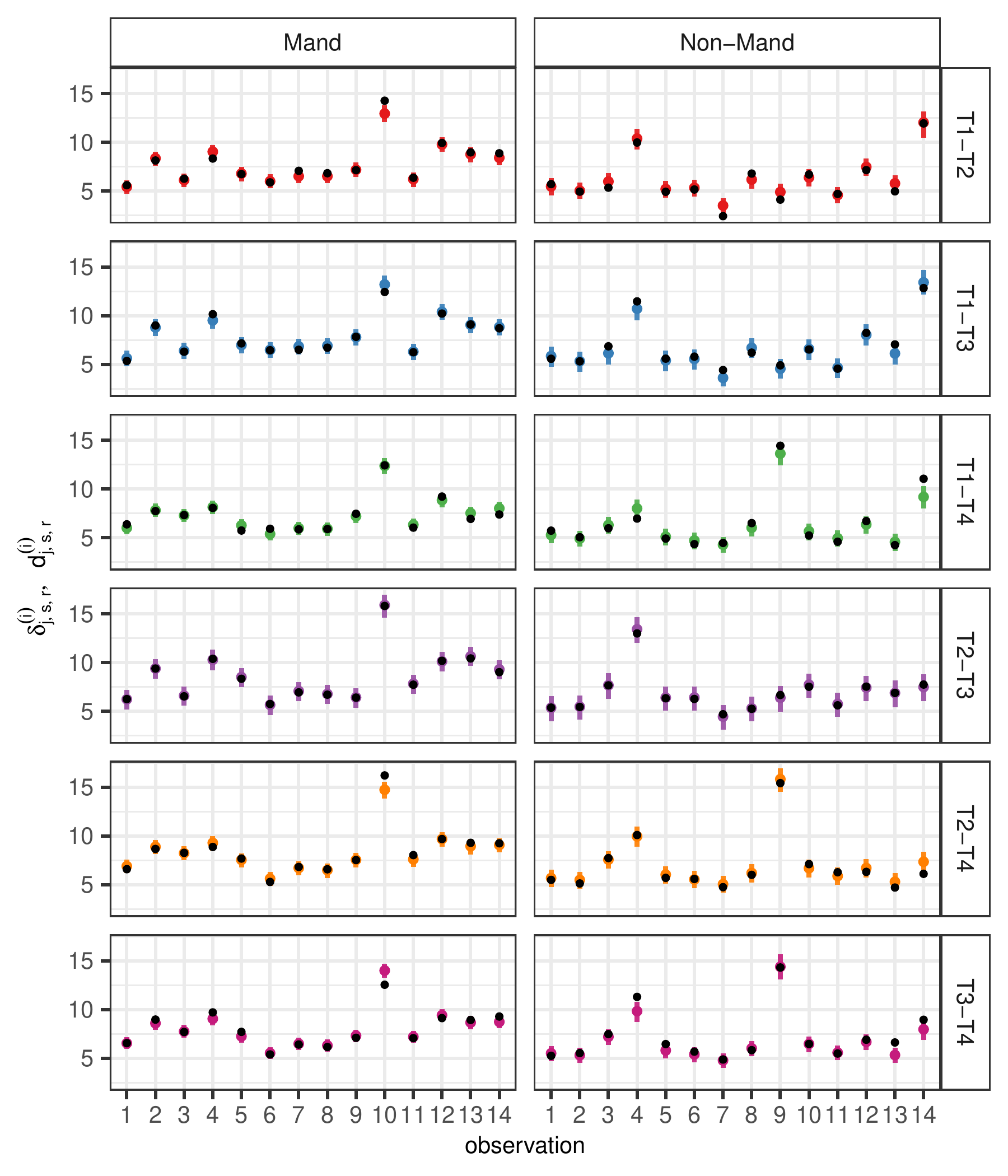}
	\vspace*{-15pt}
	\caption{Results for real data.
    Posterior medians and 90\% credible intervals of the individual latent distances $\delta_{j,s,r}^{(i)}$ between stimuli.
    Black points represent the observed distances $d_{j,s,r}^{(i)}$.
    }
	\label{fig: box_delta_s_i_real_check}
\end{figure}

Figure~\ref{fig: box_delta_s_i_real_check} shows the posterior point estimates and $90\%$ credible intervals of the individual distances $\delta_{j,s,r}^{(i)}$ as well as the corresponding observed distances.
We can see that three latent common features reconstruct the observed distances very well.
This indicates that the neural encoding of differences between multidimensional neural representations can be captured using three latent dimensions.

We also note greater uncertainty in the estimates of the denoised individual distances for the non-Mandarin subjects.
Note that our Bayesian model allows us to quantify such uncertainty via posterior credible intervals while also allowing us to perform proper individual and group comparisons.

\begin{figure}[!ht]
    \centering
	\includegraphics[width=0.6\linewidth, trim={0cm 0cm 0cm 2cm}, clip=true]{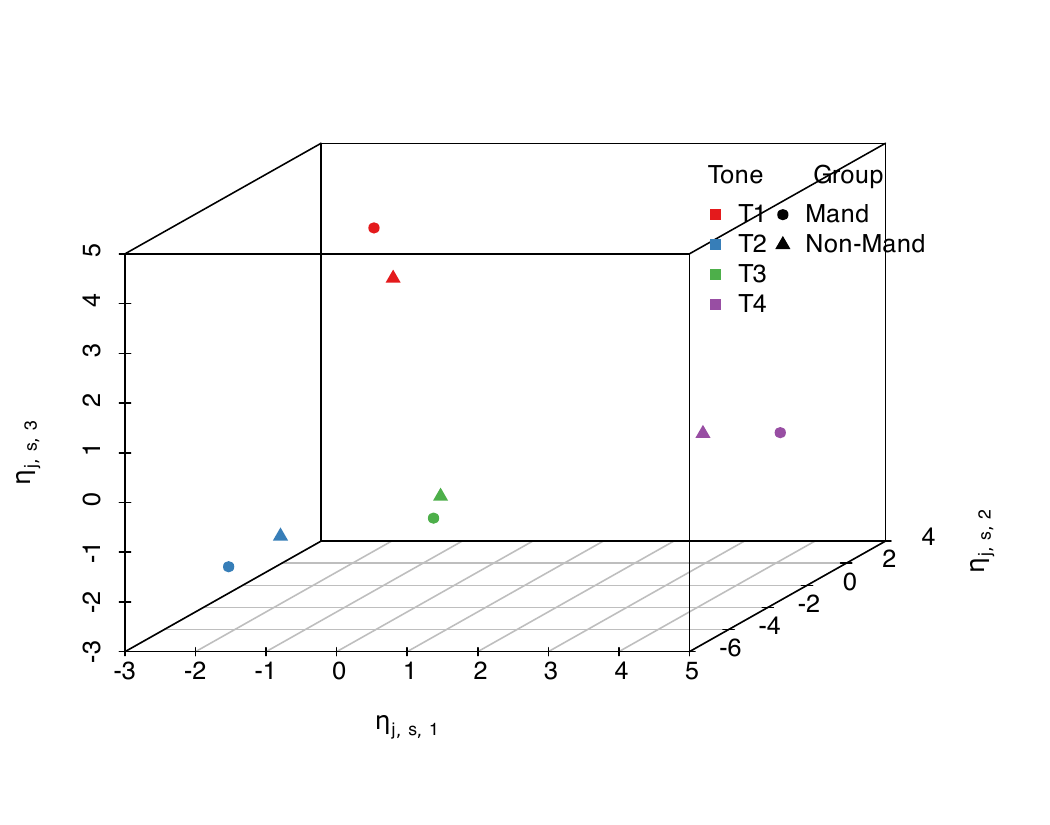}\\
	\includegraphics[height=0.28\linewidth]{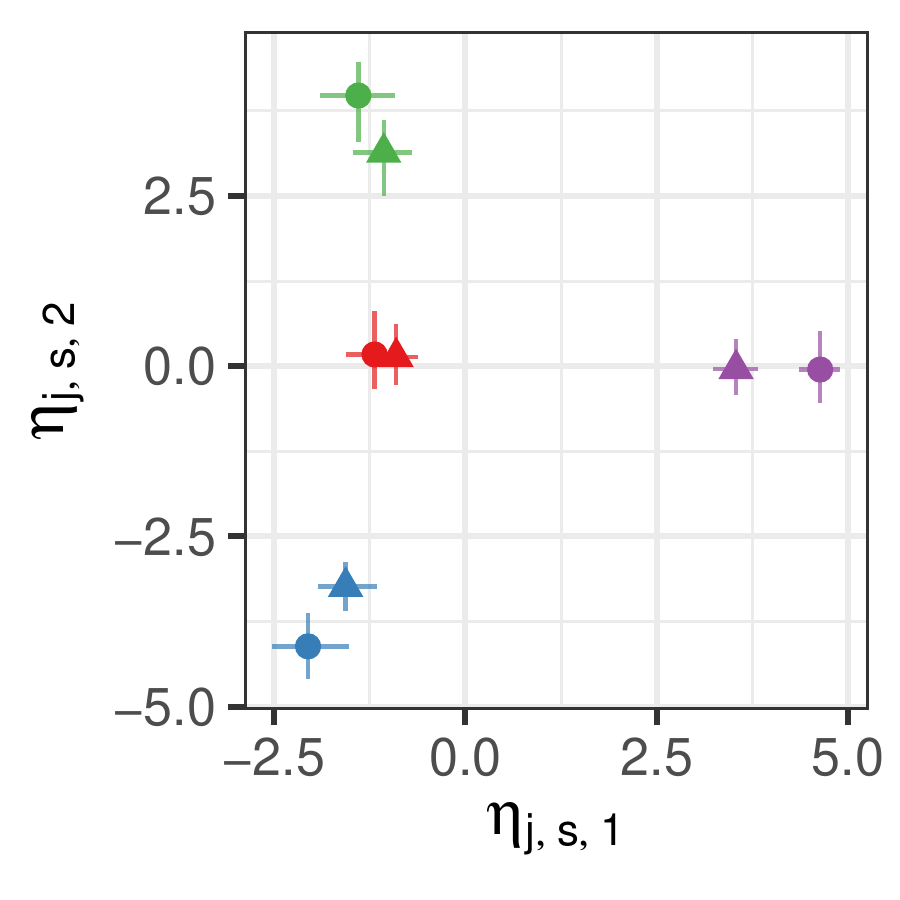}
	\includegraphics[height=0.28\linewidth]{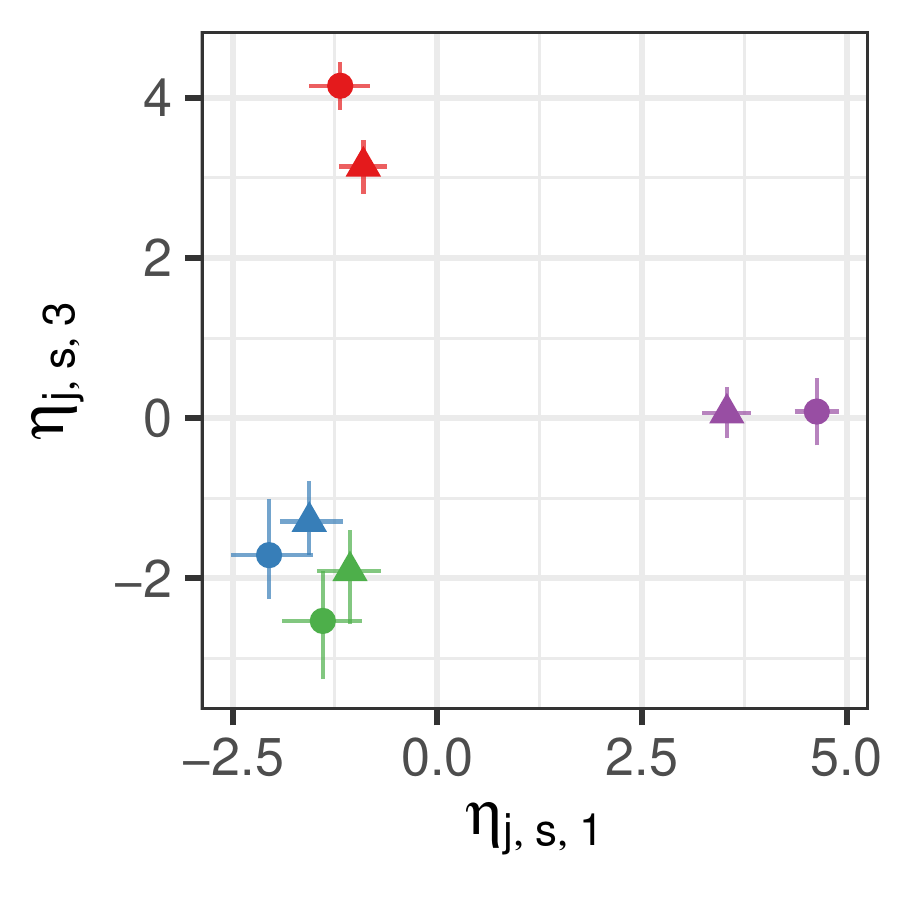}
	\includegraphics[height=0.28\linewidth]{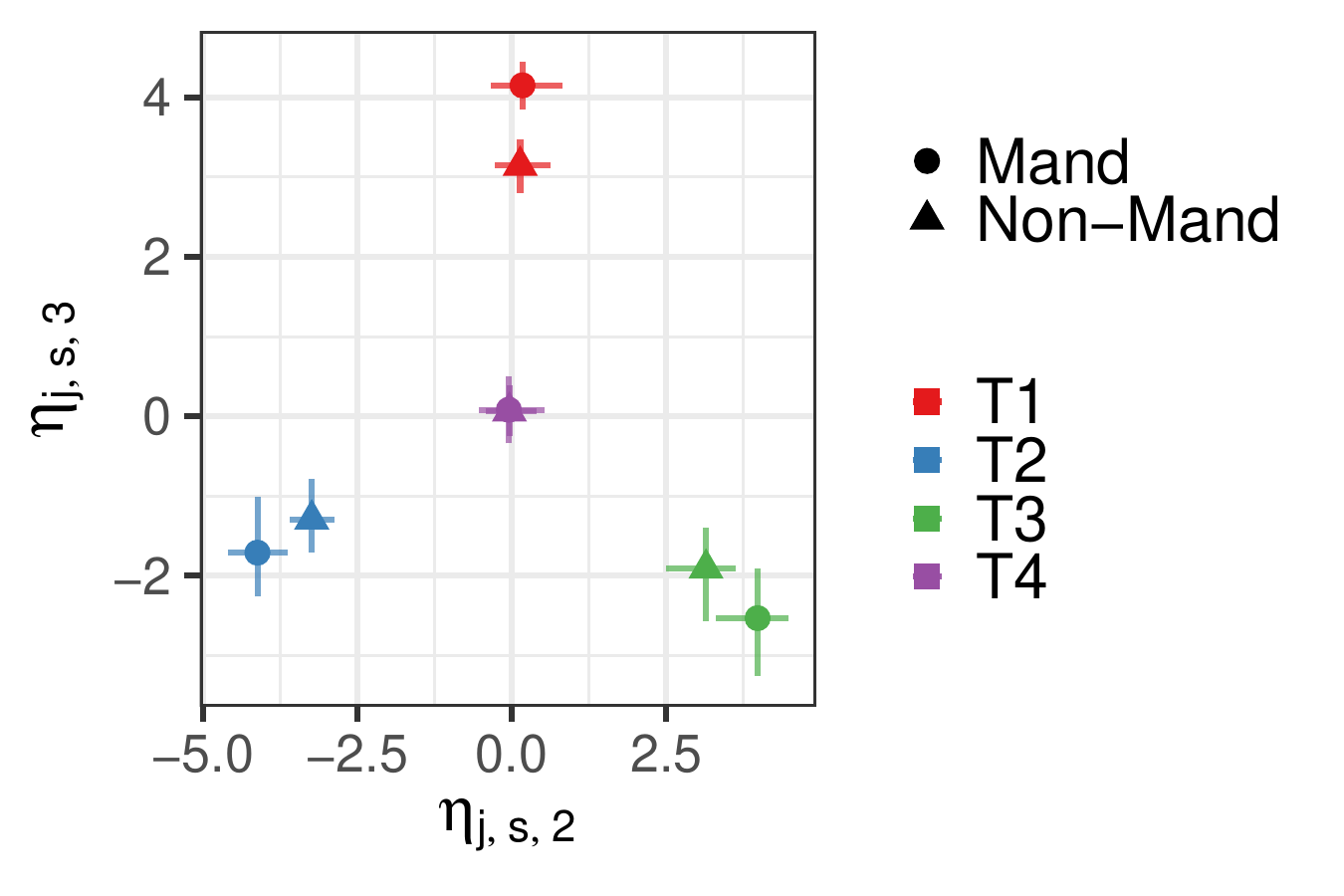}
	\vspace*{-30pt}
	\caption{Results for real data.
	Upper panel: 3-dimensional scatter plot of posterior medians of the group latent feature values $\eta_{j,s,h}$.
	Lower panel: 2-dimensional representation of the posterior medians and 90\% credible intervals of the group latent feature values $\eta_{j,s,h}$ in the different groups.	
    \label{fig: Eta_s_alig_pop_real}}
\end{figure}

\begin{figure}
	\centering
    \includegraphics[width=0.55\linewidth]{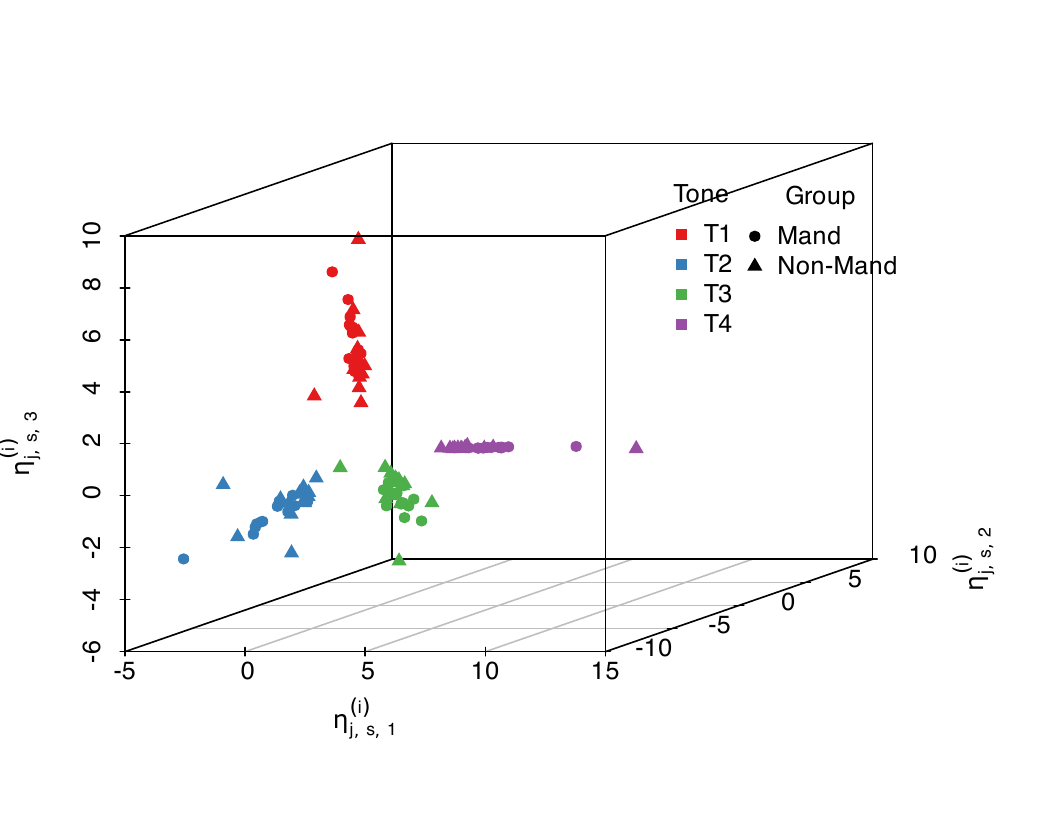}
	\includegraphics[width=0.55\linewidth]{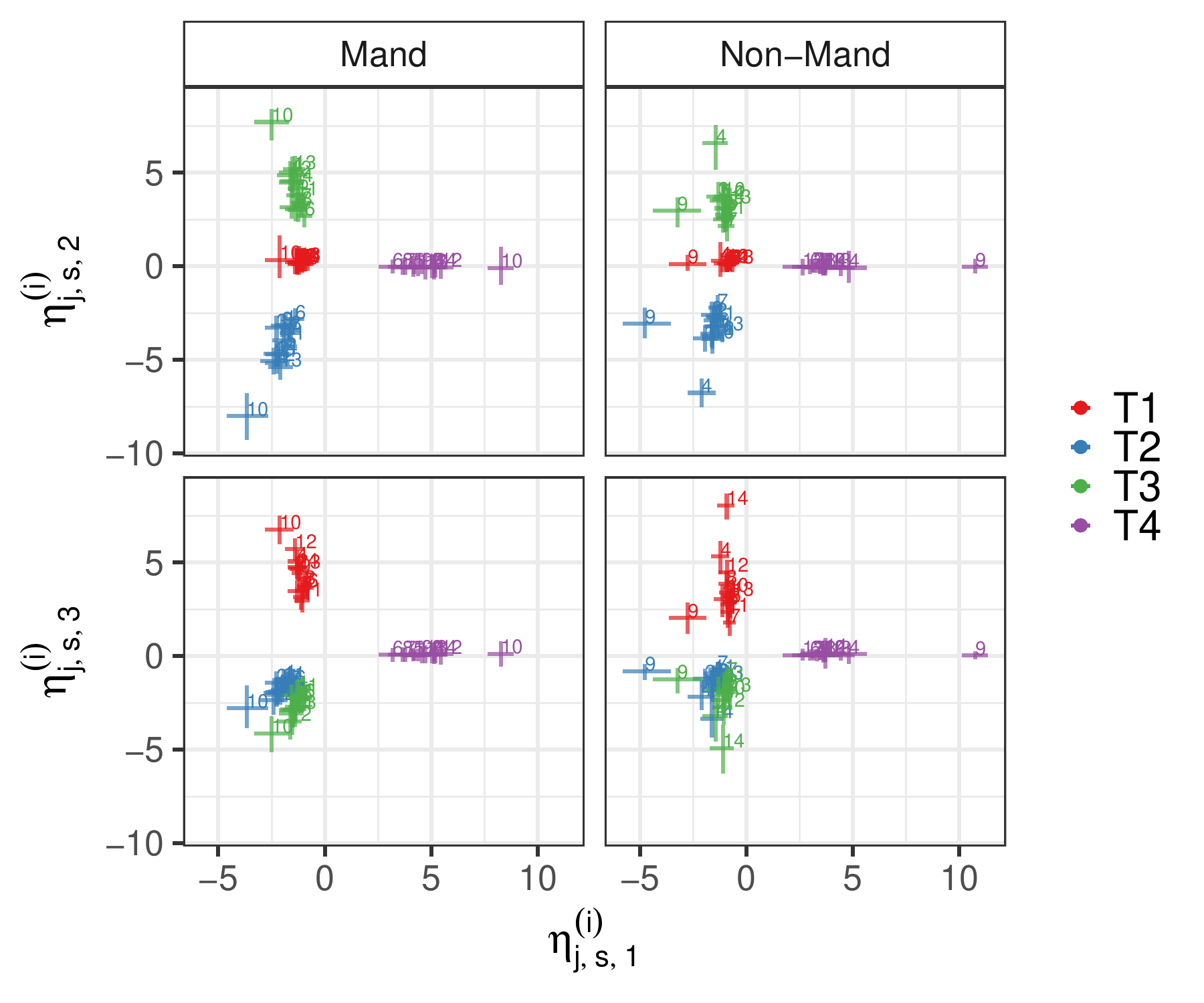}
	\hspace*{0.1cm}
	\includegraphics[width=0.55\linewidth]{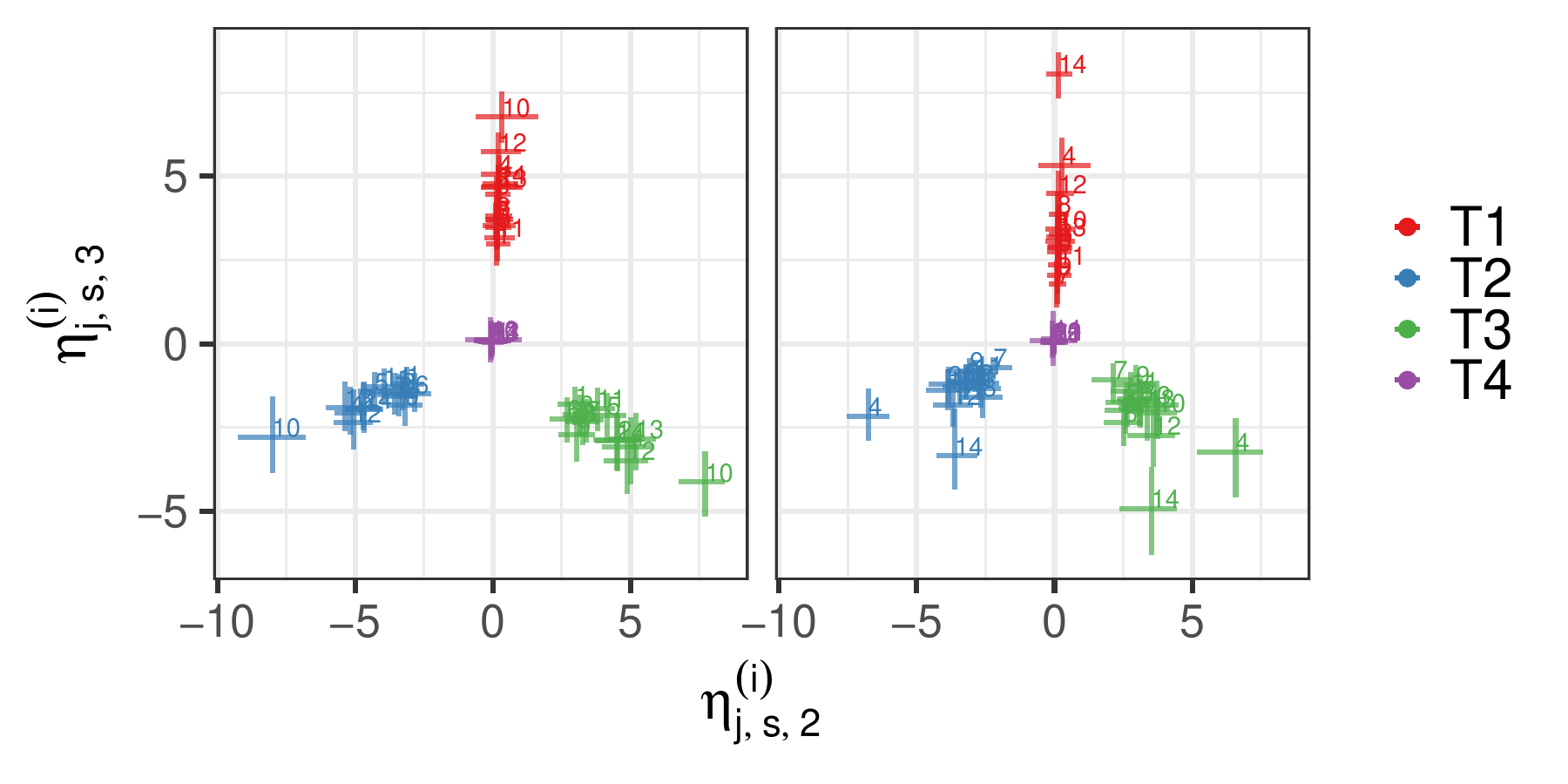}
	\caption{Results for real data.
    Upper panel: 3-dimensional scatter plot of posterior medians of the individual latent features $\eta_{j,s,h}^{(i)}$ between stimuli in the two groups.
	Lower panels: 2-dimensional representation of the posterior medians and 90\% credible intervals of the individual latent feature values $\eta_{j,s,h}^{(i)}$ in the different groups.}
	\label{fig: Eta_s_alig_all_real}
\end{figure}

Figure~\ref{fig: box_delta_s_i_real_check} also highlights the presence of some non-Mandarin speakers (e.g., subjects 7 and 11) who struggle to distinguish (i.e., smaller distances) between neural responses to specific pairs of tones (e.g., \{T1, T2\} and \{T2, T3\}) more than other subjects in the same group.
Importantly, by addressing the identifiability issues, 
we are also able to perform inference on the group and individual latent feature values that reconstruct the distances.
Notably, the model can also be used to infer structural differences at the level of the evoking stimuli with a high degree of specificity.
These differences are depicted in Figure~\ref{fig: Eta_s_alig_pop_real}.
In this figure, the first dimension encodes the difference between the neural representations of T4 (high-falling pitch) and the other tones: T1 (high-level pitch), T2 (low-rising pitch), and T3 (low-dipping pitch).
Because T4 is the high-falling tone, this dimension reflects the neural encoding of differences in pitch direction (falling vs.\ level, rising, and dipping pitch).
In contrast, the second dimension encodes the difference between the neural representations of T3 and T2.
Because T3 and T2 are the low-dipping and low-rising tones, respectively, this dimension captures neural differences in pitch direction for low-onset tones (dipping vs.\ rising pitch).
Lastly, the third dimension encodes the difference between the neural representations of T1 and T3.
Because these tones are the ones with higher (T1) and lower (T3) pitch on average, this dimension may reflect the neural encoding of differences in pitch range (high vs.\ low range).
Figure~\ref{fig: Eta_s_alig_pop_real} also highlights the fact that the three latent features play the same role in the two groups and that the tones are more distinguished (i.e., $\eta_{j,s,h}$ more distant from $0$) in all the features in Mandarin native speakers.
See also the Supplementary Materials for additional plots summarizing these results.
Figure~\ref{fig: Eta_s_alig_all_real} shows the positions of different subjects on the feature space within and across the two groups and tones and the associated $90\%$ credible intervals, presented separately for clarity.
We see that at the individual level as well: the first feature is mainly useful to distinguish tone $4$. 
Aside from a few outlying large distances, the overall reconstruction is robust, confirming that the three-dimensional latent space effectively denoises and summarizes the individual dissimilarity matrices.
The second and third features are mainly useful to represent the other tones, both at the group and individual levels.
Moreover, although we note some similarities between Mandarin and non-Mandarin speakers, as expected, there is more individual variability in the latent features for non-native speakers.
The individual latent features in Figure~\ref{fig: Eta_s_alig_all_real} also identify a few non-Mandarin speakers who struggle to distinguish the Mandarin tones, especially in the third dimension.
Detailed convergence and identifiability diagnostics for the MCMC algorithm for the real data set are provided in Section 3 in the Supplementary Materials.

\section{Discussion} \label{sec: discussion}
In this article, we proposed a novel MDS model for multi-group and multi-subject distances, motivated by auditory neuroscience research into the latent representation of speech sounds.
Applying this model to Mandarin tone perception in native speakers and tone-naive English speakers, we characterized latent features that reconstruct denoised empirical distances across both groups and individuals.
Our framework identified structural differences in neural coding linked to long-term language exposure, while also revealing shared behaviors across groups.

Our proposed approach advances the MDS literature to enable neuroscientists to perform statistical comparisons across multi-subject and multi-group brain distances by mapping them to a biologically interpretable lower-dimensional common feature space.
Our proposal allows borrowing of information across groups and subjects, providing the means for a more comprehensive understanding of how the human brain processes speech signals.
Moving beyond its role as a visualization tool, our proposed MDS framework incorporates formal uncertainty quantification and a data-driven approach for determining the optimal number of latent features.

Although our development is motivated by this auditory neuroscience application, the proposed model can be used to model structural differences between neural representations of other sensory stimuli (e.g., visual signals) captured by neuroimaging technologies such as EEG, fMRI, and MEG.
Prior neuroscience work \citep{kluender2019long, lewicki2002efficient, lotto2016speech, stilp2012efficient} suggests that the neural coding of complex multidimensional sensory stimuli is informed by principles of dimensionality reduction.
Our model could also be used to infer the latent neural dimensions needed by the human brain to encode differences between sensory signals and the impact of different kinds of experiences on the number of dimensions.
Additionally, our approach can be used to address scientific problems in other fields, such as in bioinformatics to find lower-dimensional, biologically relevant features that summarize the recorded single-cell RNA genetic expressions in several biomarkers, e.g., to extend classical MDS used in a standard dimensionality reduction pipeline in bioinformatics research \citep{perraudeau2017bioconductor}.
Such extensions might, however, require additional work to tailor our model to the specific scientific problem.
Additionally, computations may need to be scaled up when datasets with a much larger number of stimuli are available.
Efficient approximations to the posterior, e.g., the variational Bayes algorithm developed for the simpler MDS model in \citet{nguyen2017bayesian} or the Hamiltonian Monte Carlo-based sampler using representative subsamples for likelihood and gradient evaluations instead of the full distance matrix proposed recently by \citet{sheth2026sparse}, may offer viable computational templates for such high-dimensional extensions of our work.

Some topics of our ongoing exploration include extensions to nonparametric frameworks utilizing more flexible likelihood functions, clustering listeners into latent subgroups, incorporating covariates, accommodating asymmetric distance matrices, and adaptations to longitudinal experiments capturing the dynamic evolution of the latent features.

\section*{Financial Support}
This work was supported in part by the National Science Foundation grant NSF DMS-1953712, National Institute on Deafness and Other Communication Disorders grants R01DC013315 and R01DC015504.
G.\ R.\ has been partially supported by the Italian grant MUR, PRIN project 2022CLTYP4.

\baselineskip=16pt
\bibliographystyle{natbib}
\bibliography{RSA_cit}

@article{aguilar2000bayesian,
	title={Bayesian dynamic factor models and portfolio allocation},
	author={Aguilar, Omar and West, Mike},
	journal={Journal of Business \& Economic Statistics},
	volume={18},
	pages={338--357},
	year={2000},
	publisher={Taylor \& Francis}
}

@article{basu2004identifiability,
	title={Identifiability},
	author={Basu, Asit P.},
	journal={Encyclopedia of Statistical Sciences},
	year={2004},
	publisher={Wiley Online Library}
}

@article{bhattacharya2011sparse,
	title = {Sparse {B}ayesian infinite factor models},
	author = {Bhattacharya, Anirban and Dunson, David B.},
	journal = {Biometrika},
	pages = {291--306},
	volume = {98},
	year = {2011}
}

@article{bidelman2011enhanced,
	title={Enhanced brainstem encoding predicts musicians' perceptual advantages with pitch},
	author={Bidelman, Gavin M. and Krishnan, Ananthanarayan and Gandour, Jackson T.},
	journal={European Journal of Neuroscience},
	volume={33},
	pages={530--538},
	year={2011},
	publisher={Wiley Online Library}
}

@book{borg2005modern,
	title={Modern {M}ultidimensional {S}caling: {T}heory and {A}pplications},
	author={Borg, Ingwer and Groenen, Patrick J. F.},
	year={2005},
	publisher={Springer}
}

@article{bradlow2000little,
	title={The little engines that could: {M}odeling the performance of {W}orld {W}ide {W}eb search engines},
	author={Bradlow, Eric T. and Schmittlein, David C.},
	journal={Marketing Science},
	volume={19},
	pages={43--62},
	year={2000}
}

@article{brooks1998general,
	title={General methods for monitoring convergence of iterative simulations},
	author={Brooks, Stephen P. and Gelman, Andrew},
	journal={Journal of Computational and Graphical Statistics},
	volume={7},
	pages={434--455},
	year={1998},
	publisher={Taylor \& Francis}
}

@article{caclin2005acoustic,
	title={Acoustic correlates of timbre space dimensions: a confirmatory study using synthetic tones},
	author={Caclin, Anne and McAdams, Stephen and Smith, Bennett K. and Winsberg, Suzanne},
	journal={The Journal of the Acoustical Society of America},
	volume={118},
	pages={471--482},
	year={2005},
	publisher={Acoustical Society of America}
}

@article{caclin2006separate,
	title={Separate neural processing of timbre dimensions in auditory sensory memory},
	author={Caclin, Anne and Brattico, Elvira and Tervaniemi, Mari and N{\"a}{\"a}t{\"a}nen, Risto and Morlet, Dominique and Giard, Marie-H{\'e}l{\`e}ne and McAdams, Stephen},
	journal={Journal of Cognitive Neuroscience},
	volume={18},
	pages={1959--1972},
	year={2006}
}

@article{carroll1970analysis,
	title = {Analysis of individual differences in multidimensional scaling via an {N}-way generalization of ``{E}ckart--{Y}oung'' decomposition},
	author = {Carroll, J. Douglas and Chang, Jih-Jie},
	journal = {Psychometrika},
	pages = {283--319},
	volume = {35},
	year = {1970}
}

@article{carroll1980multidimensional,
	title={Multidimensional scaling},
	author={Carroll, J. Douglas and Arabie, Phipps},
	journal={Annual Review of Psychology},
	volume={31},
	pages={607--649},
	year={1980}
}

@article{chandrasekaran2007mismatch,
	title={Mismatch negativity to pitch contours is influenced by language experience},
	author={Chandrasekaran, Bharath and Krishnan, Ananthanarayan and Gandour, Jackson T.},
	journal={Brain Research},
	volume={1128},
	pages={148--156},
	year={2007},
	publisher={Elsevier}
}

@article{chandrasekaran2007neuroplasticity,
	title={Neuroplasticity in the processing of pitch dimensions: a multidimensional scaling analysis of the mismatch negativity},
	author={Chandrasekaran, Bharath and Gandour, Jackson T. and Krishnan, Ananthanarayan},
	journal={Restorative Neurology and Neuroscience},
	volume={25},
	pages={195--210},
	year={2007},
	publisher={IOS Press}
}

@misc{commandeur1993mathematical,
	title={Mathematical derivations in the proximity scaling ({PROXSCAL}) of symmetric data matrices},
	author={Commandeur, Jacques J. F. and Heiser, Willem J.},
	year={1993},
	publisher={University of Leiden}
}

@article{de1980multidimensional,
	title={Multidimensional scaling with restrictions on the configuration},
	author={De Leeuw, Jan and Heiser, Willem J.},
	journal={Multivariate Analysis},
	volume={5},
	pages={501--522},
	year={1980},
	publisher={North Holland, Amsterdam}
}

@article{desarbo1989stochastic,
	title={A stochastic multidimensional scaling vector threshold model for the spatial representation of “pick any/n” data},
	author={DeSarbo, Wayne S. and Cho, Jaewun},
	journal={Psychometrika},
	volume={54},
	pages={105--129},
	year={1989},
	publisher={Springer}
}

@article{desarbo1998bayesian,
	title={A {B}ayesian approach to the spatial representation of market structure from consumer choice data},
	author={DeSarbo, Wayne S. and Kim, Youngchan and Wedel, Michel and Fong, Duncan K. H.},
	journal={European Journal of Operational Research},
	volume={111},
	pages={285--305},
	year={1998},
	publisher={Elsevier}
}

@article{desarbo2008model,
	title={A model-based approach for visualizing the dimensional structure of ordered successive categories preference data},
	author={DeSarbo, Wayne S. and Park, Joonwook and Scott, Crystal J.},
	journal={Psychometrika},
	volume={73},
	pages={1--20},
	year={2008},
	publisher={Springer}
}

@article{di2015low,
	title={Low-frequency cortical entrainment to speech reflects phoneme-level processing},
	author={Di Liberto, Giovanni M. and O’Sullivan, James A. and Lalor, Edmund C.},
	journal={Current Biology},
	volume={25},
	pages={2457--2465},
	year={2015},
	publisher={Elsevier}
}

@article{durante2017note,
	title={A note on the multiplicative gamma process},
	author={Durante, Daniele},
	journal={Statistics \& Probability Letters},
	volume={122},
	pages={198--204},
	year={2017},
	publisher={Elsevier}
}

@article{feng2019role,
	title={The role of the human auditory corticostriatal network in speech learning},
	author={Feng, Gangyi and Yi, Han Gyol and Chandrasekaran, Bharath},
	journal={Cerebral Cortex},
	volume={29},
	pages={4077--4089},
	year={2019},
	publisher={Oxford University Press}
}

@article{feng2021emerging,
	title={Emerging native-similar neural representations underlie non-native speech category learning success},
	author={Feng, Gangyi and Li, Yu and Hsu, Shen-Mou and Wong, Patrick C. M. and Chou, Tai-Li and Chandrasekaran, Bharath},
	journal={Neurobiology of Language},
	volume={2},
	pages={280--307},
	year={2021},
	publisher={MIT Press, Cambridge, MA}
}

@article{feng2021distributed,
	title={A distributed dynamic brain network mediates linguistic tone representation and categorization},
	author={Feng, Gangyi and Gan, Zhenzhong and Llanos, Fernando and Meng, Danting and Wang, Suiping and Wong, Patrick C. M. and Chandrasekaran, Bharath},
	journal={Neuroimage},
	volume={224},
	pages={1--13},
	year={2021},
	publisher={Elsevier}
}

@article{fong2010bayesian,
	title={A {B}ayesian vector multidimensional scaling procedure for the analysis of ordered preference data},
	author={Fong, Duncan K. H. and DeSarbo, Wayne S. and Park, Joonwook and Scott, Crystal J.},
	journal={Journal of the American Statistical Association},
	volume={105},
	pages={482--492},
	year={2010},
	publisher={Taylor \& Francis}
}

@incollection{gandour1978perception,
	title={The perception of tone},
	author={Gandour, Jackson T.},
	booktitle={Tone},
	pages={41--76},
	year={1978},
	publisher={Elsevier}
}

@article{gandour1978crosslanguage,
	title={Crosslanguage differences in tone perception: a multidimensional scaling investigation},
	author={Gandour, Jackson T. and Harshman, Richard A.},
	journal={Language and Speech},
	volume={21},
	pages={1--33},
	year={1978},
	publisher={SAGE Publications Sage UK: London, England}
}

@article{gandour1983tone,
	title={Tone perception in Far Eastern languages},
	author={Gandour, Jack},
	journal={Journal of Phonetics},
	volume={11},
	pages={149--175},
	year={1983},
	publisher={Elsevier}
}

@article{gelman1992inference,
	title={Inference from iterative simulation using multiple sequences},
	author={Gelman, Andrew and Rubin, Donald B.},
	journal={Statistical Science},
	volume={7},
	pages={457--472},
	year={1992},
	publisher={Institute of Mathematical Statistics}
}

@article{holbrook2021massive,
	title={Massive parallelization boosts big {B}ayesian multidimensional scaling},
	author={Holbrook, Andrew J. and Lemey, Philippe and Baele, Guy and Dellicour, Simon and Brockmann, Dirk and Rambaut, Andrew and Suchard, Marc A.},
	journal={Journal of Computational and Graphical Statistics},
	volume={30},
	pages={11--24},
	year={2021},
	publisher={Taylor \& Francis}
}

@article{khalighinejad2017dynamic,
	title={Dynamic encoding of acoustic features in neural responses to continuous speech},
	author={Khalighinejad, Bahar and da Silva, Guilherme Cruzatto and Mesgarani, Nima},
	journal={Journal of Neuroscience},
	volume={37},
	pages={2176--2185},
	year={2017},
	publisher={Soc Neuroscience}
}

@article{kluender2019long,
	title={Long-standing problems in speech perception dissolve within an information-theoretic perspective},
	author={Kluender, Keith R. and Stilp, Christian E. and Llanos, Fernando},
	journal={Attention, Perception, \& Psychophysics},
	volume={81},
	pages={861--883},
	year={2019},
	publisher={Springer}
}

@article{krishnan2005encoding,
	title={Encoding of pitch in the human brainstem is sensitive to language experience},
	author={Krishnan, Ananthanarayan and Xu, Yisheng and Gandour, Jackson and Cariani, Peter},
	journal={Cognitive Brain Research},
	volume={25},
	pages={161--168},
	year={2005},
	publisher={Elsevier}
}

@article{krishnan2010effects,
	title={The effects of tone language experience on pitch processing in the brainstem},
	author={Krishnan, Ananthanarayan and Gandour, Jackson T. and Bidelman, Gavin M.},
	journal={Journal of Neurolinguistics},
	volume={23},
	pages={81--95},
	year={2010},
	publisher={Elsevier}
}

@article{kruskal1964multidimensional,
	title={Multidimensional scaling by optimizing goodness of fit to a nonmetric hypothesis},
	author={Kruskal, Joseph B.},
	journal={Psychometrika},
	volume={29},
	pages={1--27},
	year={1964},
	publisher={Springer-Verlag}
}

@article{kruskal1964nonmetric,
	title={Nonmetric multidimensional scaling: a numerical method},
	author={Kruskal, Joseph B.},
	journal={Psychometrika},
	volume={29},
	pages={115--129},
	year={1964},
	publisher={Springer-Verlag}
}

@article{lewicki2002efficient,
	title={Efficient coding of natural sounds},
	author={Lewicki, Michael S.},
	journal={Nature Neuroscience},
	volume={5},
	pages={356--363},
	year={2002},
	publisher={Nature Publishing Group}
}

@article{lin2019bayesian,
	title={Bayesian multidimensional scaling procedure with variable selection},
	author={Lin, Lin and Fong, Duncan K. H.},
	journal={Computational Statistics \& Data Analysis},
	volume={129},
	pages={1--13},
	year={2019},
	publisher={Elsevier}
}

@article{liu2024bayesian,
	title={Bayesian hyperbolic multidimensional scaling},
	author={Liu, Bolun and Lubold, Shane and Raftery, Adrian E. and McCormick, Tyler H.},
	journal={Journal of Computational and Graphical Statistics},
	volume={33},
	pages={869--882},
	year={2024},
	publisher={Taylor \& Francis}
}

@article{llanos2017hidden,
	title={Hidden {M}arkov modeling of frequency-following responses to {M}andarin lexical tones},
	author={Llanos, Fernando and Xie, Zilong and Chandrasekaran, Bharath},
	journal={Journal of Neuroscience Methods},
	volume={291},
	pages={101--112},
	year={2017},
	publisher={Elsevier}
}

@article{llanos2021neural,
	title={The neural processing of pitch accents in continuous speech},
	author={Llanos, Fernando and German, James S. and Gnanateja, G. Nike and Chandrasekaran, Bharath},
	journal={Neuropsychologia},
	volume={158},
	pages={1--12},
	year={2021},
	publisher={Elsevier}
}

@incollection{lotto2016speech,
	title={Speech perception: the view from the auditory system},
	author={Lotto, Andrew J. and Holt, Lori L.},
	booktitle={Neurobiology of Language},
	pages={185--194},
	year={2016},
	publisher={Elsevier}
}

@article{mesgarani2014phonetic,
	title={Phonetic feature encoding in human superior temporal gyrus},
	author={Mesgarani, Nima and Cheung, Connie and Johnson, Keith and Chang, Edward F},
	journal={Science},
	volume={343},
	pages={1006--1010},
	year={2014},
	publisher={American Association for the Advancement of Science}
}

@article{nguyen2017bayesian,
	title = {Bayesian unidimensional scaling for visualizing uncertainty in high dimensional datasets with latent ordering of observations},
	author = {Nguyen, Lan Huong and Holmes, Susan},
	issn = {14712105},
	journal = {BMC Bioinformatics},
	pages = {65--79},
	publisher = {BioMed Central Ltd},
	volume = {18},
	year = {2017}
}

@article{nili2014toolbox,
	title={A toolbox for representational similarity analysis},
	author={Nili, Hamed and Wingfield, Cai and Walther, Alexander and Su, Li and Marslen-Wilson, William and Kriegeskorte, Nikolaus},
	journal={PLoS Computational Biology},
	volume={10},
	pages={1--11},
	year={2014},
	publisher={Public Library of Science San Francisco, USA}
}

@article{oh2001bayesian,
	title = {Bayesian multidimensional scaling and choice of dimension},
	author = {Oh, Man Suk and Raftery, Adrian E.},
	journal = {Journal of the American Statistical Association},
	pages = {1031--1044},
	publisher = {Taylor {\&} Francis},
	volume = {96},
	year = {2001}
}

@article{papastamoulis2022identifiability,
	title={On the identifiability of {B}ayesian factor analytic models},
	author={Papastamoulis, Panagiotis and Ntzoufras, Ioannis},
	journal={Statistics and Computing},
	volume={32},
	pages={1--29},
	year={2022}
}

@article{park2008hierarchical,
	title={A hierarchical {B}ayesian multidimensional scaling methodology for accommodating both structural and preference heterogeneity},
	author={Park, Joonwook and DeSarbo, Wayne S. and Liechty, John},
	journal={Psychometrika},
	volume={73},
	pages={451--472},
	year={2008},
	publisher={Springer-Verlag}
}

@article{perraudeau2017bioconductor,
	title={Bioconductor workflow for single-cell {RNA} sequencing: normalization, dimensionality reduction, clustering, and lineage inference},
	author={Perraudeau, Fanny and Risso, Davide and Street, Kelly and Purdom, Elizabeth and Dudoit, Sandrine},
	journal={F1000Research},
	volume={6},
	pages={1--28},
	year={2017},
	publisher={Faculty of 1000 Ltd}
}

@article{praturu2022bayesian,
	title={A {B}ayesian approach to hyperbolic multi-dimensional scaling},
	author={Praturu, Anoop and Sharpee, Tatyana},
	journal={bioRxiv},
	pages={2022--10},
	year={2022},
	publisher={Cold Spring Harbor Laboratory}
}

@article{ramsay1977maximum,
	title={Maximum likelihood estimation in multidimensional scaling},
	author={Ramsay, James O.},
	journal={Psychometrika},
	volume={42},
	pages={241--266},
	year={1977},
	publisher={Springer-Verlag}
}

@article{ramsay1982some,
	title={Some statistical approaches to multidimensional scaling data},
	author={Ramsay, James O.},
	journal={Journal of the Royal Statistical Society: Series A},
	volume={145},
	pages={285--303},
	year={1982},
	publisher={Wiley Online Library}
}

@article{raizada2010linking,
	title={Linking brain-wide multivoxel activation patterns to behaviour: examples from language and math},
	author={Raizada, Rajeev D. S. and Tsao, Feng-Ming and Liu, Huei-Mei and Holloway, Ian D. and Ansari, Daniel and Kuhl, Patricia K.},
	journal={Neuroimage},
	volume={51},
	pages={462--471},
	year={2010},
	publisher={Elsevier}
}

@Manual{rcoreteam2024r,
	title={R: a language and environment for statistical computing},
	author={{R Core Team}},
	organization={R Foundation for Statistical Computing},
	address={Vienna, Austria},
	year={2024},
	url={https://www.R-project.org/},
}

@article{reetzke2018tracing,
	title={Tracing the trajectory of sensory plasticity across different stages of speech learning in adulthood},
	author={Reetzke, Rachel and Xie, Zilong and Llanos, Fernando and Chandrasekaran, Bharath},
	journal={Current Biology},
	volume={28},
	pages={1419--1427},
	year={2018},
	publisher={Elsevier}
}

@article{richardson1997bayesian,
	title={On {B}ayesian analysis of mixtures with an unknown number of components (with discussion)},
	author={Richardson, Sylvia and Green, Peter J.},
	journal={Journal of the Royal Statistical Society Series B: Statistical Methodology},
	volume={59},
	pages={731--792},
	year={1997},
	publisher={Oxford University Press}
}

@article{roberts2007coupling,
	title={Coupling and ergodicity of adaptive {M}arkov chain {M}onte {C}arlo algorithms},
	author={Roberts, Gareth O. and Rosenthal, Jeffrey S.},
	journal={Journal of Applied Probability},
	volume={44},
	pages={458--475},
	year={2007},
	publisher={Cambridge University Press}
}

@article{roberts2009examples,
	title={Examples of adaptive {MCMC}},
	author={Roberts, Gareth O. and Rosenthal, Jeffrey S.},
	journal={Journal of Computational and Graphical Statistics},
	volume={18},
	pages={349--367},
	year={2009},
	publisher={Taylor \& Francis}
}

@article{sheth2026sparse,
	title={Sparse {B}ayesian multidimensional scaling (s)},
	author={Sheth, Ami and Smith, Aaron and Holbrook, Andrew J.},
	journal={Computational Statistics},
	volume={41},
	pages={12},
	year={2026},
	publisher={Springer}
}

@article{shepard1962analysis,
	title={The analysis of proximities: multidimensional scaling with an unknown distance function. I.},
	author={Shepard, Roger N.},
	journal={Psychometrika},
	volume={27},
	pages={125--140},
	year={1962},
	publisher={Springer-Verlag}
}

@article{stilp2012efficient,
	title={Efficient coding and statistically optimal weighting of covariance among acoustic attributes in novel sounds},
	author={Stilp, Christian E. and Kluender, Keith R.},
	journal={PLoS One},
	volume={7},
	pages={1--13},
	year={2012},
	publisher={Public Library of Science San Francisco, USA}
}

@article{swartz2004bayesian,
	title={Bayesian identifiability and misclassification in multinomial data},
	author={Swartz, Tim B. and Haitovsky, Yoel and Vexler, Albert and Yang, Tae Y.},
	journal={Canadian Journal of Statistics},
	volume={32},
	pages={285--302},
	year={2004},
	publisher={Wiley Online Library}
}

@article{takane1977nonmetric,
	title={Nonmetric individual differences multidimensional scaling: an alternating least squares method with optimal scaling features},
	author={Takane, Yoshio and Young, Forrest W. and De Leeuw, Jan},
	journal={Psychometrika},
	volume={42},
	pages={7--67},
	year={1977},
	publisher={Springer}
}

@article{takane1981nonmetric,
	title={Nonmetric maximum likelihood multidimensional scaling from directional rankings of similarities},
	author={Takane, Yoshio and Carroll, J. Douglas},
	journal={Psychometrika},
	volume={46},
	pages={389--405},
	year={1981},
	publisher={Springer-Verlag}
}

@article{torgerson1952multidimensional,
	title={Multidimensional scaling: {I}. {T}heory and method},
	author={Torgerson, Warren S.},
	journal={Psychometrika},
	volume={17},
	pages={401--419},
	year={1952},
	publisher={Springer}
}

@book{torgerson1958theory,
	title={Theory and {M}ethods of {S}caling},
	author={Torgerson, Warren S.},
	year={1958},
	publisher={New York, Wiley}
}

@article{yanchenko2020hierarchical,
	title={Hierarchical multidimensional scaling for the comparison of musical performance styles},
	author={Yanchenko, Anna K. and Hoff, Peter D.},
	journal={Annals of Applied Statistics},
	volume={14},
	pages={1581--1603},
	year={2020},
	publisher={Institute of Mathematical Statistics}
}

@article{young1978alscal,
	title={{ALSCAL}: a nonmetric multidimensional scaling program with several individual-differences options},
	author={Young, Forrest W. and Takane, Yoshio and Lewyckyj, Rostyslaw},
	journal={Behavior Research Methods \& Instrumentation},
	volume={10},
	pages={451--453},
	year={1978},
	publisher={Springer-Verlag New York}
}

@article{zinnes1977single,
	title={Single and multidimensional same-different judgments},
	author={Zinnes, Joseph L. and Wolff, Ronald P.},
	journal={Journal of Mathematical Psychology},
	volume={16},
	pages={30--50},
	year={1977},
	publisher={Elsevier}
}

@article{zinszer2016semantic,
	title={Semantic structural alignment of neural representational spaces enables translation between {E}nglish and {C}hinese words},
	author={Zinszer, Benjamin D. and Anderson, Andrew J. and Kang, Olivia and Wheatley, Thalia and Raizada, Rajeev D. S.},
	journal={Journal of Cognitive Neuroscience},
	volume={28},
	pages={1749--1759},
	year={2016}
}

\clearpage\pagebreak\newpage
\pagestyle{fancy}
\fancyhf{}
\rhead{\bfseries\thepage}
\lhead{\bfseries SUPPLEMENTARY MATERIALS}

\baselineskip=27pt
\begin{center}
{\LARGE{Supplementary Materials for\\} 
\bf  Bayesian Mixed Multidimensional Scaling\\ for Auditory Processing
}
\end{center}

\baselineskip=12pt
\vskip 2mm

\begin{center}
        Giovanni Rebaudo$^{1}$ (giovanni.rebaudo@unito.it)\\
        Fernando Llanos$^{2}$ (fllanos@utexas.edu)\\
        Bharath Chandrasekaran$^{3}$ (bchandra@northwestern.edu)\\
        Abhra Sarkar$^{4}$ (abhra.sarkar@utexas.edu)
		\vskip 3mm
		$^{1}$ESOMAS Department, University of Torino and Collegio Carlo Alberto
		\vskip 4pt
		$^{2}$Department of Linguistics, University of Texas at Austin
		\vskip 4pt 
		$^{3}$Department of Communication Science and Disorders, Northwestern University
		\vskip 4pt $^{4}$Department of Statistics and Data Sciences, University of Texas at Austin
\end{center}

\setcounter{equation}{0}
\setcounter{page}{1}
\setcounter{table}{1}
\setcounter{figure}{0}
\setcounter{section}{0}
\numberwithin{table}{section}
\renewcommand{\theequation}{S.\arabic{equation}}
\renewcommand{\thesubsection}{S.\arabic{section}.\arabic{subsection}}
\renewcommand{\thesection}{S.\arabic{section}}
\renewcommand{\theThm}{S.\arabic{Thm}}
\renewcommand{\theCor}{S.\arabic{Cor}}
\renewcommand{\theProp}{S.\arabic{Prop}}
\renewcommand{\theLem}{S.\arabic{Lem}}
\renewcommand{\thepage}{S.\arabic{page}}
\renewcommand{\thetable}{S.\arabic{table}}
\renewcommand{\thefigure}{S.\arabic{figure}}

\baselineskip=14pt 
\vskip 10mm

Supplementary materials present additional figures of the results of the tone distances application and simulated data analysis and details on software and diagnostics.

\FloatBarrier
\section{Additional Figures: Real Data}\label{sec:Add Fig}
\begin{figure}[!ht]
	\centering
	\includegraphics[width=0.7\linewidth]{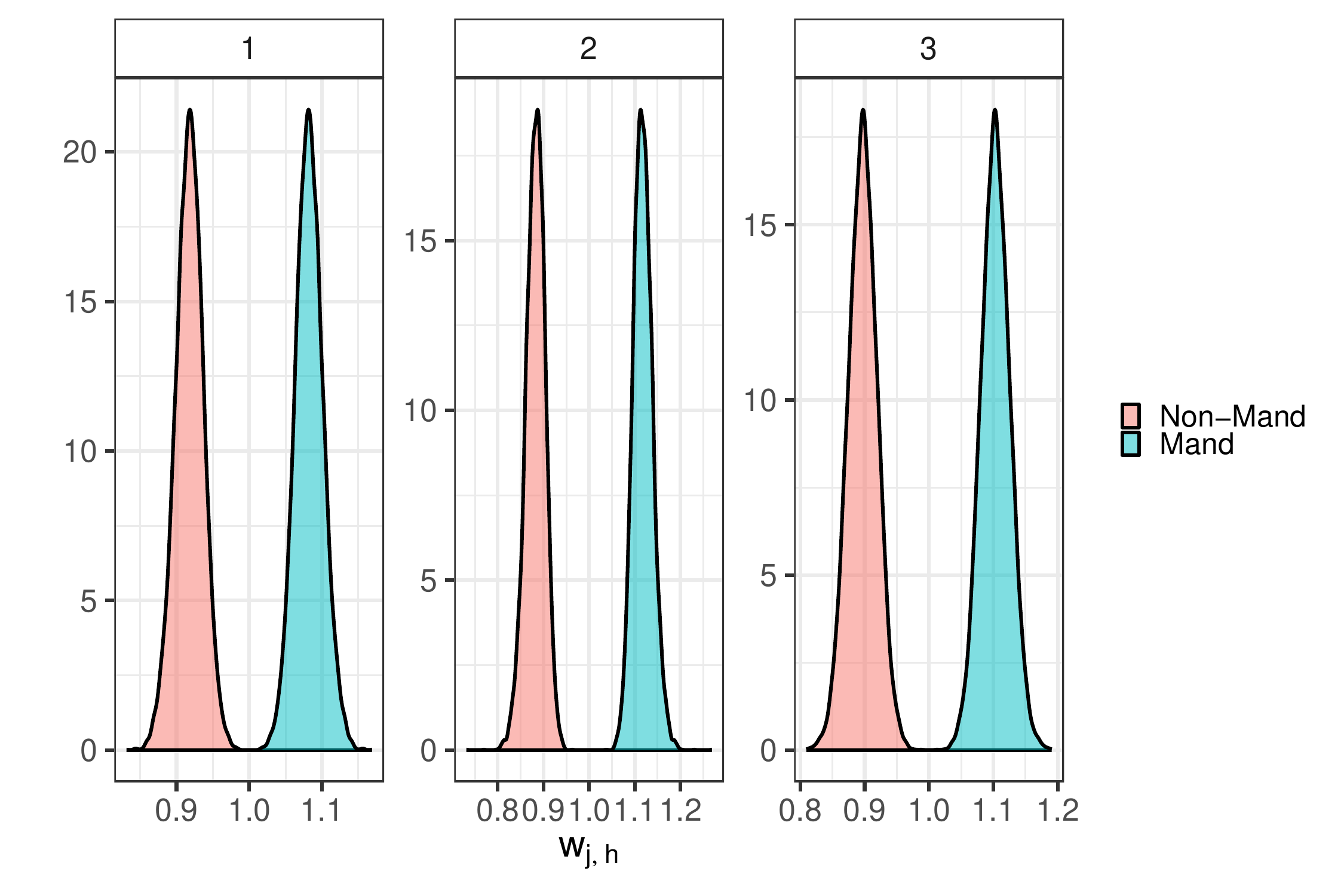}
    \vspace*{-15pt}
	\caption{Results for the tone neural distance data.
    Posterior densities of the post-processed group-and-dimension-specific weights $w_{j,h}$ in the two groups.}
	\label{fig:w_jh}
\end{figure}

\begin{figure}[!ht]
    \vspace*{-40pt}
	\centering
	\includegraphics[width=0.6\linewidth]{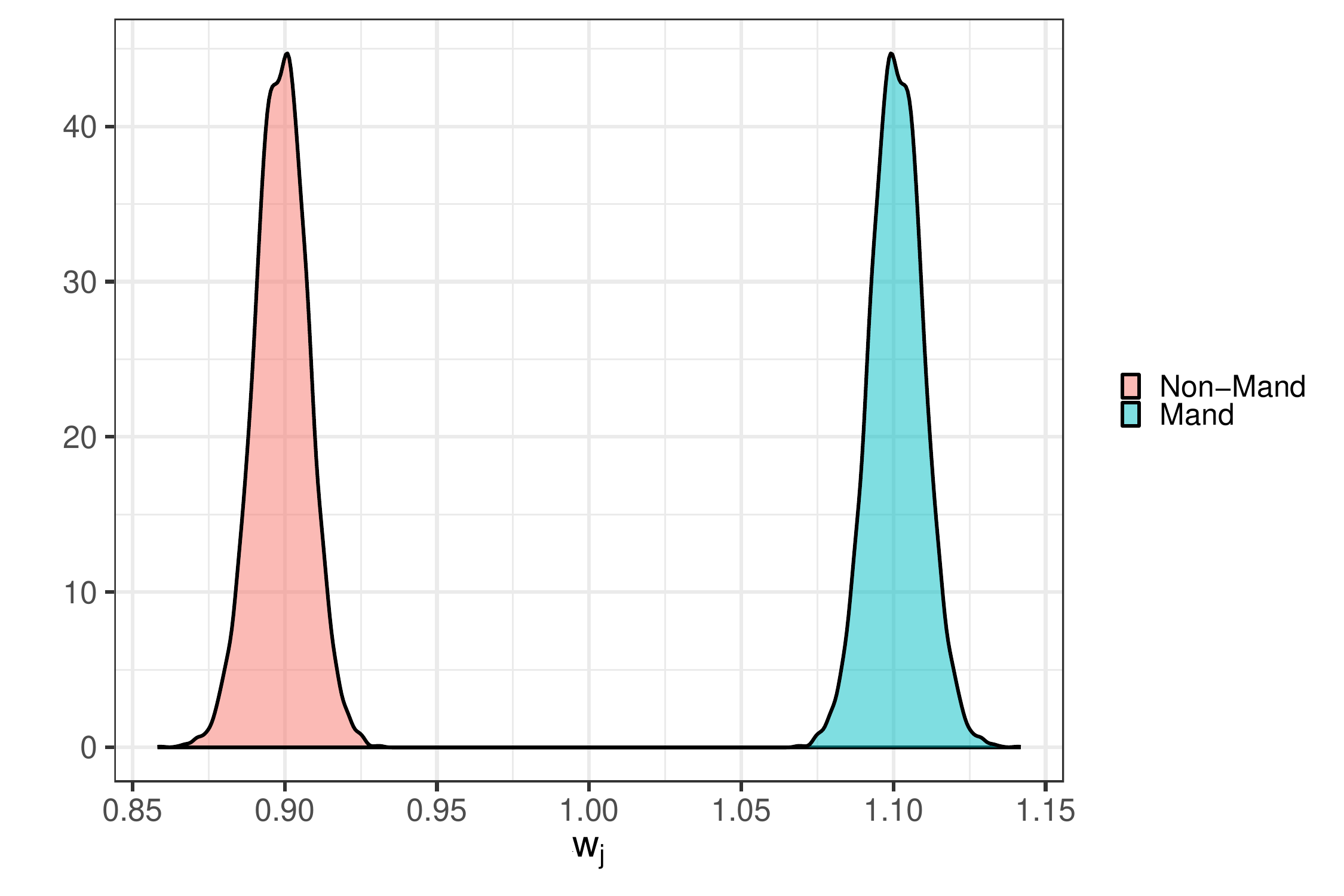}
    \vspace*{-15pt}
	\caption{Results for the tone neural distance data.
    Posterior densities of the post-processed group-level weights $w_{j}$, obtained by averaging $w_{j,h}$ over the latent dimensions.}
	\label{fig:w_j}
\end{figure}

Figures \ref{fig:w_jh} and \ref{fig:w_j} show the posterior distributions of the post-processed weights $w_{j,h}$ and of their group-level averages $w_{j} = \frac{1}{H}\sum_{h=1}^{H} w_{j,h}$, respectively, for the two groups.
Under the normalization adopted in the post-processing step, these weights are centered around $1$.
The posterior densities are fairly concentrated and clearly separated across groups.
In particular, the Mandarin-speaking group has systematically larger weights than the non-Mandarin-speaking group in all three latent dimensions, and therefore also a larger overall group-level weight $w_{j}$.
This is consistent with the larger latent distances observed for the Mandarin-speaking group in the main manuscript.
Moreover, none of the latent dimensions receives a weight equal to or close to zero in either group, indicating that all three dimensions contribute non-negligibly to the fitted latent geometry.

\begin{figure}[!ht]
	\centering
	\includegraphics[width=0.6\linewidth]{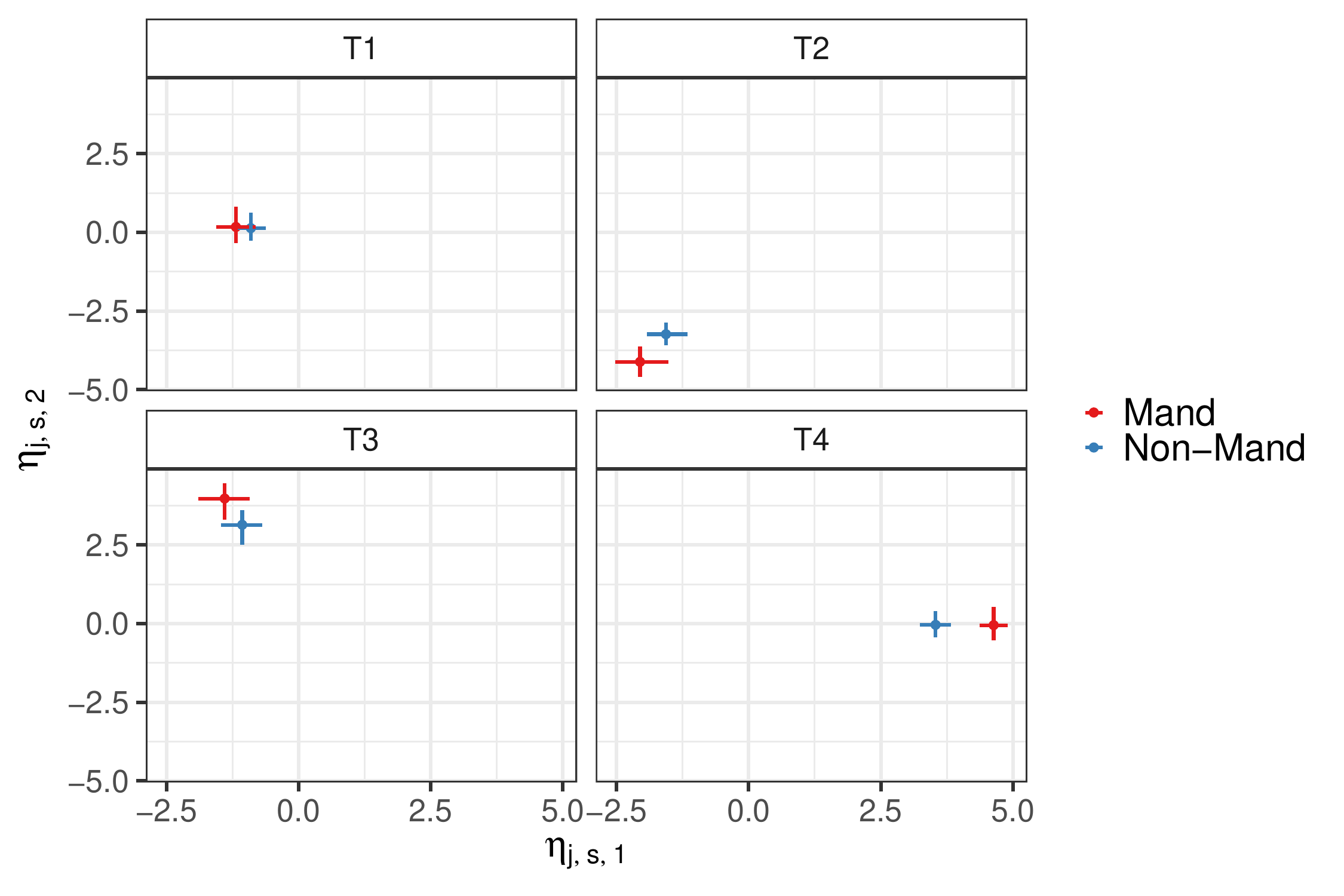}
	\includegraphics[width=0.6\linewidth]{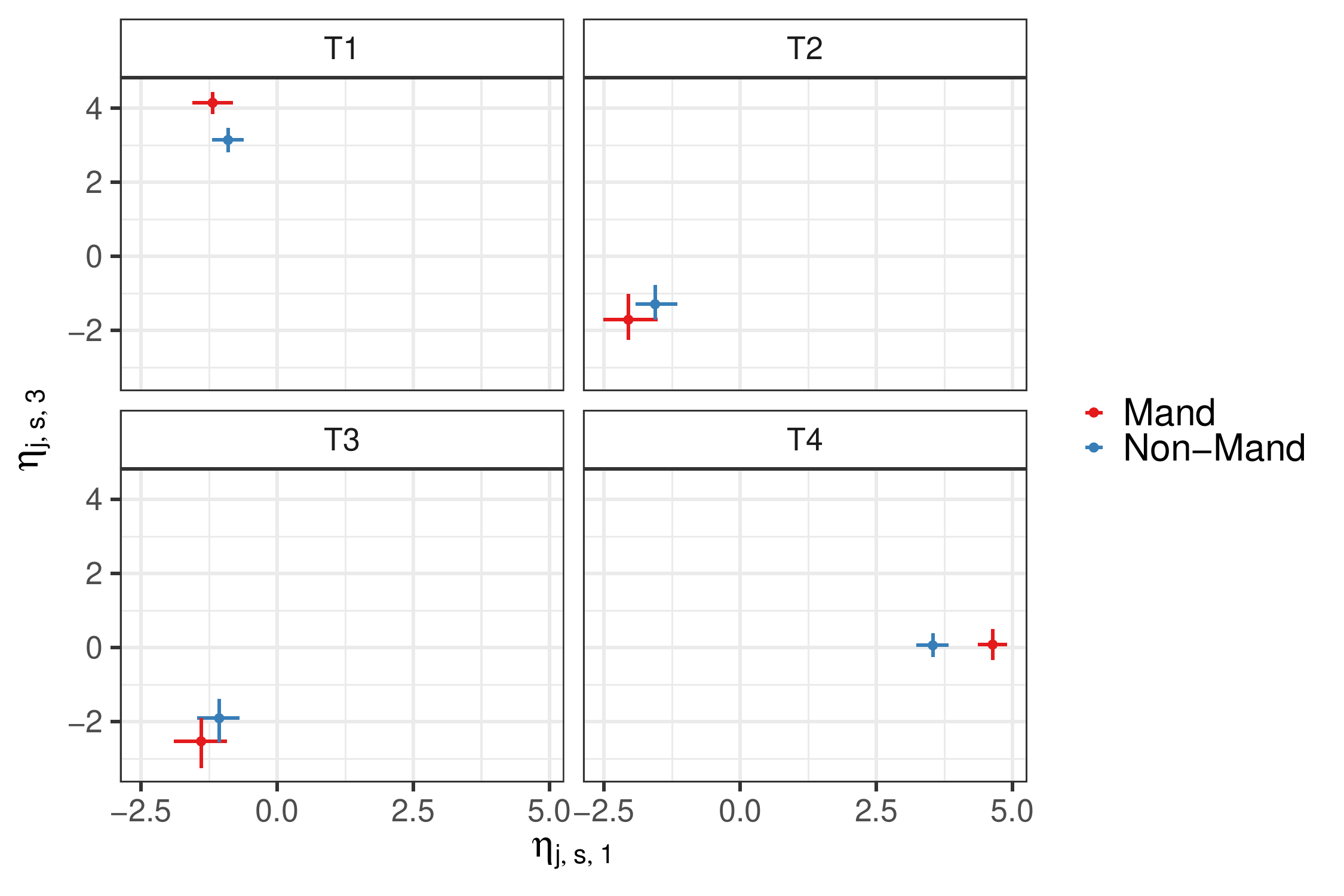}
    \vspace*{-15pt}
	\caption{Results for real data.
    Posterior medians and 90\% credible intervals of the group latent feature values $\eta_{j,s,h}$ in the different tones.}
	\label{fig: Eta_s_alig_avg_real_tone_cred}
\end{figure}

\begin{figure}
	\centering
	\includegraphics[width=\linewidth]{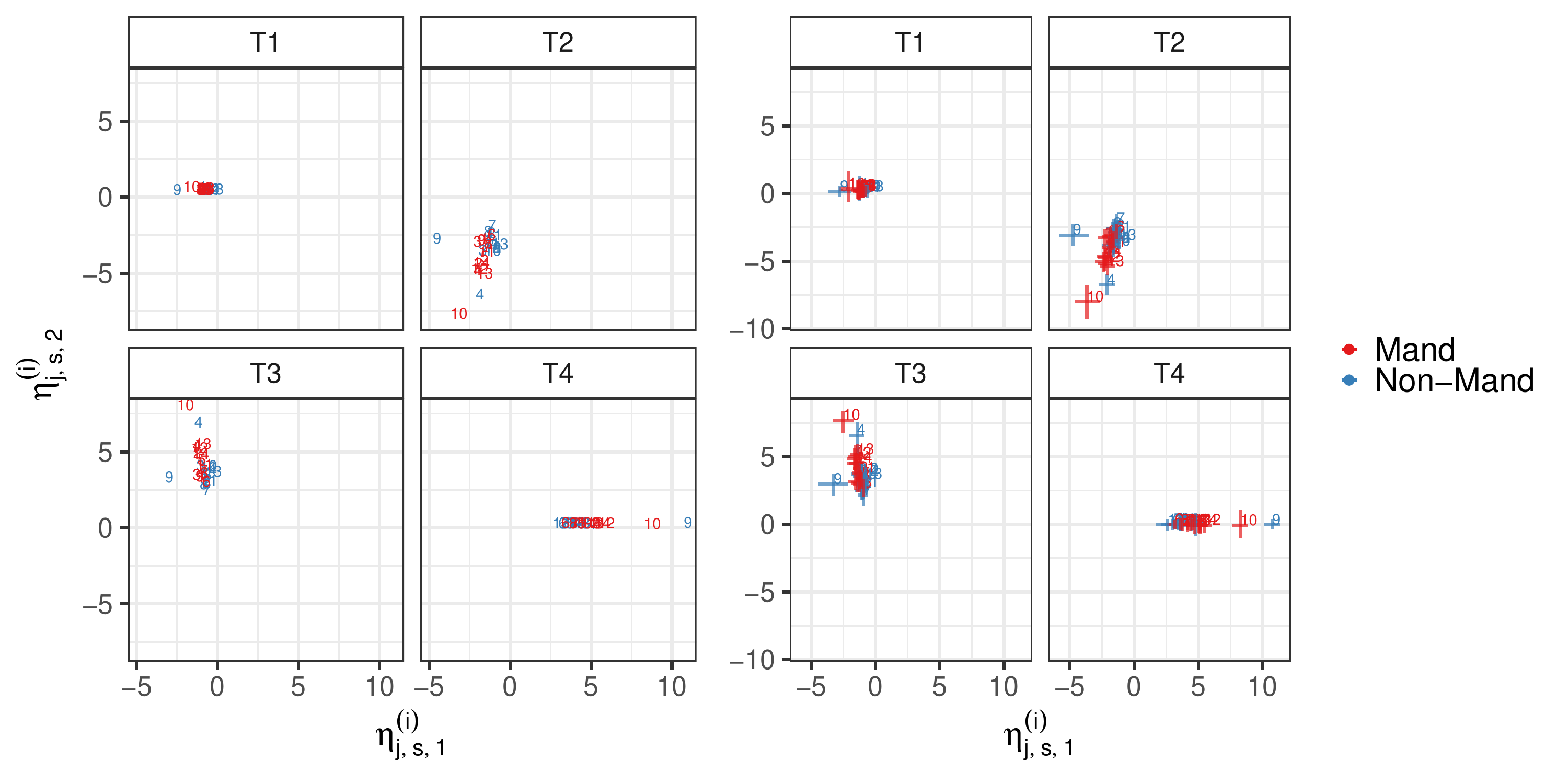}
	\includegraphics[width=\linewidth]{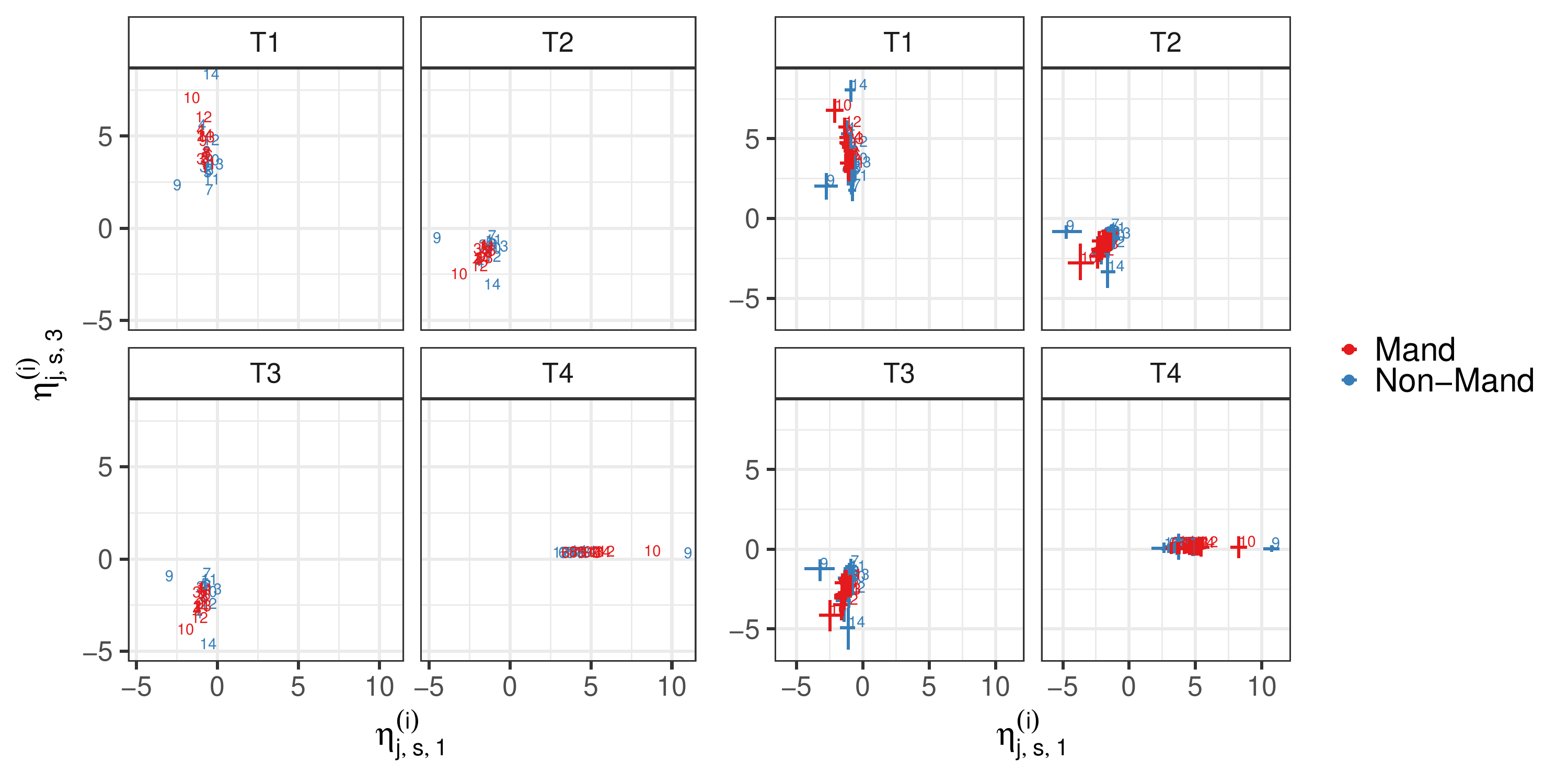}
	\caption{Results for real data.
 Left panels: Posterior medians of the individual latent features $\eta_{j,s,h}^{(i)}$ between stimuli in the different tones.
	Right panels: Posterior medians and 90\% credible intervals of the individual latent feature values $\eta_{j,s,h}^{(i)}$ in the different tones.}
	\label{fig: Eta13_i_alig_all_real_cred2}
\end{figure}

Figures \ref{fig: Eta_s_alig_avg_real_tone_cred} and \ref{fig: Eta13_i_alig_all_real_cred2} show the posterior results for the group and individual latent features, respectively.

\begin{figure}[!ht]
	\centering
	\includegraphics[width=\linewidth]{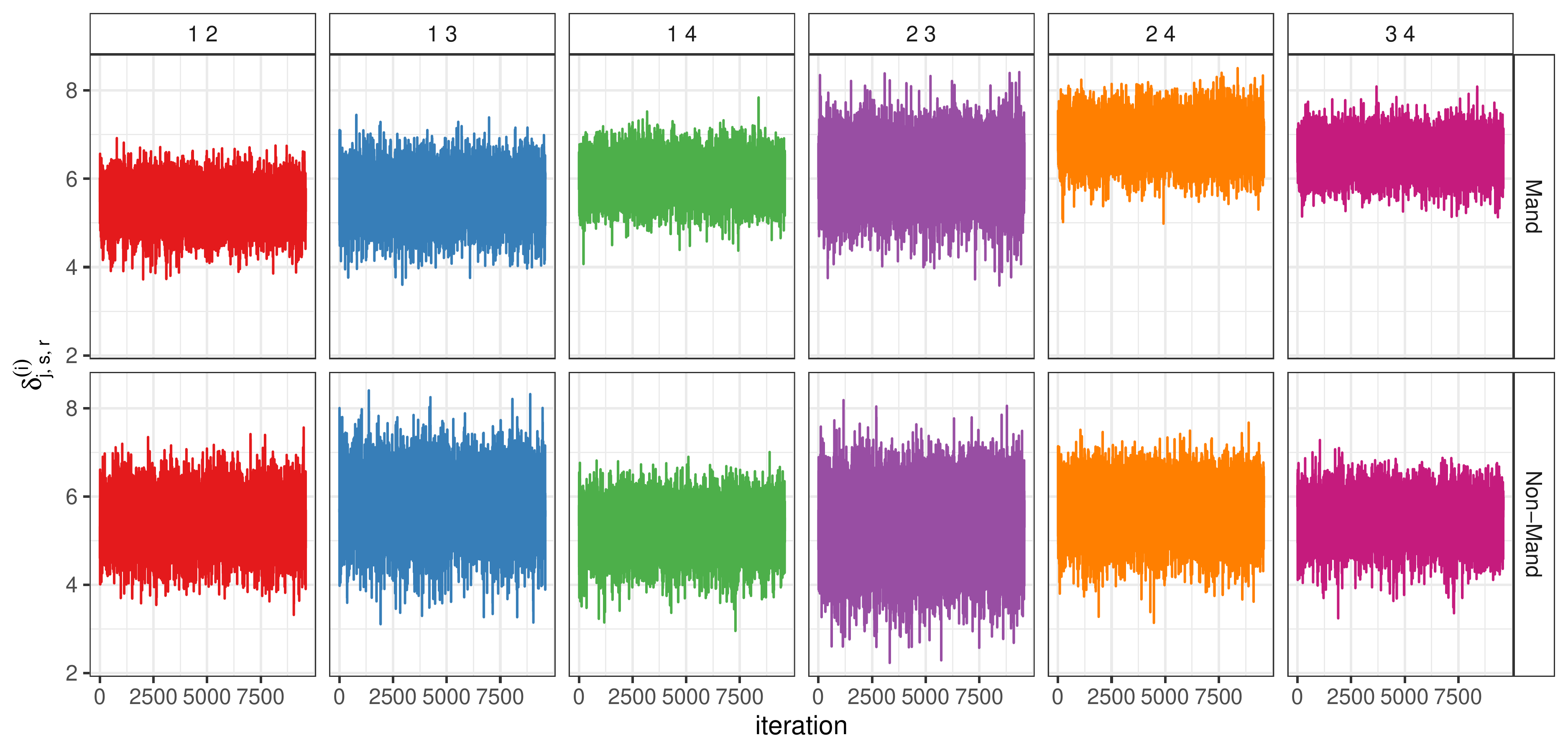}
	\caption[Trace plot $\delta_{j,s,r}^{(i)}$]{Diagnostic for tone neural distance data: Trace plot of latent individual distances between tones, $\delta_{j,s,r}^{(i)}$, in a Mandarin and a non-Mandarin speaker.}
	\label{fig: delta_s_i_Mand_Eng_real}
\end{figure}

\begin{figure}[!ht]
	\centering
	\includegraphics[width=\linewidth]{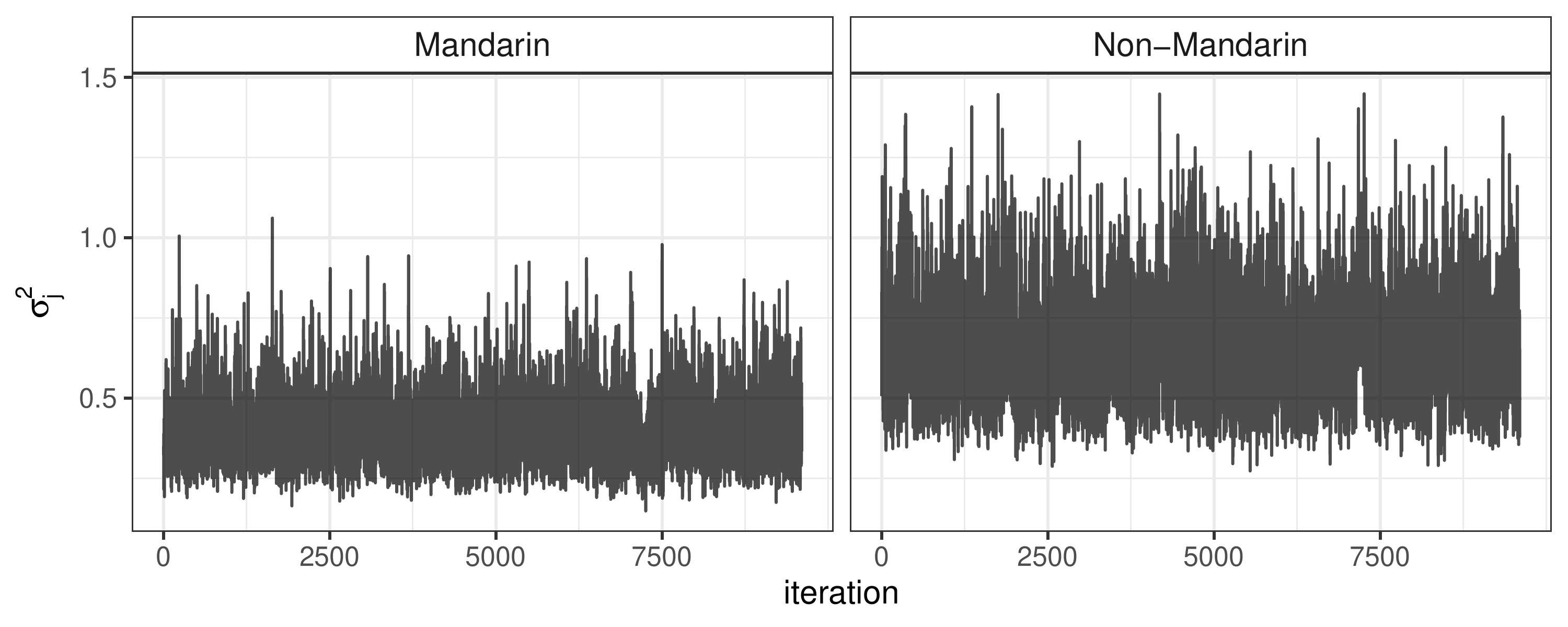}
	\caption[Trace plots of $\sigma^{2}_{j}$]{Diagnostic for tone neural distance data: Trace plots of group-specific variances $\sigma^{2}_{j}$.}
	\label{fig: sigma2_real}
\end{figure}

Finally, Figures \ref{fig: delta_s_i_Mand_Eng_real} and \ref{fig: sigma2_real} show evidence of numerical convergence of the individual distance and group variance parameters.

\section{Additional Figures: Synthetic Data}\label{sec:Add Fig sim}

\begin{figure}[!ht]
	\centering
	\includegraphics[width=\linewidth]{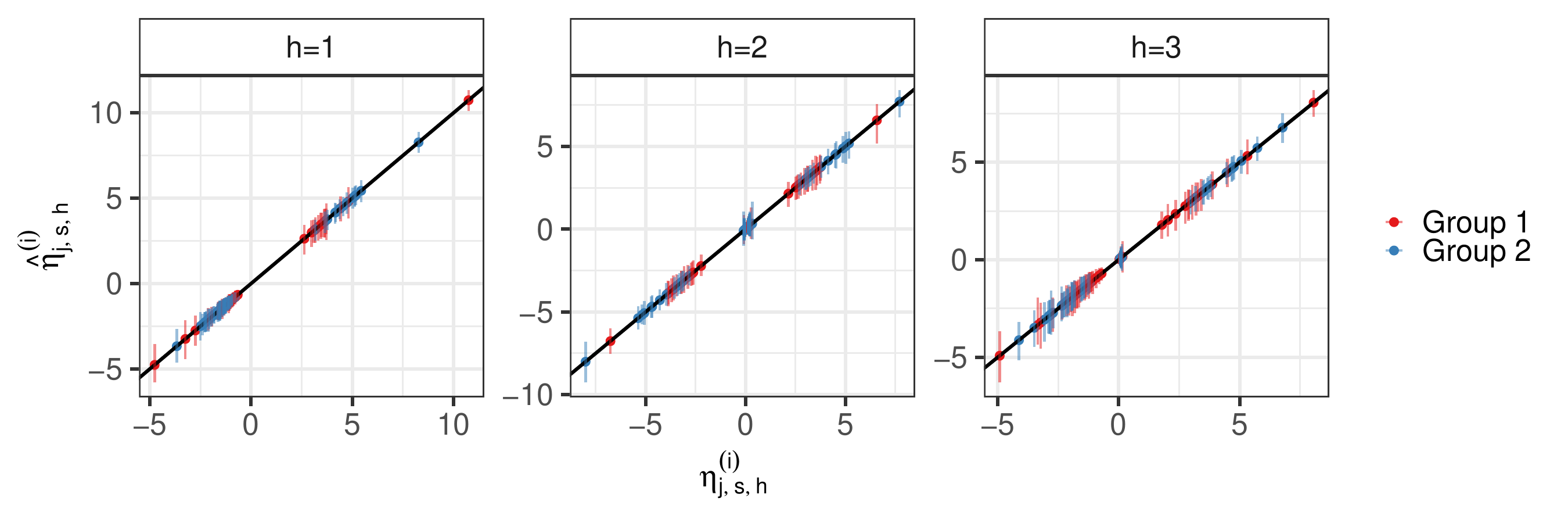}
    \vspace*{-15pt}
	\caption{Results for synthetic data.
    True features $\eta_{j,s,h}^{(i)}$ vs.\ posterior medians and 90\% credible intervals of the latent features $\wh{\eta}_{j,s,h}^{(i)}$.
	}
	\label{fig: Eta_s_i_fake_vs_obs_check_box}
\end{figure}

Figure~\ref{fig: Eta_s_i_fake_vs_obs_check_box} compares the estimated individual latent features $\eta_{j,s,h}^{(i)}$ vs the simulation truth and shows that the model recovers the underlying truth very well.

\section{Parameter Initialization, Convergence and Identifiability Diagnostics}\label{sec: convergence}

\subsection{Initialization}\label{sec:Initialization} 
The results in the main manuscript are obtained by initializing the individual latent features $\eta_{j,s,h}^{(i)}$, the group latent features $\eta_{s,h}$, and the individual weights $w_{j,h}^{(i)}$ (while $w_{j}=1$) in the MCMC chain with INDSCAL estimates obtained using the \texttt{R} package \texttt{multiway}.

\subsection{Convergence Diagnostics}\label{sec:Diagnostics} 
\begin{figure}
	\centering
	\includegraphics[width=0.52\linewidth]{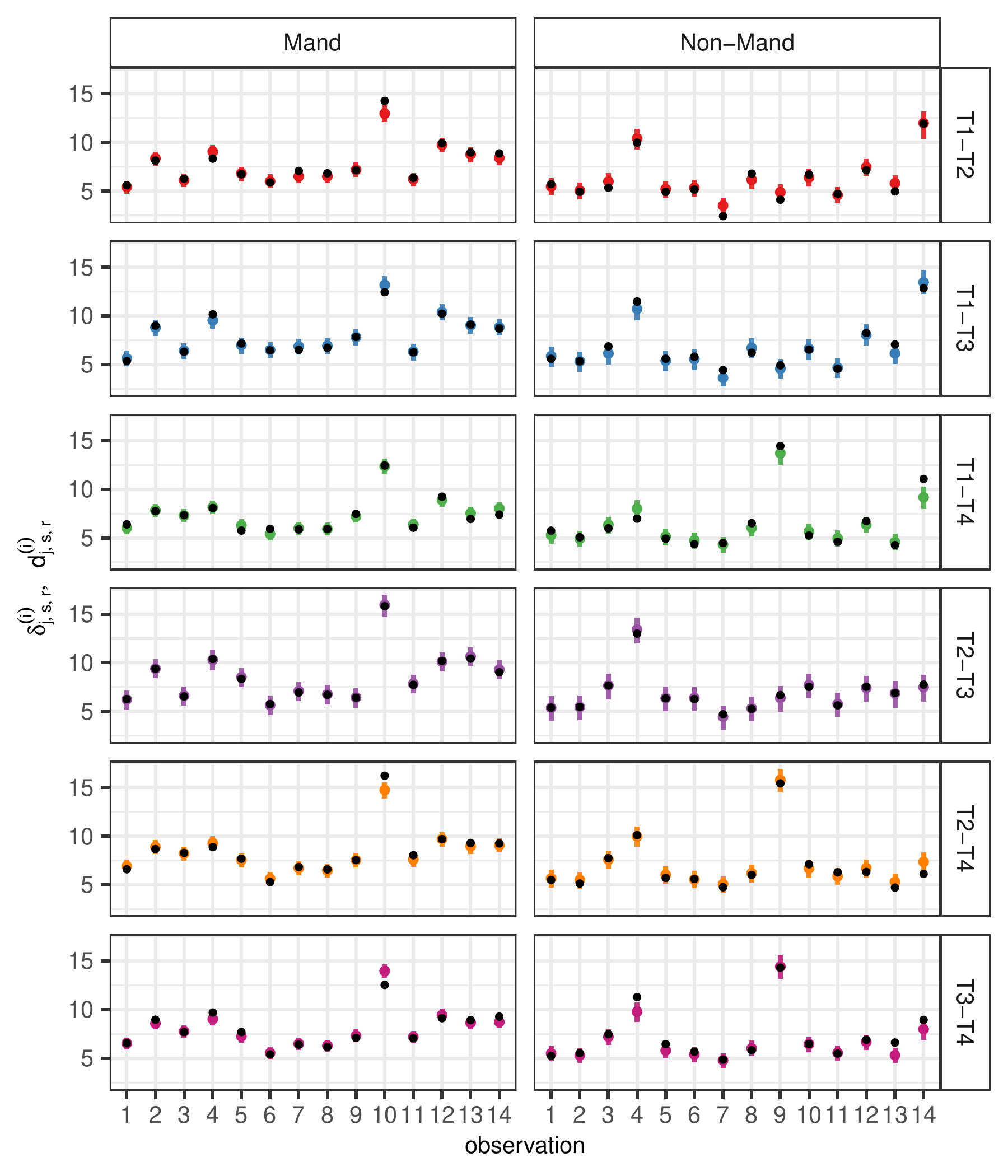}\\
	\includegraphics[width=0.52\linewidth]{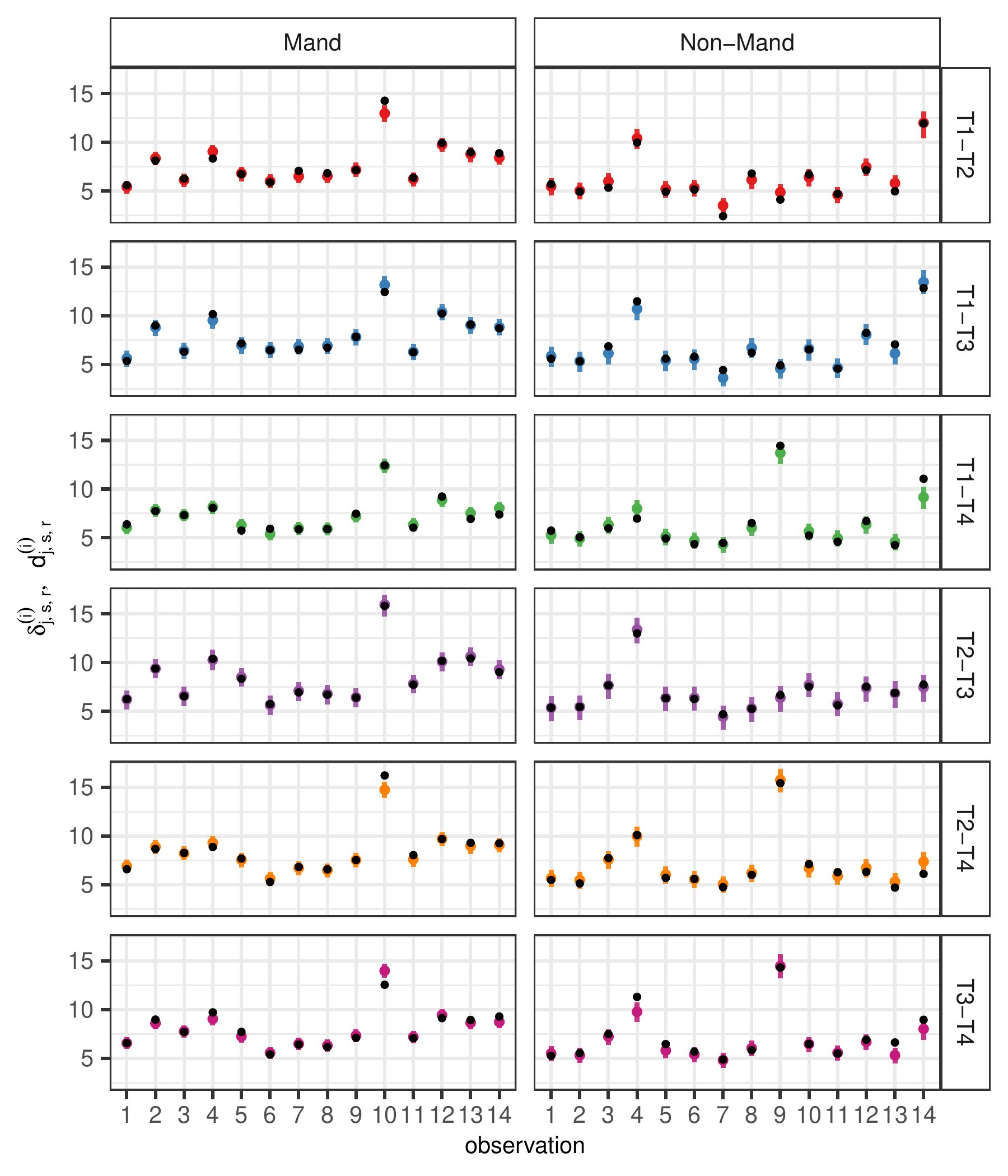}
	\caption{Results for real data.
    Posterior medians and 90\% credible intervals of the individual latent distances $\delta_{j,s,r}^{(i)}$ between stimuli with different MDS initializations.
    Black points represent the observed distances $d_{j,s,r}^{(i)}$.
    }
	\label{fig: delta_i_comp}
\end{figure}

To assess convergence, we ran the analysis also with different MDS initialization schemes  \citep[function \texttt{cmdscale} in \texttt{R},][]{rcoreteam2024r} that do not allow for individual differences, i.e., $\eta_{j,s,h}^{(i)}=\eta_{s,h}$ and $w_{j,h}^{(i)}=w_{j}=1$ for all $i,j,h$.
Specifically, we considered
\begin{itemize}
    \item Computed from the average across \textbf{all subjects'} distance matrices.
    \item Computed from the average across \textbf{Mandarin speakers'} distance matrices.
    \item Computed from the average across \textbf{non-Mandarin speakers'} distance matrices.
\end{itemize}
We show the results for the individual latent distances $\delta_{j,s,r}^{(i)}$ (which are identifiable before post-processing) with different initializations in Figure~\ref{fig: delta_i_comp}.
The results for the different MDS and INDSCAL initializations are practically the same (also compared with Figure~8 in the main manuscript).
Moreover, standard univariate \citep{gelman1992inference} and multivariate \citep{brooks1998general} convergence tests for the individual latent distances $\delta_{j,s,r}^{(i)}$ were all very close to $1$, suggesting good convergence in all cases.

\begin{figure}
	\centering
	\includegraphics[width=0.8\linewidth]{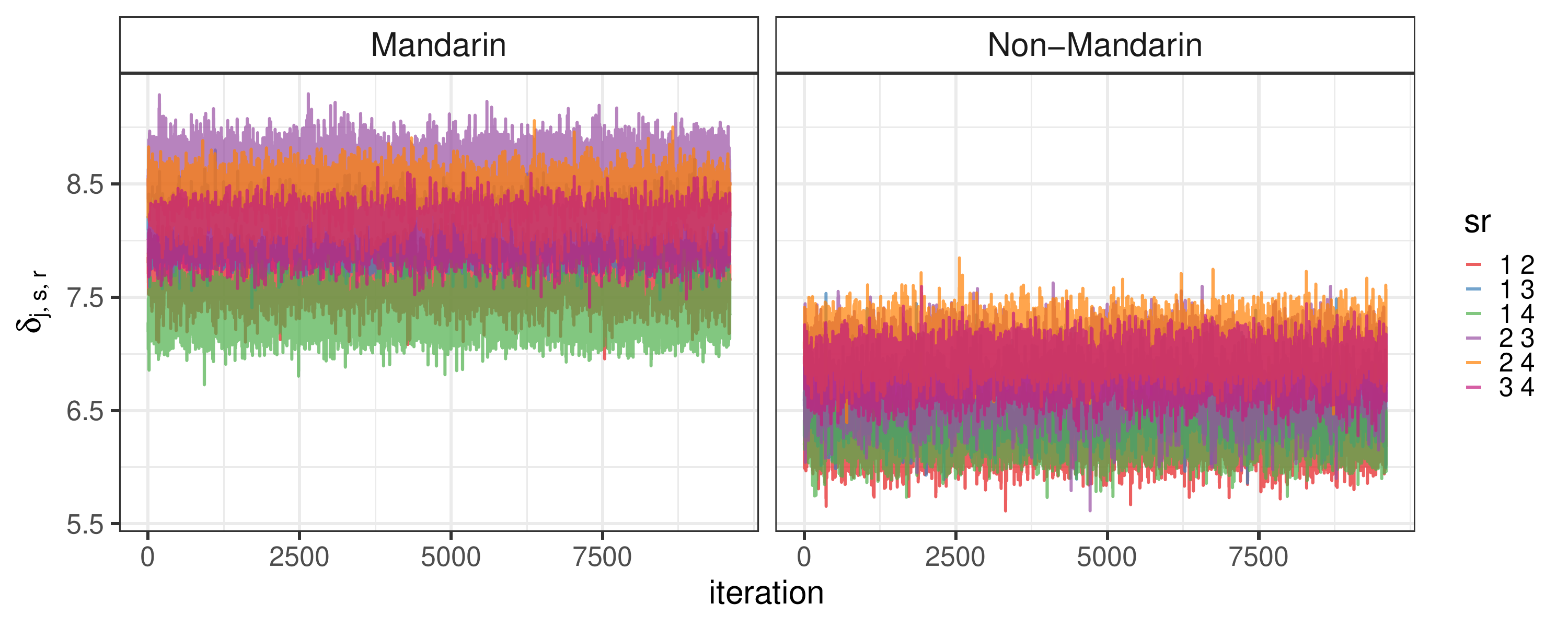}
	\caption[Trace plots of $\delta_{j,1,2}$]{Results for real data. 
    Trace plots of $\delta_{j,s,r}$, the latent distances between different pairs of tones in the two groups.}
	\label{fig:avg_delta_s_real}
\end{figure}

Next, we present some convergence diagnostics for the MCMC sampler for the Mandarin tone neural distance data analysis.
Figure~\ref{fig:avg_delta_s_real} shows the trace plots for the group-specific distances $\delta_{j,s,r}$ between the different tones.
The trace plots suggest a good performance of the MCMC algorithm in sampling from the posterior distributions.
Figure~5 shows the trace plots for the individual tone distances in a Mandarin and a non-Mandarin listener.
Finally, Figure~6 shows the trace plots of group-specific variances $\sigma^{2}_{j}$.
They do not indicate any convergence or mixing issues.

\begin{figure}
	\centering
	\includegraphics[width=0.7\linewidth]{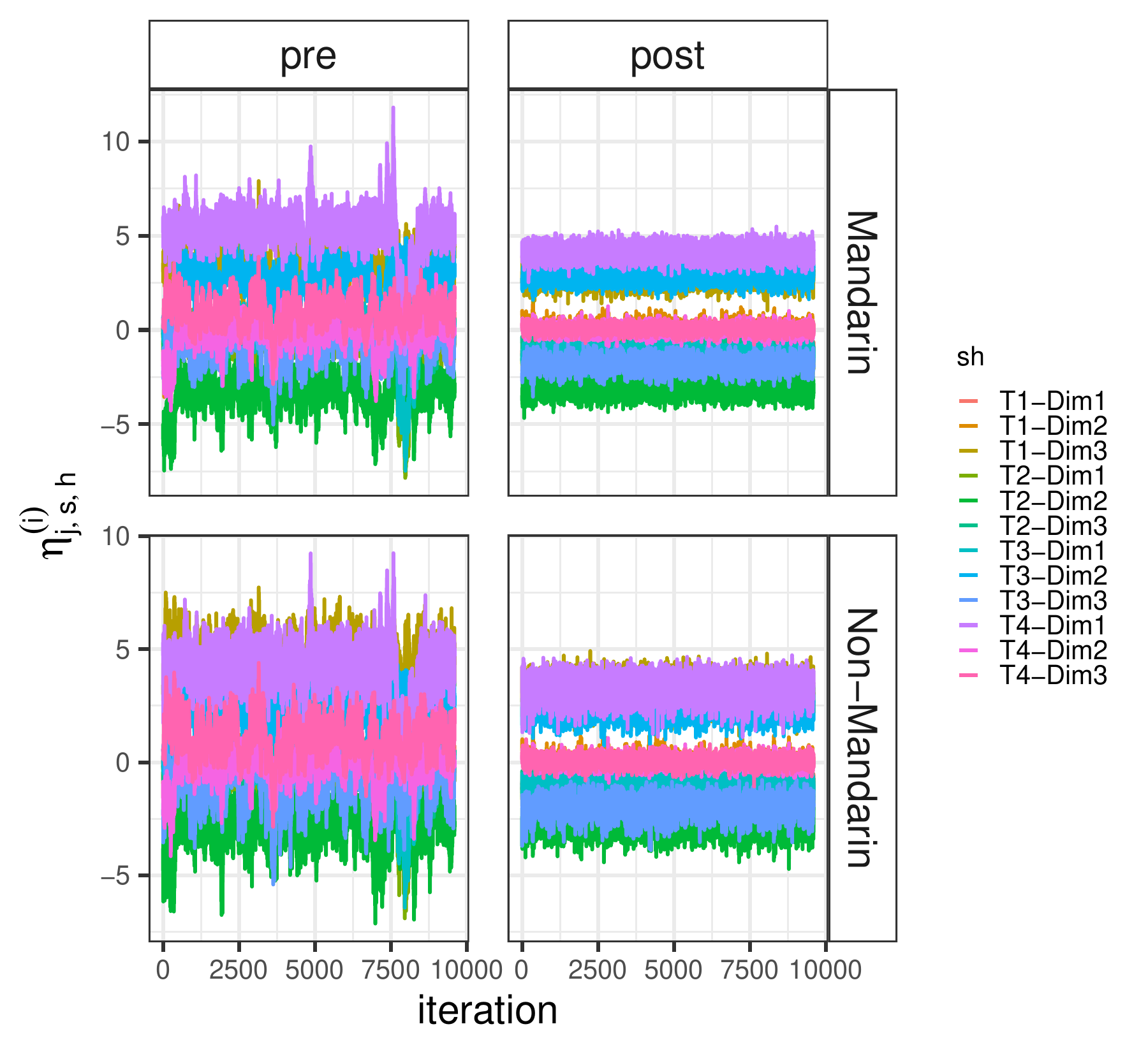}
	\caption{Diagnostic for tone neural distance data.
	Upper panel: Trace plots of the individual features $\eta_{j,s,h}^{(i)}$ for one Mandarin-speaking listener pre- and post-processing.
	Lower panel: Trace plots of the individual features $\eta_{j,s,h}^{(i)}$ for one non-Mandarin-speaking listener pre- and post-processing.}
	\label{fig: Eta_i_pre_ps_real_All}
\end{figure}

\subsection{Identifiability Diagnostics}\label{sec: ident_check}
Posterior inference on the individual latent features $\eta_{j,s,h}^{(i)}$ is only meaningful after solving the identifiability issues explained in Sections~3.1 and 4.3 of the main manuscript.

Figure~\ref{fig: Eta_i_pre_ps_real_All} shows the trace plot of the MCMC samples of the individual latent features $\eta_{j,s,h}^{(i)}$ before and after the post-processing algorithm for one Mandarin-speaking and one non-Mandarin-speaking listener.
The trace plot strongly suggests that the post-processing procedure was able to solve the identifiability issue as expected.

\section{Sensitivity to the Multiplicative Gamma Process Hyperparameters}\label{sec: sensitivity} 
The choice of the hyperparameters of the multiplicative Gamma prior is crucial to inducing dimensionality reduction in a high-dimensional sparse setting \citep{durante2017note}.
In our tone neural distance application with a small number of stimuli $S=4$, the results were very robust to their choices.
Nonetheless, we used the values suggested in \cite{durante2017note} (i.e., $a_{1}=2$ and $a_{2}=3$) to obtain the results reported in the main manuscript.
We also performed a sensitivity analysis considering two additional settings, $a_{1}=a_{2}=2$ and $a_{1}=2$, $a_{2}=1$.
In all the settings, $H=3$ latent dimensions were strongly suggested by the data.
Specifically, the adaptive criterion discussed in Section~4.2 in the main paper suggested sampling from the $H=2$ case in about $5\%$ of the MCMC iterations and from the $H=3$ case in about $95\%$ of the MCMC iterations after the burn-in.
We show the results for the individual latent distances $\delta_{j,s,r}^{(i)}$ (which are identifiable without post-processing) with the two additional hyperparameter settings (i.e., $a_{1}=a_{2}=2$ and $a_{1}=2$, $a_{2}=1$) in Figure~\ref{fig: delta_i_comp_MGP} below.
These results are practically identical and consistent with those reported in Figure~8 of the main manuscript.

\begin{figure}[!ht]
	\centering
\includegraphics[width=0.52\linewidth]{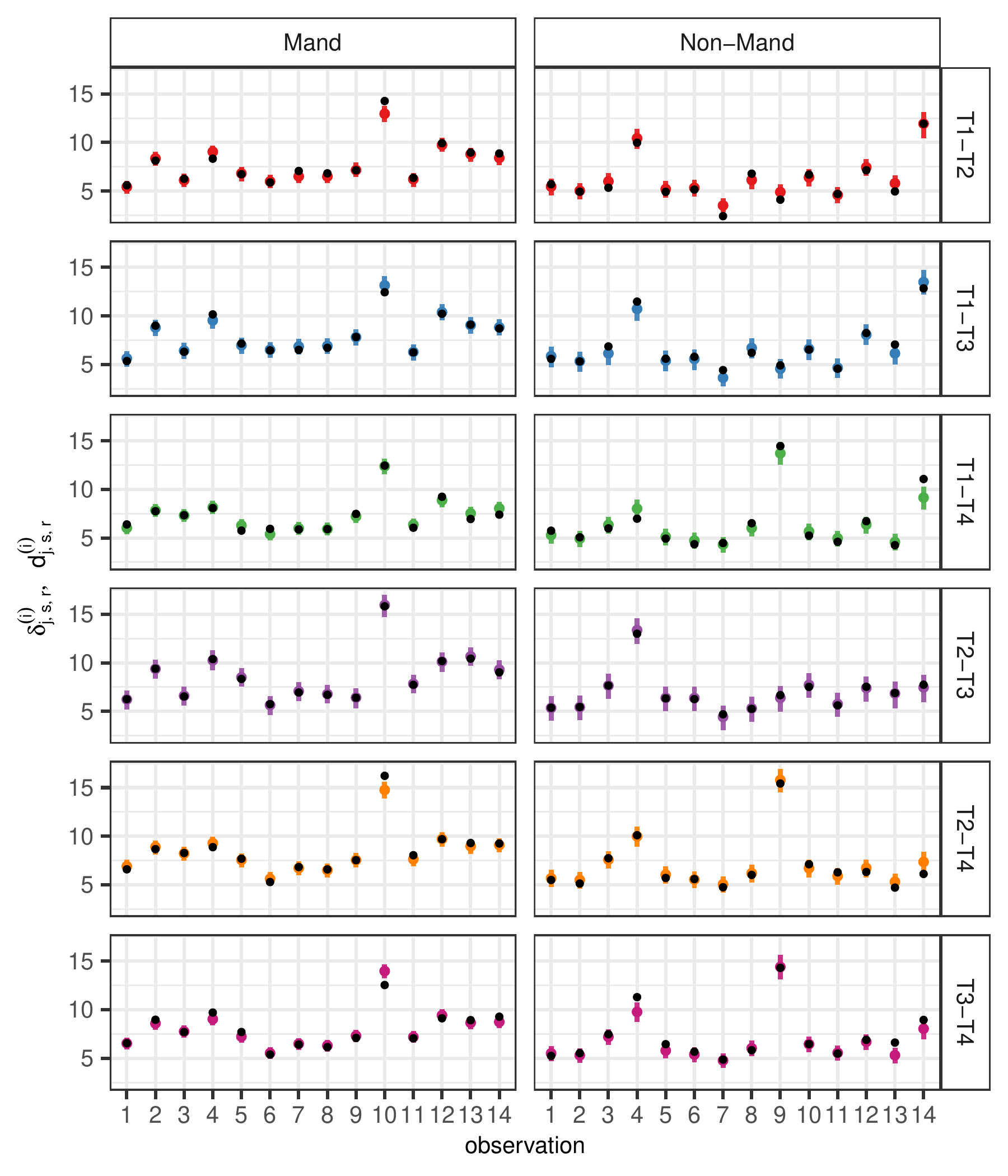}\\
\includegraphics[width=0.52\linewidth]{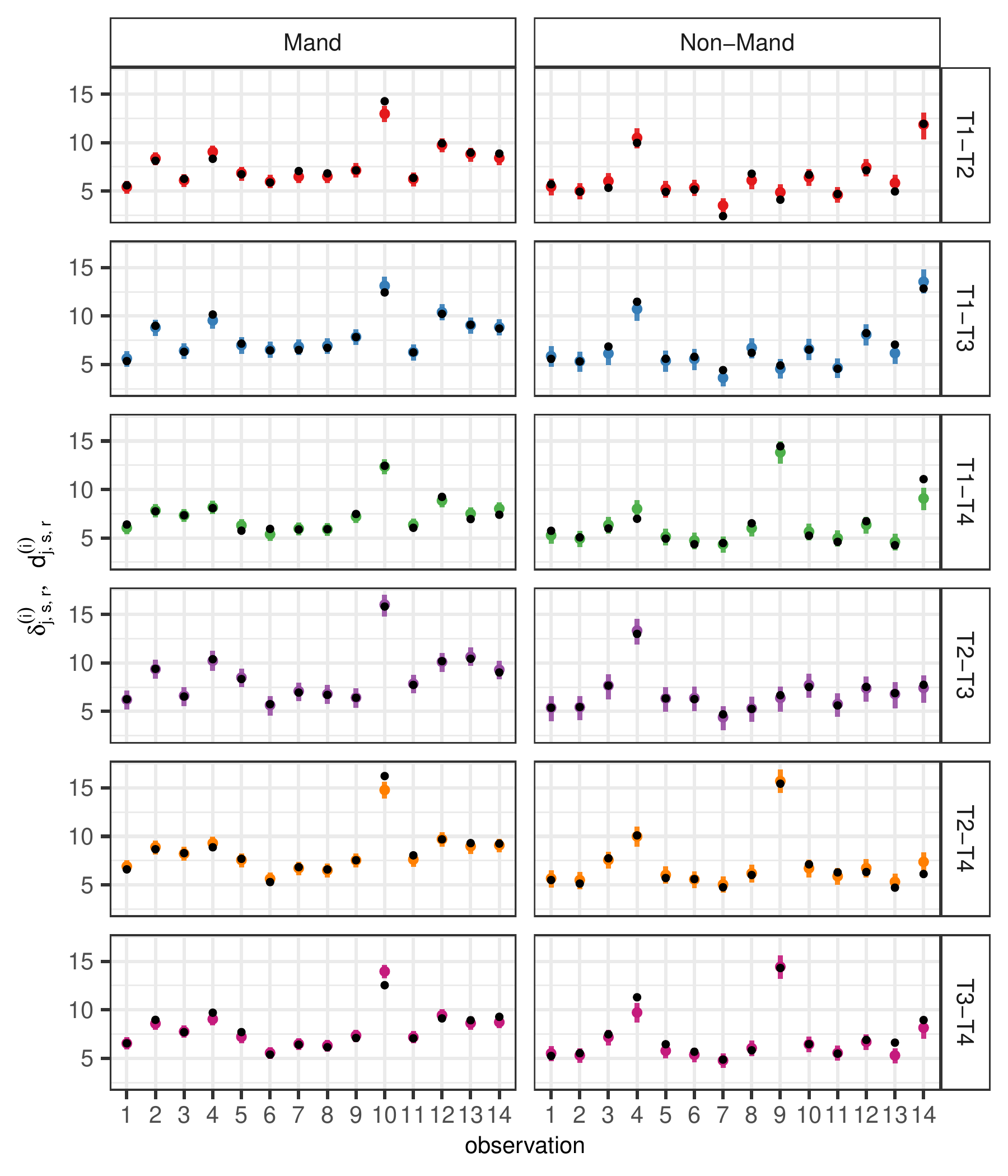}
	\caption{Results for real data. 
    Posterior medians and 90\% credible intervals of the individual latent distances $\delta_{j,s,r}^{(i)}$ between stimuli with different MGP hyperparameters.
    Black points represent the observed distances $d_{j,s,r}^{(i)}$.}
	\label{fig: delta_i_comp_MGP}
\end{figure}

\section{Additional Simulation Scenarios}
\label{sec:supp_sim_additional}

We considered additional simulation scenarios to further assess the recovery of the latent distances and of the true latent dimension $H$.
In all settings, we fixed the number of groups to $J=2$ and the group sizes to $n_{1}=n_{2}=14$, as in our motivating application.
We considered two numbers of stimuli, $S \in \{4,8\}$, and two values for the true latent dimension, $H_{\mathrm{true}} \in \{2,3\}$.
Data were generated from the proposed mixed MDS model, so that only the first $H_{\mathrm{true}}$ latent dimensions were active and the remaining coordinates were set to zero.
For each scenario, we generated $20$ replicated datasets.

The shared latent coordinates were specified to ensure separable stimuli coherently with the motivating application.
For $H_{\mathrm{true}}=3$, the stimuli were placed on vertices of the cube $\{-3,3\}^{3}$: when $S=8$, all eight cube vertices were used, whereas when $S=4$, a subset of four vertices was selected, namely $(3,3,3),\ (3,-3,-3),\ (-3,3,-3),\ (-3,-3,3)$.
For $H_{\mathrm{true}}=2$, the construction was analogous, using the same configurations after dropping the third coordinate.
In all cases, inactive coordinates were set to zero.

Group-specific weights were fixed at $w_{1}=0.95$ and $w_{2}=1.05$.
For each group $j$, the individual-specific weights $w_{j,h}^{(i)}$ were set to be equally spaced between $0.7$ and $1.3$ across the $n_{j}=14$ individuals, and this same sequence was replicated across all latent dimensions $h$.
Given these true parameter values, the latent distances were defined as in the model
\[
    \delta_{j,s,r}^{(i)} = \left\{ \sum_{h=1}^{H_{\mathrm{true}}} \left[ w_{j}\, w_{j,h}^{(i)}\bigl(\eta_{s,h}-\eta_{r,h}\bigr) \right]^{2} \right\}^{1/2},
\]
and the observed dissimilarities were then sampled independently from Gamma distributions with mean $\delta_{j,s,r}^{(i)}$ and variance $\sigma_{\epsilon,j}^{2}=0.1$, for $j=1,2$.

For each replicate, we first estimated the latent dimension using the adaptive model and then fitted the model again by fixing $H$ at the posterior mode selected by the adaptive run, exactly as in the application.
Performance was summarized by the proportion of replications in which the selected dimension matched the truth, $\Pr(\widehat H = H_{\mathrm{true}})$, by the root mean squared error (RMSE) for the latent distances, and by the relative Frobenius error computed from the latent distances.
The latter is also the criterion used in the adaptive step, where dimensions are selected to target a relative Frobenius error smaller than $0.1$.

\begin{table}[ht]
\centering
\begin{tabular}{lcc|cc}
\hline
 & \multicolumn{2}{c|}{$S=4$} & \multicolumn{2}{c}{$S=8$} \\
Metric & $H_{\mathrm{true}}=2$ & $H_{\mathrm{true}}=3$ & $H_{\mathrm{true}}=2$ & $H_{\mathrm{true}}=3$ \\
\hline
Prop$(\widehat H = H_{\mathrm{true}})$& 0.950 & 0.650 & 0.950 & 0.900 \\
RMSE latent distances              & 0.193 & 0.682 & 0.093 & 0.482 \\
Relative Frobenius error           & 0.026 & 0.079 & 0.016 & 0.057 \\
\hline
\end{tabular}
\caption{Simulation summary for the additional well-specified scenarios.
Each entry is averaged over $20$ replicated datasets.}
\label{tab:supp_sim_additional}
\end{table}

Overall, Table~\ref{tab:supp_sim_additional} shows that the model is consistently able to achieve excellent recovery of the latent distances while selecting a parsimonious number of dimensions in a fully adaptive manner.
Indeed, in all scenarios, the true latent distances are reconstructed with a relative Frobenius error below $0.1$, which is precisely the threshold used in the adaptive procedure to select $H$ in a data-driven way.
In all scenarios considered, the model identifies the true latent dimension in the majority of simulations.
These results indicate that the adaptive procedure successfully recovers the true latent dimension in most cases, and selects lower-dimensional representations when they are sufficient to achieve essentially the same geometric recovery.

\section*{References}
Brooks, S.\ P.\ and Gelman, A.\ (1998). General methods for monitoring convergence of\\ \hspace*{0.2cm} iterative simulations. \emph{Journal of Computational and Graphical Statistics}, {\bf{7}}, 434--455.\\
\\
Durante, D.\ (2017). A note on the multiplicative gamma process. \emph{Statistics \& \mbox{Probability}} \hspace*{0.2cm} \emph{Letters}, {\bf{122}}, 198--204.\\
\\
Gelman, A.\ and Rubin, D.\ B.\ (1992). Inference from iterative simulation using multiple\\ \hspace*{0.2cm} sequences. \emph{Statistical Science}, {\bf{7}}, 457--472.\\
\\
R Core Team (2024). \emph{R: a language and environment for statistical computing}. R Foun-\\
\hspace*{0.2cm} dation for Statistical Computing, Vienna, Austria.

\end{document}